\documentclass[aps,prd,twocolumn,a4paper,preprintnumbers,superscriptaddress,10pt,floatfix,nofootinbib]{revtex4-2}
\usepackage{graphicx}
\usepackage{subcaption}
\usepackage{placeins}
\usepackage{multirow}
\usepackage{latexsym}
\usepackage{hyperref}
\usepackage[british]{babel}
\usepackage{amsmath}
\usepackage{float}
\usepackage{xcolor}
\usepackage{dcolumn}
\usepackage{bm}

\usepackage{soul} 

\usepackage{dcolumn} 

\newcommand{\bv}[1]{\mathbf{#1}}

\def\ga{\gamma}
\def\de{\delta}
\def\ep{\epsilon}
\def\ze{\zeta}

\def\la{\lambda}

\def\om{\omega}

\def\pa{\partial}
\def\half{\frac{1}{2}}

\def\bk{{\mathbf{k}}}

\def\bq{{\mathbf{q}}}

\def\bv{{\mathbf{v}}}
\def\bx{{\mathbf{x}}}

\def\bU{{\mathbf{U}}}

\def\mcP{{\mathcal{P}}}

\newcommand{\ben}{\begin{equation}}
\newcommand{\een}{\end{equation}}
\newcommand{\bea}{\begin{eqnarray}}
\newcommand{\eea}{\end{eqnarray}}
\newcommand{\ba}{\begin{array}}
\newcommand{\ea}{\end{array}}
\newcommand{\bit}{\begin{itemize}}
\newcommand{\eit}{\end{itemize}}

\newcommand{\vev}[1]{\langle #1 \rangle}

\newcommand{\aln}{\alpha_\text{n}}
\newcommand{\cs}{c_\text{s}}

\newcommand{\eThermal}{\epsilon_Q}

\newcommand{\Hn}{H_\text{n}}
\newcommand{\GeoFac}{\Gamma}

\newcommand{\Omgw}{\Omega_\text{gw}}

\newcommand{\mcPgw}{\mcP_\text{gw}}

\newcommand{\rinit}{r_\text{i}}

\newcommand{\sinit}{s_\text{i}}

\newcommand{\tbar}{\bar{t}}
\newcommand{\tsh}{t_\text{sh}}
\newcommand{\teddy}{t_\text{ed}}
\newcommand{\tend}{t_\text{end}}
\newcommand{\tdecay}{t_\text{d}}
\newcommand{\tmax}{t_\text{max}}
\newcommand{\tref}{t_\text{ref}}
\newcommand{\tinit}{t_\text{i}}
\newcommand{\Tc}{T_\text{c}}
\newcommand{\Tn}{T_\text{n}}

\newcommand{\Upar}{\bar{U}_{\parallel}}

\newcommand{\UparMax}{\bar{U}_{\parallel,*}}
\newcommand{\UperpMax}{\bar{U}_{\perp,*}}

\newcommand{\Vsw}[1]{\overline{V}_{\!#1}}
\newcommand{\vw}{v_\text{w}}

\graphicspath{{./figures}}

\begin{document}

\title{ Gravitational waves from strong first order phase transitions}
\author{José Correia}
\affiliation{
 Department of Physics and Helsinki Institute of Physics, PL64, FI-00014 University of Helsinki, Finland
}%

\author{Mark Hindmarsh}

\affiliation{
 Department of Physics and Helsinki Institute of Physics, PL64, FI-00014 University of Helsinki, Finland
}%
\affiliation{
 Department of Physics and Astronomy, University of Sussex, Falmer, Brighton BN1 9QH, U.K
}%

\author{Kari Rummukainen}
\affiliation{
 Department of Physics and Helsinki Institute of Physics, PL64, FI-00014 University of Helsinki, Finland
}%

\author{David J. Weir}
\affiliation{
 Department of Physics and Helsinki Institute of Physics, PL64, FI-00014 University of Helsinki, Finland
}%

\date{\today}

\preprint{HIP-2025-17/TH}

\begin{abstract}
We study gravitational wave production at strong first order phase transitions, with large-scale, long-running simulations of a system with a scalar order parameter and a relativistic fluid. 
One transition proceeds by detonations 
with asymptotic wall speed $\vw=0.92$ and transition strength $\aln=0.67$, 
and the other by deflagrations, 
with a nominal asymptotic wall speed $\vw=0.44$ and transition strength $\aln=0.5$.
We investigate in detail the power spectra of velocity and shear stress and - for the first time in a phase transition simulation - their time decorrelation, which is essential for the understanding of gravitational wave production.  In the detonation, the decorrelation speed is larger than the sound speed over a wide range of wavenumbers in the inertial range, supporting a visual impression of a flow dominated by supersonic shocks. 
Vortical modes do not contribute greatly to the produced gravitational wave power spectra even in the deflagration, where they dominate over a range of wavenumbers.   In both cases, we observe dissipation of kinetic energy by acoustic turbulence, and in the case of the detonation an accompanying growth in the integral scale of the flow. The gravitational wave power approaches a constant with a power law in time, from which can be derived a gravitational wave production efficiency. For both cases this is approximately  $\tilde{\Omega}^\infty_\text{gw} \simeq 0.017$, even though they have quite different kinetic energy densities.
The corresponding fractional density in gravitational radiation today, normalised by the square of the mean bubble spacing in Hubble units, for flows which decay in much less than a Hubble time, is
$\Omega_{\text{gw},0}/(\Hn R_*)^2=(4.8\pm1.1)\times 10^{-8}$ for the detonation, and 
$\Omega_{\text{gw},0}/(\Hn R_*)^2=(1.3\pm0.2)\times 10^{-8}$ for the deflagration. 
\end{abstract}

\maketitle


\section{\label{sec:intro}Introduction}

The detection of gravitational waves from black hole and neutron star mergers~\cite{PhysRevLett.116.061102,PhysRevLett.119.141101,PhysRevLett.119.161101,KAGRA:2022twx} has inaugurated the age of gravitational wave astronomy. 
In addition, the detection of signals compatible with a nanohertz stochastic gravitational wave background (SGWB) by the NANOGrav~\cite{NANOGrav:2023hvm} and EPTA~\cite{EPTA:2023fyk} pulsar timing array collaborations, has renewed the focus on gravitational waves as a  window through which to study early universe phenomena~\cite{NANOGrav:2023hvm}.

A possible source of primordial gravitational waves would come from anisotropic stresses generated by first-order phase transitions. Although the Standard Model of particle physics has only cross-overs at the QCD~\cite{Borsanyi:2016ksw} and electroweak~\cite{Kajantie:1996mn,Kajantie:1996qd}  transitions, and would therefore generate no visible imprint on the SGWB, many Beyond the Standard Model theories 
allow first-order phase transitions at - and above - the electroweak scale (see Ref.~\cite{Caprini:2019egz} for an overview, or for some examples~\cite{Profumo:2007wc,Cline:2012hg,Beniwal:2018hyi,Dorsch:2016nrg,Basler:2016obg,Megias:2018sxv}) or at the QCD scale~\cite{Aoki:2017aws,vonHarling:2017yew}; possibly also in a dark sector 
(see e.g.~\cite{Schwaller:2015tja,Addazi:2017gpt,Breitbach:2018ddu,Croon:2018erz}).

Such phase transitions proceed by the nucleation, expansion and merger of bubbles of the low-temperature phase, leaving behind perturbations in the cosmic fluid (see e.g.~\cite{Hindmarsh:2020hop} for a review). The process from bubble collisions onwards generates shear stresses and hence gravitational waves, but the dominant source in transitions which are not extremely supercooled is the emission from long-lived acoustic waves: the sound of the phase transition~\cite{Hindmarsh:2013xza}.

First-order phase transitions at energy scales between $100$~GeV and $1$~TeV are expected to generate SGWB imprints in the mHz frequency range, making them detectable by the future Laser Interferometer Space Antenna (LISA)~\cite{2017arXiv170200786A,Caprini:2019egz,LISACosmologyWorkingGroup:2022jok}. Pulsar timing array experiments could detect gravitational waves from phase transitions at lower energy scales of $100\,$MeV -- $1\,$GeV~\cite{Witten:1984rs} (see also the latest pulsar timing Array constraints in the Erratum of Ref.~\cite{Afzal_2024}).

Emission by acoustic waves in first order phase transitions has been studied with 3-dimensional numerical simulations~\cite{Hindmarsh:2015qjv,Hindmarsh:2017gnf,Auclair:2022jod,Cutting:2019zws,Jinno:2022mie,Dahl:2024eup,Caprini:2024gyk} over the past decade, leading to key insights on the interplay and relative importance of each contributing effect to the expected imprints. These are especially well understood for weak to intermediate strength phase transitions, where these insights have been incorporated into a simple model, the sound shell model~\cite{Hindmarsh:2016lnk,Hindmarsh:2019phv,Sharma:2023mao,RoperPol:2023dzg,Giombi:2024kju} (see also Refs.~\cite{Jinno:2017fby,Konstandin:2017sat}).

The model is not expected to work well for strong phase transitions, where fluid velocities become relativistic. The main reason lies with 
the inaccuracy of the assumption that the fluid velocities and density perturbations are small and that the resulting hydrodynamic flows can be treated in the linear approximation. 
As future observational facilities are more likely to observe stronger transitions than weaker ones, studies of non-linear flows are crucial for improving the quality of current predictions. Numerical studies of transitions with root mean square (RMS) fluid velocities up to around $0.3$ have already revealed important effects, such as suppression of the kinetic energy and GW power, reheating of the metastable phase, the generation and turbulent decay of vorticity, and the dissipation of acoustic modes by shock formation ~\cite{Cutting:2019zws,Dahl:2021wyk,Auclair:2022jod,Dahl:2024eup,Caprini:2024gyk}.

When there is extreme supercooling, the vacuum pressure accelerates the bubble wall to ultrarelativistic speeds and the fluid velocity can also become highly relativistic, and kinetic energy is concentrated in a very thin shell around the wall. No 3-dimensional simulations of this challenging regime have yet been performed, but preliminary modelling in Refs.~\cite{Jinno:2019jhi,Lewicki:2022pdb} suggests that the power spectrum will look similar to that found from collisions of bubble walls in a vacuum.

Characterising non-linear hydrodynamic processes is difficult because of the computational requirements of 3D simulations. One needs to accurately resolve multiple length scales all the way from the bubble wall thickness to the size of sound horizon, which requires a large and finely discretised lattice. Additionally, the whole simulated system must evolve for a sufficiently long time to allow the non-linear processes to operate. 
In the case of transitions proceeding by detonations, where there is no interaction of the pressure waves surrounding the expanding bubbles before they collide, significant savings in memory and time requirements can be found by modelling the order parameter dynamics in the advancing phase front as a discontinuous change in the equation of state, in so-called Higgsless simulations~\cite{Jinno:2022mie,Caprini:2024gyk}.  However, in deflagrations, where the pressure waves in front of expanding bubbles interact with each other~\cite{Cutting:2019zws}, Higgless simulations miss important physical effects such as slowing down of the bubble walls. Still, the latest results from these simulations (Ref.~\cite{Caprini:2024gyk}) highlight the importance of understanding kinetic energy decay for calculations of the  gravitational wave spectrum, especially for strong first order phase transitions.

In this paper, we select two of the strongest phase transitions studied in Ref.~\cite{Cutting:2019zws}, a deflagration and a detonation, and study them in a larger volume and for a longer time. 
The increased volume enables an increased scale separation between the characteristic length scale of the flow and the shock thickness. The longer time gives a clearer view of the decay of the flow, which is due to turbulent energy transport to small scales where energy is dissipated by viscosity. 

We assume that there is no seed magnetic field generated before or during the phase transition, so that the magnetic field may be neglected. Gravitational waves from magnetohydrodynamic turbulence have been studied in Refs.~\cite{RoperPol:2019wvy,Brandenburg:2021bvg}.

We study power spectra of the important quantities of the flow, including the velocity and the shear stress, and the resulting gravitational wave power spectrum, with the aim of laying the groundwork for improved modelling of strong phase transitions. We also study the unequal time correlators (UETCs), which are important for understanding gravitational wave production.

Non-linearities show their importance in several places. 
As already identified~\cite{Cutting:2019zws}, significant vorticity is generated by the sound shell collisions in deflagrations. Further, the compressional velocity fields in both the detonation and the deflagration contain prominent shocks.

The shear stress power spectrum shows the non-linearity in a departure from Gaussianity. Although its shape is well described by a convolution of the two-point functions of the velocity field, the amplitude is underpredicted, significantly so in the case of the detonation where shocks are strongest. The presence of the shocks, where the enthalpy density is positively correlated with velocity, and the relativistic RMS velocities achieved after the phase transition, motivate an analysis with the 4-velocity, weighted by the enthalpy density. This improves the agreement between the convolution and the measured shear stress, but the amplitude is still too low, by a factor of two in the case of the detonation.

The UETCs also show the effect of non-linearities.  We see the expected Gaussian decorrelation with time difference in the vortical flow of the deflagration, as predicted by the Kraichnan random sweeping model~\cite{kraichnan:1964}. In this model, short-distance modes are ``swept'' by the average velocity at the dominant scale of the flow.
We also see sinusoidal behaviour of the compressional mode in both the deflagration and detonation, as would be expected from acoustic oscillations.  However, the sinusoid is modulated by a Gaussian envelope, and the correlations die out after a few oscillations. Furthermore, the oscillation period for a given wavenumber is faster than would be predicted for waves travelling at the sound speed, indicating that the decorrelation is caused by the motion of the shocks. The time decorrelation of the subdominant vortical mode of the deflagration is also interesting, showing a combination of Gaussian and oscillatory behaviour.  We can model this in terms of random sweeping by acoustic oscillations. 

Despite the non-linearities, we show that the shapes of the gravitational wave power spectra are well described by a Gaussian approximation for the weighted 4-velocity field, in which the power spectrum is given by a convolution of velocity power spectra weighted by certain kernels. The rate of growth of the amplitude is also quite well reproduced. 
As shown in Ref.~\cite{Niksa:2018ofa}, compressional and vortical components of the velocity field generate gravitational waves in three modes, according to the combinations of the velocity power spectra which appear in the averaging of the shear stress: fully compressional, fully vortical, and mixed compressional-vortical. We show that even when vorticity comprises approximately half the kinetic energy, as in the deflagration, the gravitational wave power is dominated by fully compressional production, with the mixed mode next in importance.  Purely vortical production is negligible in comparison. The relative efficiency of acoustic production is consistent with the observations in Ref.~\cite{Brandenburg:2021bvg}.

Our simulations are long enough to see the square of the kinetic energy decline by $80\%$, and hence the rate of gravitational wave production, which is proportional to the square of the kinetic energy, is then significantly lower by the end.  At frequencies higher than the peak, the gravitational wave power spectrum saturates, and we are able to make a firm prediction of the final power spectrum.  The gravitational wave power is still growing around the peak frequency, but we can extrapolate the observed growth of the power spectrum to make a more uncertain, but still useful prediction.  We find that the total gravitational power can be expressed in terms of a dimensionless efficiency factor of $\mathrm{O}(10^{-2})$, indicating that a good estimate of the effective lifetime of the flow is given by taking it to last for one shock decay time, computed from the mean bubble spacing divided by the initial RMS velocity. A similar efficiency factor has also been observed in smaller Higgsless simulations~\cite{Caprini:2024gyk}.

Further work is needed to characterise the power spectrum around the peak, and at lower frequencies, where the growth in the characteristic flow length scale which accompanies the kinetic energy decay is expected to have an impact~\cite{Dahl:2021wyk,Auclair:2022jod,Dahl:2024eup}.

This paper is divided as follows: 
a summary of the acoustic generation of gravitational waves and key observables (Section \ref{sec:grav}); 
methodology of the simulation setup, with additional details added in appropriate appendices (Section \ref{sec:sim_setup});
the main results (Section \ref{sec:results});
analysis of the results in terms of a model of acoustic production (Section \ref{s:GWpsVelps}); 
 and a conclusion summarising the manuscript and outlining possible next steps (Section \ref{sec:conc}). Scripts used for the figures in this manuscript, and the accompanying data (sans simulation slices) are available at ~\cite{correia_2025_16760997} and ~\cite{correia_2025_16760912}, respectively.

\section{\label{sec:grav}First-order phase transitions }

We begin by reviewing the thermodynamics and hydrodynamics of a thermal first-order phase transition as might have taken place in the early universe (see e.g.~\cite{Mazumdar:2018dfl,Hindmarsh:2020hop,Athron:2023xlk}). 
The transition is driven by a scalar field, for example the Higgs field, or an effective scalar field, functioning as an order parameter. This is coupled to other fields, such as those of the Standard Model. The excitations of these fields can be approximated as a relativistic fluid whose equation of state depends on the order parameter.

As the universe cools, the free energy of the system develops two minima as a function of the order parameter, corresponding to the two possible phases of the system. 
At the critical temperature $\Tc$, the free energies  of the two phases are equal. 
When the temperature drops below the critical temperature $\Tc$, an energy barrier prevents the homogeneous evolution from one phase to another. 
The transition proceeds instead by the nucleation of bubbles of the stable low-temperature  phase in the metastable high-temperature phase. The bubbles expand and collide, and eventually the stable phase fills the entire volume.

As the bubbles expand, excess free energy in the metastable phase is transferred to shells of moving and heated fluid via the coupling between the order parameter and the fluid. This  effective dissipative coupling models the friction as particles change their masses when crossing the wall~\cite{Bodeker:2017cim,Bodeker:2009qy}, which causes a departure from equilibrium. The coupling dictates the expansion speed of the bubble wall. If this coupling is negligible and therefore the frictional force on the fluid is insufficient to stop the walls from becoming ultrarelativistic, the transition is known as a ``runaway''~\cite{Bodeker:2017cim,Bodeker:2009qy}. This scenario has been simulated in Refs.~\cite{Cutting:2018tjt,Lewicki:2020jiv,Cutting:2020nla,Lewicki:2020azd}, but is not considered here.  

If the wall speed is subsonic, the  hydrodynamic solution is a compression wave in front of the bubble wall with a leading shock, known as a deflagration~\cite{landau1987fluid}. For walls which are supersonic, but whose speed is lower than a critical threshold (known as the Chapman-Jouguet speed), a rarefaction wave develops behind the wall, while the compression wave in front still exists.  Past this limiting speed, the wall catches up with the leading shock, and thus for faster wall speeds the fluid disturbance is a detonation, a pure rarefaction wave behind the bubble wall. 

After the transition is complete, the shells of compression and rarefaction continue to propagate as acoustic waves.
 The interaction of the shells and, after the completion of the transition, the acoustic waves, produces gravitational waves. 
Unless the RMS fluid velocities in the fluid are of order unity, the acoustic waves last much longer than the duration of the phase transition itself and are the dominant source of gravitational waves in a first order phase transition~\cite{Hindmarsh:2013xza,Hindmarsh:2015qjv,Hindmarsh:2015qta}. 

The interaction of pairs of shells also produces vorticity in the form of vortex rings near their intersection~\cite{Cutting:2019zws}.  For deflagrations with RMS velocities around $0.3$ the vortical component of velocity can be as large as the compressional component.

The subsequent evolution of the velocity field is determined by non-linear effects, which transfer power to short distance scales, in the form of shocks for the compressional modes, and turbulent eddies for the vortical modes. Both types of energy transfer are often referred to as turbulence. The timescales for the development of both kinds of turbulence is the ratio of the dominant length scale of the fluid flow to the relevant RMS velocity.  We will refer to these timescales as the shock time for acoustic turbulence, and the eddy turn-over time for vortical turbulence.  We define them more precisely at the end of this Section.

The generation of gravitational waves by acoustic modes and by turbulent flows are sometimes viewed as separate mechanisms (often termed acoustic and Kolmogorov turbulence), and turbulence is sometimes implicitly viewed as entirely vortical. 
The separation is somewhat artificial, and overlooks the fact that the evolution from the acoustic phase to a turbulent one is a continuous process. 

As mentioned in the Introduction, studying the onset and evolution of turbulence requires letting simulations evolve for a sufficiently long time (i.e. sufficient shock or eddy turn-over times). Valuable insights have already been gained from long-running simulations in Refs.~\cite{Dahl:2021wyk,Dahl:2024eup,Auclair:2022jod,Brandenburg:2017neh,Jinno:2022mie,Caprini:2024gyk}, but without seeding from a phase transition mediated by a dynamical scalar field as in this work.

In the following sections we define the physical system we are simulating, 
and the observables we measure in order to characterise the fluid flows and understand the gravitational wave production.

\subsection{Effective theory of a first order phase transition}

We use the ``bag model'' equation of state for the fluid, which gives a very simple realisation of a system with a first order phase transition.  
It has the advantage that it can achieve any hydrodynamically allowed combination of transition strength $\alpha_\text{n}$ and wall speed $\vw$. The sound speed $\cs$ is always  $1/\sqrt{3}$, which is unrealistic, but it allows one to decouple the effect of changing sound speed.  Another unrealistic feature is that both phases exist at all temperatures. We postpone studies of equations of state with different sound speeds and a limited metastability temperature range in order to focus on the non-linear fluid dynamics. 

We denote the order parameter of the transition, assumed to be a single real scalar field, as $\phi$.
Following Ref.~\cite{Cutting:2019zws}, we will use the effective potential (the field-dependent part of the free energy density)
\begin{equation}
    V(\phi, T) = V_0(\phi) - (a(\phi)-a_0)T^4  \; ,
\end{equation}
where $a(\phi)$ is a function modelling the effective relativistic degrees of freedom as a function of the field value.

In this expression, 
$a_0 \equiv a(0)$, $a_0 = g_* \pi^2 / 30$, and $g_*$ being the effective number of relativistic degrees of freedom in the symmetric phase. 

The zero-temperature effective potential is given by
\begin{equation}
V_0(\phi) = \frac{1}{2} M^2 \phi^2 - \frac{1}{3}\mu \phi^3 + \frac{1}{4}\lambda \phi^4 -V_c \;,
\end{equation}
where $M$, $\mu$ and $\lambda$ are parameters that control the phase transition, and $V_c$ is a constant which is chosen to set the potential at zero temperature in the broken phase to vanish. 
In the model the degrees of freedom function is 
\begin{equation}
a(\phi) = a_0 - \frac{\Delta V_0}{T_c^4} \bigg[ 3 \bigg( \frac{\phi}{\phi_b} \bigg)^2  - 2 \bigg( \frac{\phi}{\phi_b} \bigg)^3 \bigg] ,
\end{equation}
where 
$\phi_b$ is the order parameter at the $\phi \ne 0$ minimum of $V_0$, and $\Delta V_0 = V_0(0) - V_0(\phi_b)$. Hence $V(0,\Tc) = V(\phi_b,\Tc)$, as required. By construction then, $\Delta V_0 = V_c$. The consequences of this thermodynamic model for phase transition simulations were briefly examined in Ref.~\cite{Cutting:2022zgd}.

The fluid energy density and pressure receive a contribution from the scalar field driving the phase transition, via the effective potential,
\begin{align}
\epsilon(T, \phi) &= 3a(\phi) T^4 + V_0(\phi)\; , \\
p(T, \phi) &= a(\phi) T^4 - V_0(\phi)\; .
\end{align}
We define the thermal energy density 
$\eThermal = 3 w / 4$, where $w = \ep + p$ is the enthalpy density,
and the trace anomaly as 
\begin{equation}
\theta(\phi,T) = \frac{1}{4} \bigg( \epsilon(\phi,T) - 3p(\phi,T) \bigg)\; ,
\label{e:TraAno}
\end{equation}
which is equal to $V_0(\phi)$ in the bag model. 
We define the transition strength parameter as the ratio of the trace anomaly difference between the two phases to the thermal energy, or 
\begin{equation}
    \alpha_\text{n} = \frac{1}{\eThermal (\Tn)} \bigg( \theta(0, \Tn) - \theta(\phi_b, \Tn) \bigg) \; ,
\end{equation}
where $\Tn$ is the nucleation temperature, which we define as the temperature at the maximum volume-averaged bubble nucleation rate.\footnote{Other choices of notation exist. A popular definition of the nucleation temperature is the temperature at which on average one bubble is nucleated per Hubble volume. Furthermore, some authors denote by $T_*$ what we call $\Tn$ in the present work.} In exponential nucleation, where the nucleation rate density increases exponentially with time after the critical temperature is reached,  this coincides with there being a volume fraction of $1/e$ in the broken phase. In the simulations described herein, all bubbles are nucleated simultaneously at $t=0$. 

The bubble nucleation process results in a length scale $R_*$, equal to the inverse of the bubble number density, counted after the transition has completed and nucleation has stopped.  It can be viewed as the mean spacing between the bubble centres. 

We denote the Hubble rate at the nucleation temperature $\Hn$, related to the energy density through the Friedmann equation 
\begin{equation}
\Hn^2 = \frac{8\pi G}{3} \bar{\ep}_0,
\end{equation}
where $\bar{\ep}_0  \equiv \epsilon(0, \Tn)$ is the initial rest energy density in the metastable phase. 
Our assumption that the transition is rapid implies that the mean spacing is much less than the Hubble length, or $\Hn R_* \ll 1$.

Now that the thermodynamics of the phase transition have been described, we can move towards the equations of motion of the field-fluid system.  We use the metric convention $(-\, +\, +\, +)$. If the fluid and the field are uncoupled, we can define corresponding energy-momentum tensors
\begin{align}
T^{\mu \nu}_\text{f} &= w u^\mu u^\nu + g^{\mu \nu}p \label{eq:em_tensor_fluid} \\
T^{\mu \nu}_\phi &= \partial^\mu \phi \partial^\nu \phi + 
    \frac{1}{2}g^{\mu\nu} (\partial \phi)^2\,, \label{eq:em_tensor_field}
\end{align}
where $u^\mu$ is the fluid 4-velocity. 
When we introduce coupling between the field and the fluid, the total energy-momentum conservation implies
\begin{align}
\partial_\mu T^{\mu\nu} = \partial_\mu T^{\mu \nu}_\phi + \partial_\mu T^{\mu \nu}_\text{f} =0 \,.
\end{align}
Following Refs.~\cite{Ignatius:1993qn,Kurki-Suonio:1995yaf}, we introduce 
a phenomenological coupling between the fluid and the field with no temperature dependence on the function of the coupling, whose purpose is to enable energy transfer from the field to the fluid, and hence drives the bubble walls to their asymptotic velocities. This is such that 
\begin{align}
\partial_\mu T^{\mu\nu}_\phi = -\partial_\mu T^{\mu\nu}_\text{f} = \eta u^\mu \partial_\mu \phi \partial^\nu \phi  .
\label{eq:coupling2}
\end{align}
Here $\eta$ is a tunable coupling parameter. 
The equation of motion for the field is 
\ben
\Box \phi - V_{,\phi}(\phi, T) = \eta u^\mu  \pa_\mu \phi ,
\een
consistent with the energy-momentum conservation equations. 
This simple model has been successfully used in previous simulations~\cite{Cutting:2019zws}.

Note that in the simulations, the gravitational waves are sourced by the fluid only.  The sourcing by the scalar field is negligible for macroscopic bubbles, unless the transition is so strongly supercooled that the available energy of the transition is entirely converted into the accelerated motion of the bubble wall~\cite{Cutting:2018tjt,Cutting:2020nla,Lewicki:2019gmv,Lewicki:2020jiv,Lewicki:2020azd}.

\subsection{\label{sec:uetc}(Un)equal time correlators}

In order to relate the longitudinal and transverse fluid velocities to the generation of shear stress and conversely gravitational waves, we will define several two point correlation functions. Throughout this manuscript we will adopt the convention that $P_Y (k,t)$ of a field variable $Y(\bx,t)$ corresponds to the spectral density of the spatial Fourier transform of the field, which is assumed to be a random variable drawn from a homogeneous and isotropic distribution. For example, the spectral density of the 3-velocity field with components $v^i(\bx,t)$ is defined through
\begin{equation}
(2\pi)^3 \delta(\bk-\bk') P_v (k, t) = \langle v^{i}(\bk,t) v^{*i}(\bk',t) \rangle.
\end{equation}
We define the power spectrum through 
\ben
\mathcal{P}_Y(k,t) = \frac{k^3}{ 2\pi^2 } P_Y(k, t) 
\een
Hence Parseval's theorem is expressed as $\int \mathrm{d}k\, \mathcal{P}_Y(k,t)/k = \langle Y^2(\bx,t) \rangle$. 
We we also refer to the spectral density as an equal time correlator (ETC).  Note that the power spectrum has the same dimensions as $Y^2$, that is $[\mcP_Y] = [Y]^2$, while $[P_Y] = {\sf L}^3 [Y]^2$.

Vector correlators can be decomposed into longitudinal (compressional) and transverse (vortical) components with the projectors
\begin{equation}
\la^\parallel_{ij}(\bk) = \hat{k}_i \hat{k}_j, \quad  
\lambda^\perp_{ij}(\bk) = \delta_{ij} - \hat{k}_i \hat{k}_j .
\end{equation}
Hence the spectral densities of the components of the 3-velocity field are 
\begin{equation}
(2\pi)^3 \delta(\bk-\bk') P_{v_A} (k, t) = \langle v^{i}(\bk,t) \la^A_{ij}(\bk) v^{*j}(\bk',t) \rangle,
\end{equation}
with $A = \, \parallel$ denoting the compressional component and $A=\perp$ the vortical.

We define the unequal time correlators (UETCs) of a field $Y$ through 
\begin{equation}
(2\pi)^3 \delta(\bk-\bk') P_{Y} (k, t, t') = \langle Y(\bk,t) Y^*(\bk',t') \rangle .
\end{equation}
So defined, the UETC is symmetric on interchange of the two times,  $P_{Y} (k, t', t) = P_{Y} (k, t, t')$.
In the simulations, 
$t'$ is fixed at some reference time $t_\text{ref}$. 
This reference time is chosen to be the time when 95\% of the simulation volume is in the broken phase.  

We define the decorrelation function of the field $Y$ as 
\begin{equation}
D_Y(k, t, t_\text{ref}) = \frac{ P_Y (k, t, t_\text{ref})} { \sqrt{ P_Y (k, t) P_Y (k, t_\text{ref}) }  }.
\label{e:DecFunDef}
\end{equation}

We measure transverse and longitudinal projected power spectra and UETCs of the 3-velocity $v^i$.  We also measure the power spectra of the spatial components of the enthalpy-weighted 4-velocity 
\begin{equation}
U^i = \sqrt{\frac{w}{\bar w}}\,\gamma v^i, 
\label{e:Udef}
\end{equation}
where $\bar w$ is the volume-averaged enthalpy density.  For brevity we will call $U^i$ the weighted 4-velocity.
This quantity has advantages over the 3-velocity for relativistic flows, as we show below.

The same exercise can be carried out for the anisotropic shear stress $\Pi$, obtained by projecting out the transverse traceless spatial part of the energy-momentum tensor $\Pi_{ij} = \Lambda_{ij,kl} T^{kl}$, 
and taking two-point correlations, 
\begin{equation}
(2\pi)^3 \delta(\bk-\bk') P_\Pi (k, t, t_\text{ref}) = \langle \Pi^{ij} (\bk,t) \Pi^{ij} (\bk',t_\text{ref}) \rangle
\end{equation}
where $\Lambda_{ij,kl}$ is the transverse-traceless projector,
\begin{equation}
\Lambda_{ij,kl} = \lambda^\perp_{ik} \lambda^\perp_{jl}  - \frac{1}{2} \lambda^\perp_{ij} \lambda^\perp_{kl}.
\end{equation}

The spatial components of the energy momentum tensor for the fluid, i.e. Eq.~\ref{eq:em_tensor_fluid}, can be written
\ben
T^{ij} = \bar{w} U^i U^j + \de^{ij} p.
\een
Hence they are quadratic in the weighted 4-velocity components $U^i$, which is a useful property when considering correlations of the energy-momentum tensor.

We measure the UETCs of the compressional and vortical velocity components $P_{v_\parallel}(k, t_1, t_2)$, $P_{v_\perp}(k, t_1, t_2)$ and also the shear stress UETC $P_\Pi(k, t_1, t_2)$, relative to a single fixed reference time $t_2 = \tref$, with $t_1 > t_2$. 
We assume that the decorrelation functions for the weighted 4-velocity are close to those of the 3-velocity, which we write $D_\parallel$ and $D_\perp$.

To find the gravitational wave power spectrum, we will assume a weak gravity regime, where metric perturbations are linearly sourced by the fluid. 
Gravitational waves are the transverse traceless parts of the spatial metric perturbations 
$h_{ij}$, which are sourced by the shear stress $\Pi_{ij}$.

The spectral density of the canonical momentum of the transverse-traceless metric perturbations  
\begin{equation}
(2\pi)^3 \delta(\bk-\bk') P_{\dot{h}} (k, t) = \langle \dot{h}_{ij}(\bk,t) \dot{h}_{ij}(\bk',t) \rangle
\end{equation}
can in turn be used to define a fractional energy density spectrum for the gravitational waves
\begin{equation}
\mcPgw(k, t) =  \frac{1}{\bar\epsilon_0} \frac{1}{32\pi G} \mathcal{P}_{\dot{h}}(k, t),
\end{equation}
where $\bar\ep_0$ is the volume averaged initial energy density,
which under our assumptions is constant.  We will call $\mcPgw$ the gravitational wave power spectrum.

\subsection{Analytical expectations}

In this section, we give general arguments for the expected form of the power spectra and decorrelation functions for mildly relativistic velocity fields.

Both velocity power spectra are expected on causality grounds~\cite{Durrer:2003ja} to go as $k^5$ at wavenumbers much less than the inverse bubble spacing, $kR_* \ll 1$.  The argument is that bubble nucleation is a causal process, and so the resulting two-point equal-time velocity correlations must vanish for separations greater than the causal horizon. The spectral density is therefore analytic in $k$.  The velocity is a vector, and the universe is held to be isotropic. Therefore the spectral density of the velocity field cannot be a constant as $k \to 0$.  Hence the spectral density goes as $k^2$, and the power spectrum as $k^5$.   

At high wavenumbers, $kR_* \gg 1$, the compressional and vortical correlators are expected to behave differently.  The compressional modes have shocks which appear already at the bubble expansion stage, which continue to propagate after the transitions are complete.  The power spectrum of a velocity field with discontinuities goes as $k^{-1}$ at large $k$~\cite{Dahl:2021wyk,Dahl:2024eup}: this is therefore the expected form for the compressional modes at all times.

The vortical modes should develop turbulence and evolve towards a Kolmogorov spectrum, which for our definition of the velocity power spectrum is $k^{-2/3}$ \cite{Auclair:2022jod}.  However, this takes several eddy turn-over times.  As far as we are aware, there are no theoretical predictions for the initial power spectrum of the vortical modes laid down by bubble collisions.

The decorrelation functions of the compressional modes are expected to be dominated by their oscillatory behaviour, taking the form \cite{Hindmarsh:2019phv}
\begin{equation}
D_{\parallel} (k, t, t_{\text{ref}}) = \cos \bigg( \cs  k(t-t_\text{ref}) \bigg) .
\label{e:DparExp}
\end{equation}
For purely vortical modes, the expectation for the decorrelation function is that it follows the Kraichnan sweeping model~\cite{kraichnan:1964}
\begin{equation}
D_\perp (k, t, t_{\text{ref}}) = \exp\bigg(-\frac{1}{2} \Vsw{\perp}^2 k^2 (t-t_{\text{ref}})^2 \bigg),
\end{equation}
which has been verified by numerical simulations of freely decaying vortical modes 
as a good approximation at high wavenumbers~\cite{Auclair:2022jod}.  We will see that the expectations will need revision 
when the sweeping velocity field contains compressional modes. 

Kinetic energy is expected to decay due to non-linear (turbulent) transport of energy to small length scales, where it is dissipated by viscosity.  In our simulations, the viscosity is numerical, and is a result of the finite resolution of the advection scheme (see Appendix \ref{ap:advect}).
The dissipation of energy in the compressional modes occurs at shocks, which 
decay over a timescale 
\begin{equation}
\tsh = R_* / \UparMax ,
\end{equation} 
where $\UparMax = \max(U_\parallel)$. Note that this decay timescale is independent of viscosity. 
The analogous timescale for the evolution of vortical turbulence (also called the eddy turn-over time) is 
\begin{equation}
\teddy = R_* / \UperpMax
\end{equation}
where $\UperpMax = \max(U_\perp).$ 
Our simulations run for at least one shock decay time.

\section{\label{sec:sim_setup}Simulation setup}

We simulate in a Minkowski background,  which is valid on all scales 
if the equation of state of the fluid is pure radiation, due to the scale invariance of the system in an expanding universe~\cite{Brandenburg:1996fc,Hindmarsh:2015qta}. 
Departures from scale invariance during the transition can be neglected 
if the mean bubble spacing $R_*$ is much less than the Hubble length, $\Hn R_* \ll 1$.  This is achieved if the transition completes in much less than a Hubble time $\Hn^{-1}$. 

By applying standard variational techniques, we can obtain the equations of motion of the system. 
The equations of motion of the field are then,
\begin{equation}
-\ddot{\phi} + \nabla^2 \phi - \frac{\partial V}{\partial \phi} = \eta \gamma ( \dot{\phi} + v^i \partial_i \phi) \;,
\end{equation}
while for the fluid we obtain,
\begin{multline}
\dot{E} + \partial_i (Ev^i) + p[\dot{\gamma} + \partial_i(\gamma v^i)] - \frac{\partial V}{\partial \phi} \partial_i \phi \\
= \eta \gamma^2 (\dot{\phi} + v^i \partial_i \phi)^2
\end{multline}
and
\begin{multline}
\dot{Z}+\partial_j (Z_i v^j) + \partial_i p + \frac{\partial V}{\partial \phi} \partial_i \phi \\
= -\eta \gamma (\dot{\phi} + v^j \partial_j \phi) \partial_i \phi \;,
\end{multline}
where $E=\gamma \epsilon$ and $Z^i = \gamma^2 w v^i$ are fluid energy and momentum state variables, $\gamma$ is the Lorentz boost factor and $v^i$ the three-velocity. These serve as the basis of the discretization scheme on a 3-dimensional grid. We will use the field-fluid simulation code SCOTTS~\cite{Hindmarsh:2013xza}. Here the scalar field is evolved via standard leapfrog with a 7-point Laplacian stencil, and the fluid variables via standard operator splitting~\cite{2003rnh..book.....W}, where state variables are updated term-by-term in the equations above. The damping term is handled via a Crank-Nicholson scheme~\cite{Crank_Nicolson_1947}.

\begin{table}
    \begin{ruledtabular}
    \begin{tabular}{c D{.}{.}{-1}}
    Parameter & \multicolumn{1}{c}{Value}  \\
    \hline
    \rule{0pt}{3ex}$M^2/T_c^2$ &  0.0427 \\ 
    $\mu/T_c$ &  0.168 \\ 
    $\lambda$ & 0.0732  \\ 
    $g_*$ & 106.75  \\ 
    $G$ & 1
    \end{tabular}
  \end{ruledtabular}
    \caption{\raggedright A summary of the scalar field potential parameters $M^2, \lambda, \mu$, number of effective relativistic degrees of freedom of the fluid $g_*$ and gravitational constant $G$ adopted for the simulations. 
    \label{tab:parameters}}
\end{table}

For computing the gravitational wave power spectra, 
we evolve a 6-component auxiliary symmetric tensor $u_{ij}$, obeying the equation \cite{PhysRevD.77.043517},
\begin{equation}
\ddot{u}_{ij} - \nabla^2 u_{ij} = 16 G\bar{w} \left(U_i U_j - \frac{1}{3}\de_{ij}\bU^2\right),
\end{equation}
again with leapfrog and a 7-point Laplacian stencil. The gravitational wave modes are recovered from projections of the Fourier transforms $\tilde{u}_{ij}(\bk,t)$ with the transverse-traceless projector, $\tilde{h}_{ij} = \Lambda_{ij, kl} \tilde{u}_{kl}$. 

Our simulation grid has spacing $\delta x = 1.0 T_c^{-1}$, with timesteps of size $\delta t = 0.2 T_c^{-1}$. In previous works (e.g. \cite{Cutting:2019zws}) these choices demonstrated reasonable energy conservation (see Appendix \ref{ap:Econs}) while capturing the dynamical features of bubble collision and the variation of gravitational wave energy density. In particular, this choice was seen to be sufficient to accurately describe the gravitational wave power spectra for our chosen values of $v_w$ and $\alpha_n$.

The phase transition is initiated by creating 64 Gaussian bumps in $\phi$ whose centre is in the broken phase $\phi_b = 2.0 T_c$ at random positions on the simulation grid, 
while avoiding overlaps.  This results in a mean bubble spacing  of $R_* = 1024$. These bumps are large enough that they are guaranteed to expand, and during their evolution they relax towards a self-similar velocity and enthalpy profile (see Appendix \ref{ap:profile}). The fluid, initially at rest, is driven by the pressure of expanding scalar field bubbles to form outward-moving fluid shells.

For consistency with previous works, we will use the same scalar field potential parameters, the same number of effective relativistic degrees degrees of freedom $g_*$ and gravitational constant $G$. These are summarised in Table~\ref{tab:parameters}.
\begin{table}
    \begin{ruledtabular}
    \begin{tabular}{c D{.}{.}{-1} D{.}{.}{-1}} 
    Phase transition  & \multicolumn{1}{c}{$\Tn/\Tc$} & \multicolumn{1}{c}{$\eta/\Tc$} \\
    \hline
    \rule{0pt}{3ex}{$\aln = 0.67$, $v_w=0.92$} & 0.23 & 0.15 \\ 
    {$\aln = 0.50$, $v_w=0.44$} & 0.25 & 3.5  
    \end{tabular}
  \end{ruledtabular}
    \caption{\raggedright A summary of nucleation temperature $\Tn$ and field-fluid coupling parameter $\eta$ chosen to set the transition strength $\aln$ and wall speed $\vw$ in our simulations. Upper row corresponds to the detonation, the lower one to the deflagration.
    \label{tab:scenarios}}
\end{table}

In order to continue where Ref.~\cite{Cutting:2019zws} left off, we select two test cases, the strongest transitions studied in that paper, one proceeding by deflagrations and the other by detonations. 
The deflagration arises from $\aln=0.5$ and $\vw=0.44$, and the detonation from $\aln=0.67$ and $\vw=0.92$.
The choices of nucleation temperature $\Tn$ and coupling $\eta$ that lead to such $(\aln, \vw)$ values are summarised in Table~\ref{tab:scenarios}.

\begin{figure*}[p]
        \begin{subfigure}[b]{0.49\textwidth}
            \includegraphics[width=1.0\textwidth]{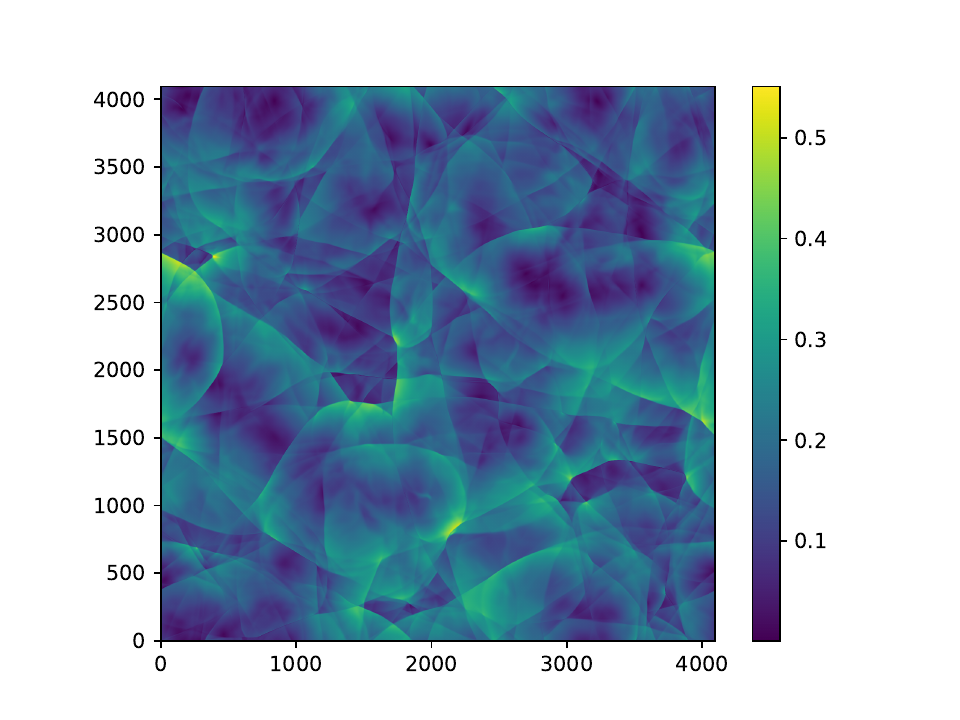}
            \caption{$|\bv|(x,y,z=0)$}
        \end{subfigure}
        \begin{subfigure}[b]{0.49\textwidth}
            \includegraphics[width=1.0\textwidth]{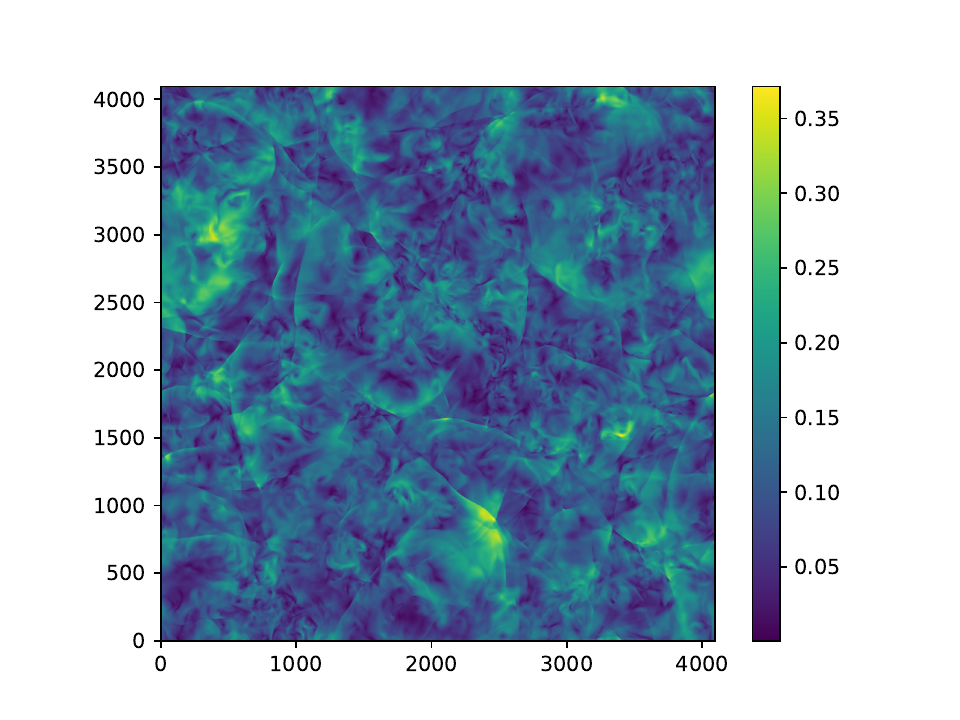}
            \caption{$|\bv|(x,y,z=0)$}
        \end{subfigure}
        \begin{subfigure}[b]{0.49\textwidth}
            \includegraphics[width=1.0\textwidth]{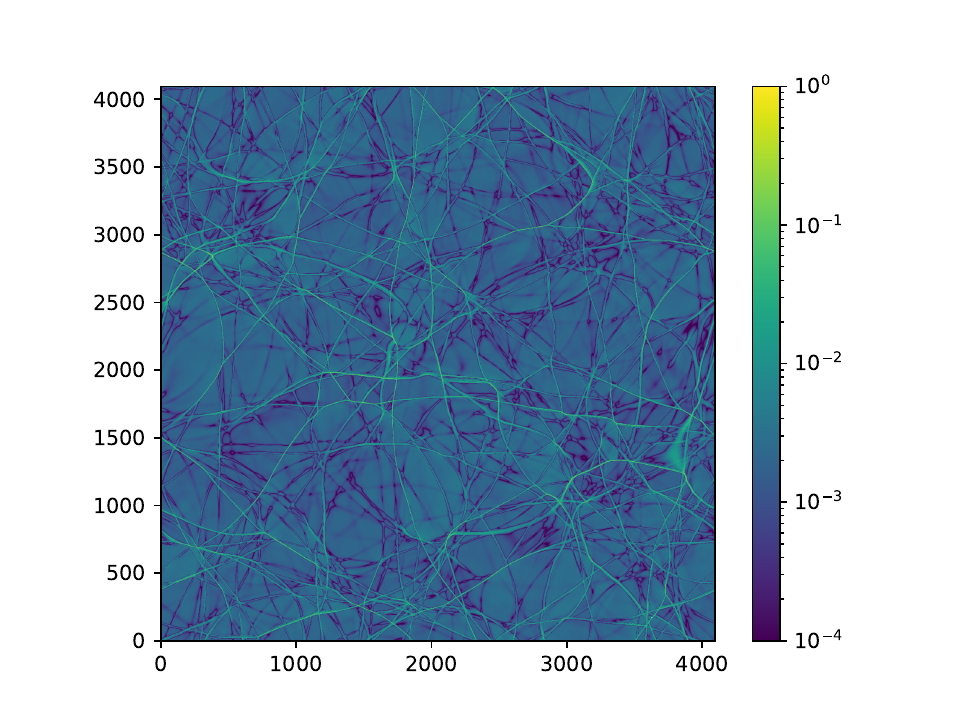} 
            \caption{$| \nabla \cdot \bv|(x,y,z=0)$}
        \end{subfigure}
        \begin{subfigure}[b]{0.49\textwidth}
            \includegraphics[width=1.0\textwidth]{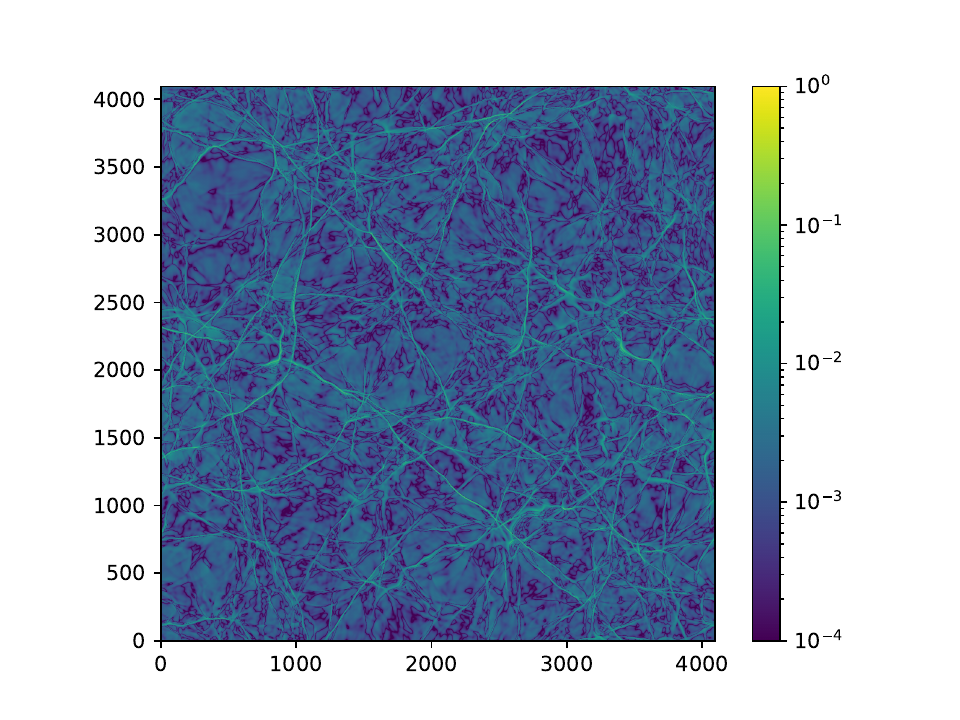}
            \caption{$| \nabla \cdot \bv|(x,y,z=0)$}
        \end{subfigure}
        \begin{subfigure}[b]{0.49\textwidth}
            \includegraphics[width=1.0\textwidth]{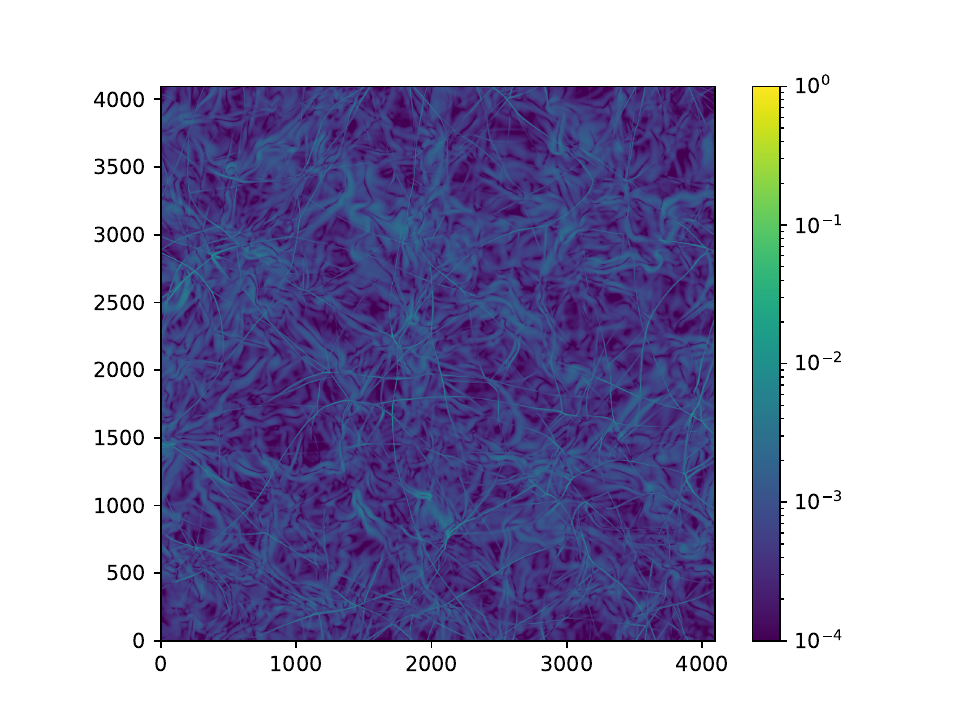} 
            \caption{$| \nabla \times \bv|(x,y,z=0)$}
        \end{subfigure}
        \begin{subfigure}[b]{0.49\textwidth}
            \includegraphics[width=1.0\textwidth]{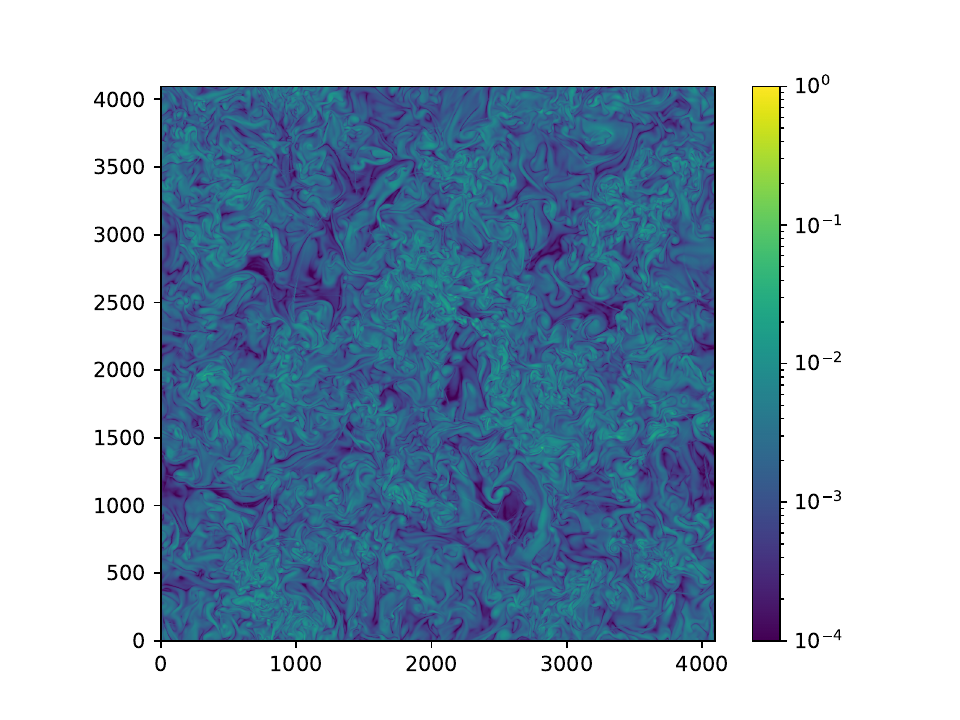}
            \caption{$| \nabla \times \bv|(x,y,z=0)$}
        \end{subfigure}
\caption{\raggedright Slices through simulations. Left: detonation ($\alpha_n=0.67$, $v_w=0.92$) at time $t=5000/T_c$. Right: deflagration ($\alpha_n=0.50$, $v_w=0.44$) at time $t=7700/T_c$. Top panels show the magnitude of the 3-velocity $|\bv|$, middle panels the magnitude of the divergence of spatial components of the  3-velocity $|\nabla \cdot \bv|$, and the bottom panels showcase the magnitude of the curl $|\nabla \times \bv|$. 
\label{fig:figureSlices}}
\end{figure*}

We now list the main observables, which consist of volume-averaged quantities, power spectra, and unequal-time correlators. 

We have two estimators of the average wall speed \cite{Cutting:2022zgd}. The first is obtained from the volume-averaged kinetic and gradient energies of the scalar field, 
\bea
e_{\dot\phi} =\frac{1}{\mathcal{V}} \int \mathrm{d}^3 x \half {\dot\phi^2}, \quad 
{e_D} = \frac{1}{\mathcal{V}} \int \mathrm{d}^3 x \half ( \nabla\phi)^2, 
\eea
as
\begin{equation}
    \label{eq:vw_estimator_energies}
    v_{w,\phi} = \bigg( \frac{e_{\dot\phi}}{e_D} \bigg)^{1/2}
\end{equation}
The second comes from the rate of change of the total number of points in the broken phase,
\begin{equation}
    \label{eq:vw_estimator_links}
    v_{\text{w,b}}  = \frac{1}{A} \frac{\mathrm{d} \mathcal{V}_\text{b}}{\mathrm{d}t}
\end{equation}
where $A$ is the total number of sites where the scalar field changes sign (ie. sites where $\phi_{i}\phi_{i+1}<0$).

Other volume-averaged quantities include the rest energy density $\bar\ep$, the pressure  $\bar{p}$, from which the enthalpy density $\bar{w} = \bar{\ep} + \bar{p}$ is derived.  

We also output the volume-averaged weighted 4-velocity,
\begin{equation}
    \bar{U}^2 = \frac{1}{\bar{w} \mathcal{V}} \int \mathrm{d}^3 x \, U^i U^i .
\end{equation}    
Average fluid transverse and longitudinal velocities are reconstructed from their power spectra. For example, the root mean square longitudinal (compressional, $A=\parallel$) and transverse (vortical, $A=\perp$) components of the weighted 4-velocity are 
\begin{equation}
    \bar{U}_{A}^2 = \int \frac{\mathrm{d}k}{k} \mathcal{P}_{U_{A}}(k)
\end{equation}    
where  $\mathcal{P}_{U_{A}} (k)$ is the power spectrum of projection $A$ of the weighted 4-velocity field. 
Analogously, we construct the RMS transverse and longitudinal 3-velocities,
$\overline{v}_\perp$ and $\overline{v}_\parallel$ from their power spectra $\mathcal{P}_{v_{A}}(k)$.
Likewise, we measure the shear stress power spectrum $\mcP_\Pi(k)$.  The RMS shear stress is closely related to the volume-averaged kinetic energy $\bar{w}\bar{U}^2$.

We measure the UETCs of the longitudinal and transverse components of the 3-velocity and the shear stress.  The selection is memory-limited: in order to measure UETCs one needs to keep the full Fourier transform array at the reference time in memory. The fields for which power spectra are computed are $U_{\parallel,\perp}$, $v_{\parallel,\perp}$ and $\Pi$; for UETCs only $v_{\parallel,\perp}$ and $\Pi$ are measured.

From the velocity power spectra, we construct the integral scales, which characterise length scales for the fluid flow for each velocity field.  For example, the weighted 4-velocity integral scales are
\begin{equation}
    \xi({U_A}) = \frac{1}{\bar{U}_{A}^2}\int \frac{\mathrm{d} k}{k^2} \mathcal{P}_{U_{A}}(k) \, ,
    \label{e:IntScaDef}
\end{equation}
and we also use the notation $\xi_A \equiv \xi(U_A)$ for brevity.
The integral scales are expected to be of order $R_*/2\pi$ just after the phase transition, 
and to increase with time as the RMS velocity components decay \cite{Dahl:2021wyk,Dahl:2024eup,Auclair:2022jod}, 
due to turbulent transport of energy to small length scales.

Simulations start at time $t=0$ and run to final times $t_\text{end} = 4990/\Tc$ (detonation) and $8000/\Tc$ for the main deflagration run.  A separate run with the same initial conditions and $t_\text{end} = 5880/\Tc$ was used to measure UETCs in the deflagration. 

Volume-averaged quantities were recorded from $t=0$, in intervals of $10/\Tc$ time units. 
Power spectra were also recorded from $t=0$, at intervals given in Table~\ref{tab:UETCinfo}.
The UETC reference times, given in the same table, were chosen near the times when the simulation box is $95\%$ converted to broken phase.

\begin{table}
    \begin{ruledtabular}
    \begin{tabular}{c D{.}{.}{0} D{.}{.}{0}}
    Phase transition & \multicolumn{1}{c}{Interval [$\Tc^{-1}$]} & \multicolumn{1}{c}{$t_\text{ref} \Tc$} \\
    \hline
    \rule{0pt}{3ex}{$\alpha_\text{n} = 0.67$, $v_w=0.92$} & 40 & 1000 \\ 
    {$\alpha_\text{n} = 0.50$, $v_w=0.44$} & 100 & 4500  
    \end{tabular}
  \end{ruledtabular}
    \caption{\raggedright Intervals between power spectra measurements, and reference times for unequal time correlators. 
    \label{tab:UETCinfo}}
\end{table}

A selection of slices from each of the simulations is shown in Fig.~\ref{fig:figureSlices}. 
They show pseudocolour plots of magnitude of  3-velocity $|\bv|$, divergence of 3-velocity $|\nabla \cdot \bv|$ and its curl $|\nabla \times \bv|$. The latter two are presented in logarithmic scale to highlight the presence of relevant non-linearities (shocks, which appear as bright streaks, and vorticity in the form of eddies). These are presented for times $t=5000/T_c$ in the detonation, and $t=7700/T_c$ for the deflagration. An animation for the deflagration is available at~\cite{correia_2025_15481610}.

A noteworthy feature in the detonation is the appearance of three-way junctions of shocks, termed $\lambda$-configurations~\cite{courant1999supersonic}.  These appear between shocks colliding at less than a certain critical angle, which is a function of the Mach number and the adiabatic index~\cite{courant1999supersonic,PhysRevA.45.6130}, and can be seen in the laboratory and in astrophysical flows~\cite{2016ApJ...823..148H}.
The  $\lambda$-configurations already indicate non-linear interactions in the fluid flow.

\section{\label{sec:results}Results}

\begin{figure*}[ht]
    \includegraphics[width=1.0\columnwidth]{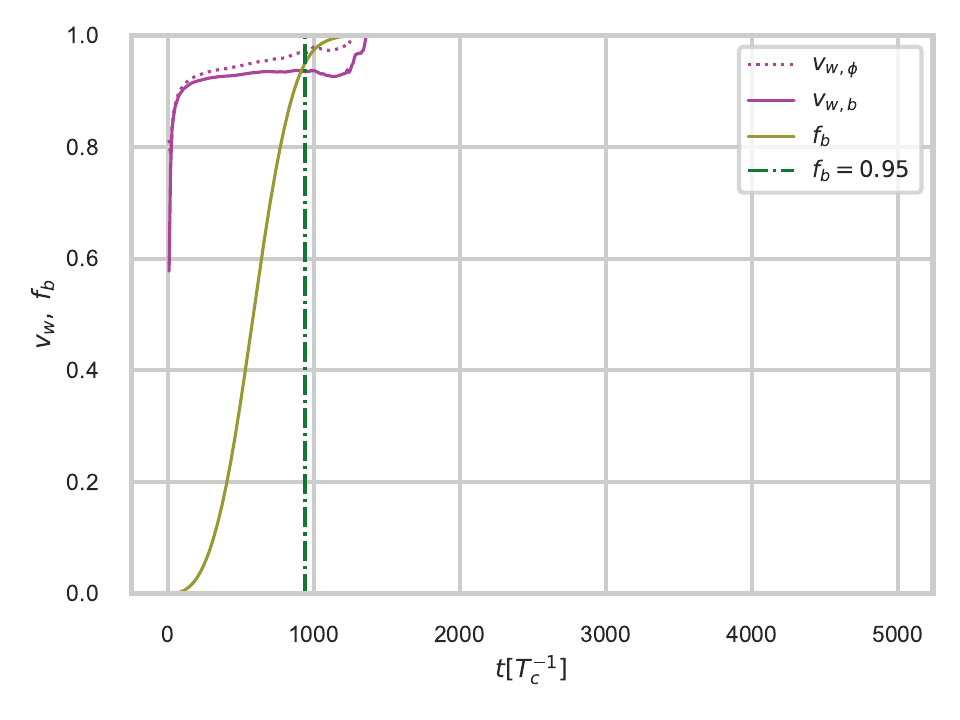}
    \includegraphics[width=1.0\columnwidth]{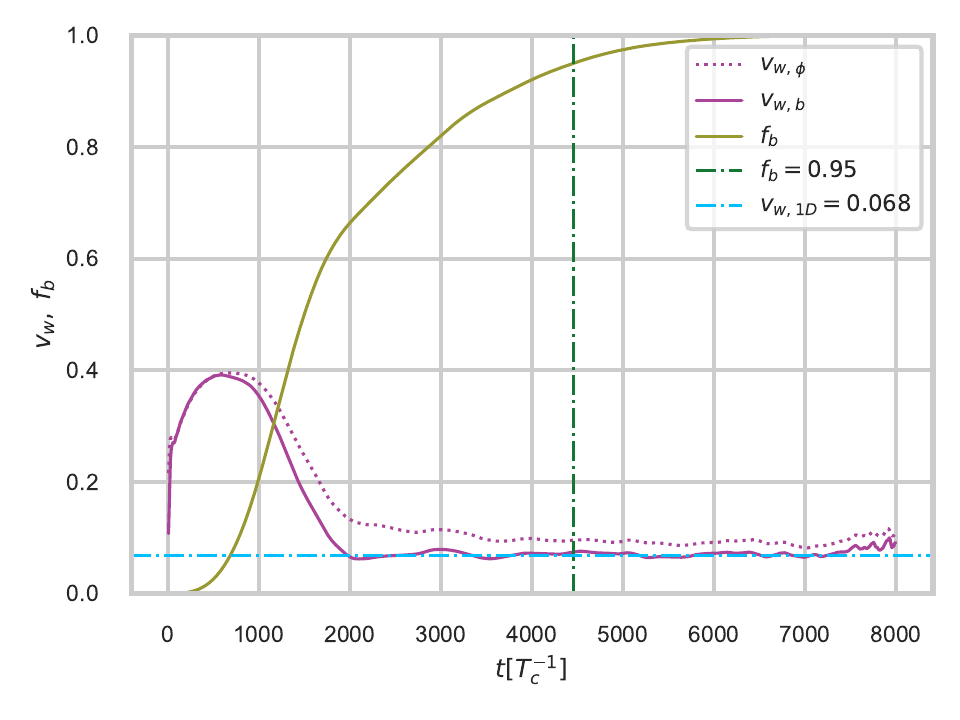}
    \includegraphics[width=1.0\columnwidth]{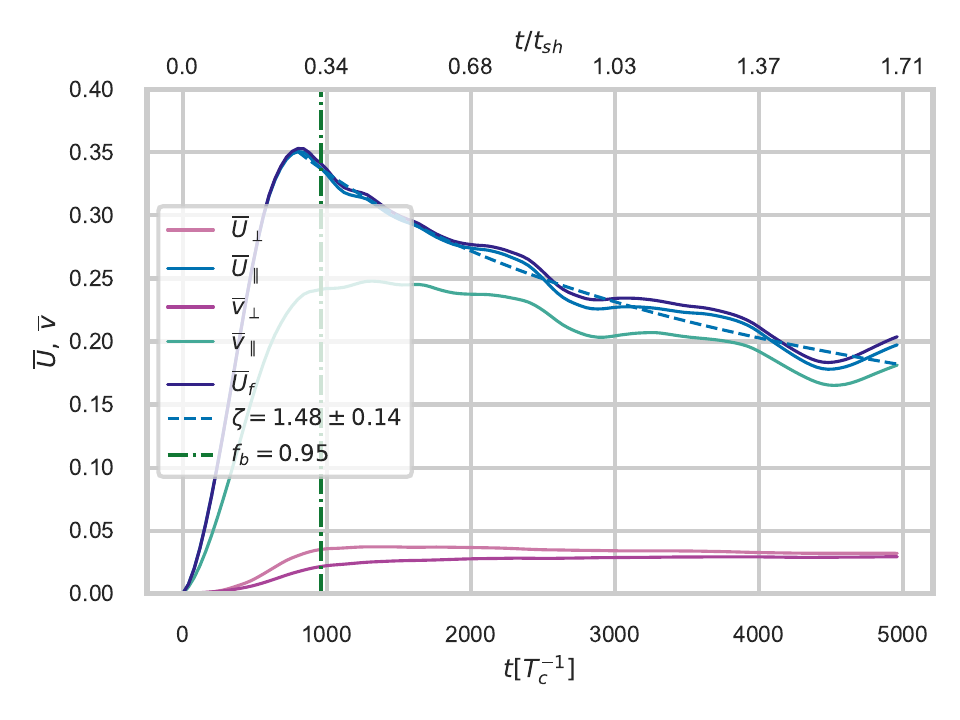}
    \includegraphics[width=1.0\columnwidth]{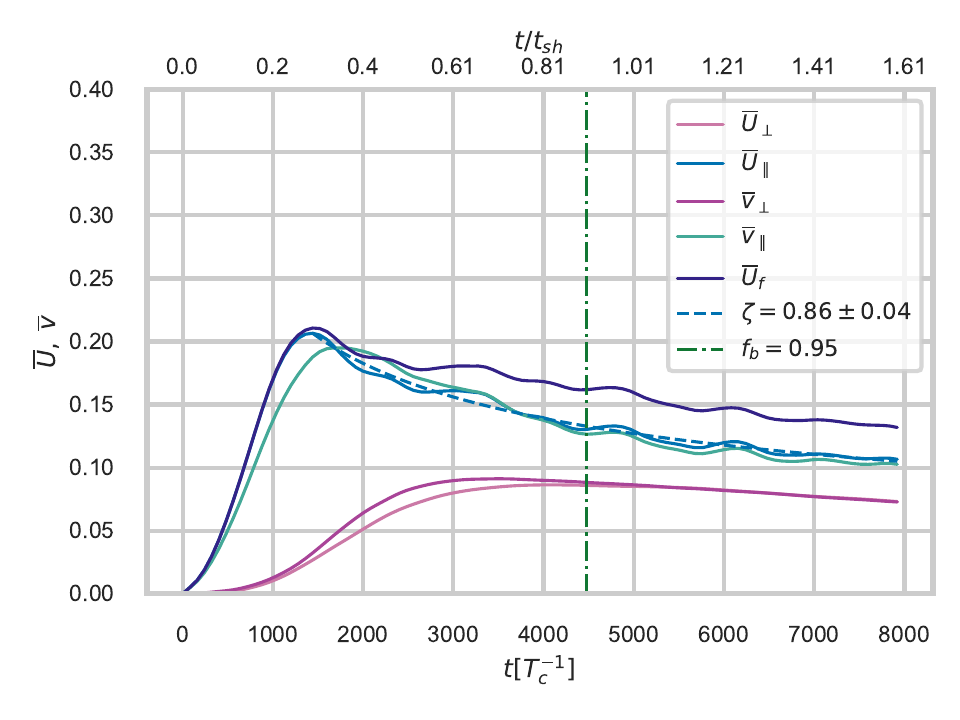}
    \includegraphics[width=1.0\columnwidth]{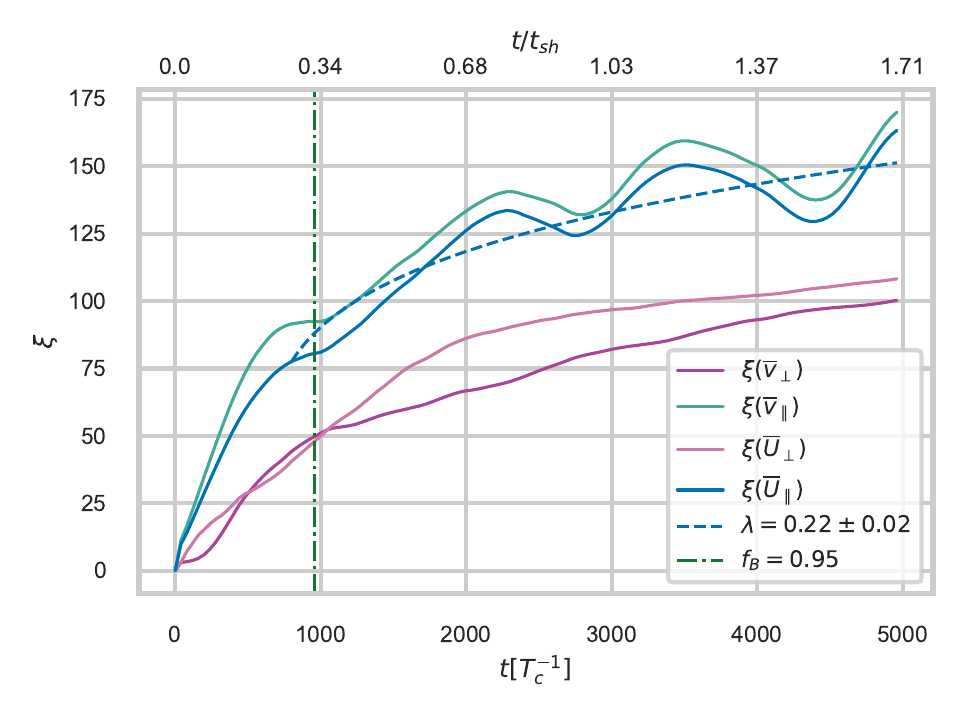}
    \includegraphics[width=1.0\columnwidth]{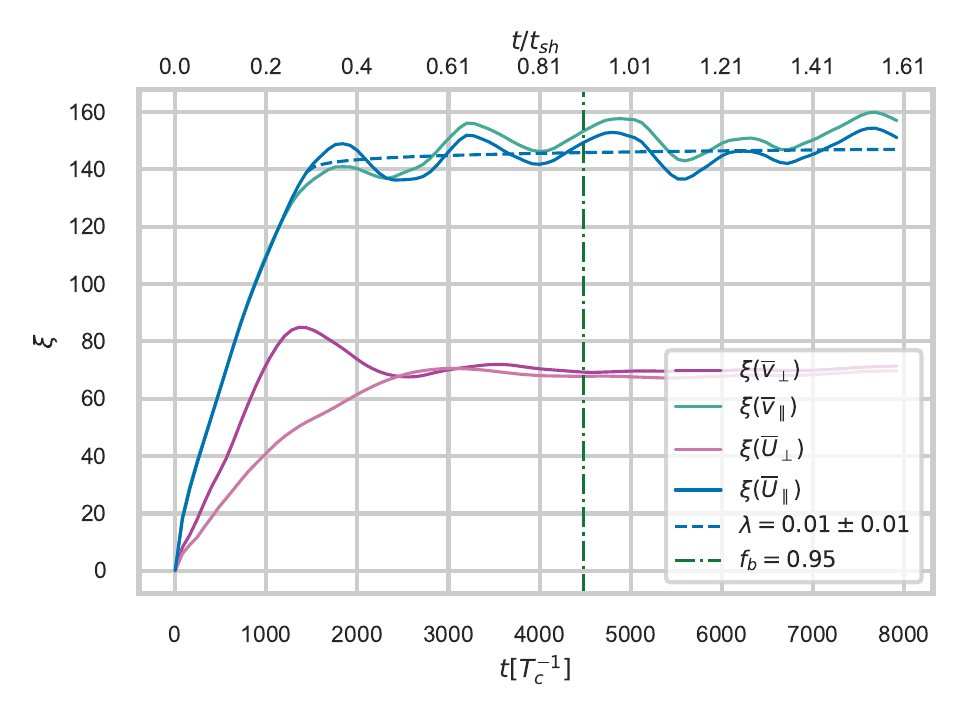}
    \caption{\raggedright
    Selected volume-averaged quantities for the detonation ($\aln=0.67$, $\vw=0.92$, left-hand-side) and deflagration ($\aln=0.50$, $\vw=0.44$, right-hand-side).
    Top: wall speed estimates $\vw$. The velocity estimate $v_{w,\phi}$ is computed from the ratio of kinetic and
        gradient energy of the scalar field Eq.~(\ref{eq:vw_estimator_energies}) (purple dotted line), and $v_{w,N}$ through the rate of change of number of sites in broken phase Eq.~(\ref{eq:vw_estimator_links}) (purple solid line). The fraction of sites in the broken phase $f_\text{b}$ is plotted in olive green, and a dot-dashed vertical green line indicates when 95\% of the sites are in the broken phase (this also marks $t_{\text{ref}}$, the reference time for unequal time correlators). 
Middle: root mean square enthalpy-weighted 4-velocities for compressional $\bar{U}_\parallel$ (in red) and vortical $\bar{U}_\perp$ (in blue) components, along with their non-relativistic equivalents $\bar{v}_{\parallel,\perp}$. 
The total RMS weighted 4-velocity $\bar{U}$ is also shown.
Bottom: integral scales, for the compressional and vortical modes, defined in Eq.~\eqref{e:IntScaDef}. We also display fitted decay and growth indexes ($\zeta, \lambda$, respectively) for compressional kinetic energy and integral scale.
\label{fig:U_vel_xi}}
\end{figure*}

\subsection{Volume-averaged quantities}

We begin with the fractional volume in the broken phase and the wall velocities, shown in Fig.~\ref{fig:U_vel_xi} (top). 

In the detonation (left panel) the fraction of broken phase reaches the 95\% at a time  $t \approx 950/\Tc$
with the transition completing shortly thereafter, at $t \approx 1400/T_c$. The wall speed quickly grows to its expected asymptotic limit of $\vw=0.92$ within $200/T_c$ time units from the start of the simulation. 
This rise at the end is merely a consequence of velocity estimators breaking down when the transition is nearly completed: for example the wall velocity estimator Eq.~(\ref{eq:vw_estimator_links}) divides by the inferred area of the bubble walls $A$, which by this time will be close to zero. 

The deflagration (right-hand-side panel of figure~\ref{fig:U_vel_xi}) however, is rather different: the velocity starts to rise towards the expected wall speed of $\vw=0.44$, but never reaches it, as reheating of the surrounding plasma slows down bubble wall velocities to $\vw \approx 0.05-0.1$. 
This slow-down begins at roughly $t \approx 600/T_c$ or alternatively, when the fraction of broken sites $f_B$ is roughly $5\%$ and stops when the broken fraction is $65-70\%$, as can be seen on  Fig.~\ref{fig:U_vel_xi}. Also shown is the asymptotic wall velocity from one-dimensional spherical droplet simulations of Ref.~\cite{Cutting:2022zgd}. The two velocity estimates are in excellent agreement. 

The transition then proceeds rather slowly (at least when compared to the initial times) and the time when the box is at $95\%$ in the broken phase $t_\text{ref}=4460/T_c$, with the transition fully completing at the very end of the run $t_\text{end}=8000/T_c$. This reheating process was first shown in Ref.~\cite{Cutting:2019zws} and marks an important difference between deflagrations in weak and strong phase transitions, as it results in a reduction of the efficiency of kinetic energy production and therefore in a suppression of the resulting gravitational wave spectra.

We now turn our attention to fluid velocities. Fig.~\ref{fig:U_vel_xi} (middle row) shows projected and total fluid 3-velocities and weighted 4-velocities.
The maximum fluid velocities for each of the components of the weighted 4-velocity are given in Table~\ref{tab:tsh_teddy}, along with the inferred shock and eddy turn-over times, which are defined using $R_*$ as the length scale.

We see that in the detonation, the compressional modes dominate, and show a marked decrease from the maximum, decaying by about 40\% by the end of the simulation. The vortical velocity is much smaller: at most $U_\perp \simeq 0.05$, and decays by about 15\% by the end of the simulation. 
One also sees that the weighted 4-velocity is around 40\% larger than the 3-velocity. This is too large a difference to be accounted for by the Lorentz $\ga$ factor, which is around 1.03 for the maximum RMS 3-velocity.  Instead it points to strong correlations between the enthalpy and the velocity. This is a feature of a shock, rather than a sound wave, where they are in antiphase. Along with the strong visual impression of shocks in Fig.~\ref{fig:figureSlices}, we can infer that the detonation flow is shock-dominated.

By contrast, the deflagration generates a significant amount of vorticity. The flow is predominantly compressional when the weighted 4-velocity $U$ is at  its maximum (at $t \approx 1500/T_c$). However, vorticity continues to rise with $\bar{U}_\perp$ reaching roughly $0.08$ at $t=3000/T_c$. Given the initially faster decay rate of $\bar{U}_\parallel$, 
these end up being of similar magnitude at the end of the run, with $\bar{U}_\parallel \approx 0.11$ and $\bar{U}_\perp \approx 0.08$ at $t=8000/T_c$.  
Indeed, the compressional and vortical modes appear to be reaching a constant ratio: after a time around $5000/\Tc$, the ratio between $U_\parallel/U_\perp $ fluctuates in the range 1.4 -- 1.5. The ratio is not equipartition, which would be $U_\perp \approx \sqrt{2}U_\parallel$, but it does suggest some exchange between the components which merits further exploration with longer simulations.

Another contrast is that here there is much less difference between the weighted 4-velocity and the 3-velocity, indicating that shocks are less important (although they are still clearly visible in Fig.~\ref{fig:figureSlices}).

The accompanying animation (Ref.~\cite{correia_2025_15481610}) shows that vortical modes are generated by the interactions of compression waves as bubbles collide, particularly as the leading shock crosses the phase boundary into the broken phase. 
Vorticity can also be generated by shock collisions ~\cite{Pen:2015qta}: a possible sign of this effect is the slight growth in the RMS 3-velocity following the completion of the transition in the detonation.  The weighted 4-velocity however decreases.

The integral scales computed from compressional and vortical components of the weighted 4-velocity and 3-velocity fields can be found in the bottom row of Fig.~\ref{fig:U_vel_xi}. One can see a clear growth tendency in the integral scales following the detonation, but the scales remain approximately constant after the deflagration. The strong oscillations are present because the integral scale is sensitive to the longest wavelengths in the flow, where there are few wavenumbers contributing to the power spectrum. 

\begin{table}
    \begin{ruledtabular}
    \begin{tabular}{c D{.}{.}{-1} D{.}{.}{-1} D{.}{.}{0} D{.}{.}{0}}
    Phase transition & \multicolumn{1}{c}{$\text{max}(U_\parallel)$} & \multicolumn{1}{c}{$\text{max}(U_\perp)$} & \multicolumn{1}{c}{$t_\text{sh} [T_c^{-1}]$} & \multicolumn{1}{c}{$t_\text{ed} [T_c^{-1}]$} \\
    \hline
    \rule{0pt}{3ex}{$\alpha_\text{n} = 0.67$, $v_w=0.92$} & 0.35 & 0.04 & 2920 & 27535 \\ 
    {$\alpha_\text{n} = 0.50$, $v_w=0.44$}  & 0.21 & 0.09 & 4958 & 11865  
    \end{tabular}
  \end{ruledtabular}
    \caption{\raggedright Maximum values of each compressional ($U_{\parallel}$) and vortical ($U_\perp$) four-velocities, along with the associated shock and eddy formation times ($t_\text{sh}$ and $t_\text{ed}$, respectively).
    \label{tab:tsh_teddy}}
\end{table}

The decay of RMS velocity is a power law~\cite{Auclair:2022jod,Dahl:2021wyk,Dahl:2024eup}.  
For gravitational wave production, we are particularly interested in the compressional modes,  and we fit the longitudinal weighted 4-velocity $U_\parallel$ as 
\begin{equation}
    U_\parallel = \max(U_\parallel)\left[1 +  \frac{\Delta t}{t_*} \right]^{-\zeta/2}
    \label{e:ZetDef}
\end{equation}
where $\Delta t = t - \tmax$ and $\zeta$ is the decay index of the mean square, which is closely related to the decay index of the kinetic energy fraction in compressional modes, 
\begin{equation}
K_{\parallel} = \Gamma U^2_\parallel ,
\end{equation}
where $ \Gamma = {\bar{w}}/{\bar{\ep}_0}  $, 
as the enthalpy density is almost constant after the completion of the transition. This is demonstrated in Appendix~\ref{ap:Econs}, where we plot the thermal energy $e_Q \equiv 3\bar{w}/4$.
Note that, as defined, $\Gamma$ is not exactly equal to the adiabatic index of an ultrarelativistic fluid, as total energy $\bar{\ep}_0$ is divided between kinetic and thermal energy.

In connection with the decay of kinetic energy, the integral scale is expected to grow. We also provide fits of the integral scale to the following form,
\begin{equation}
    \xi_\parallel = \xi_\parallel(\tmax)\left[1 + \frac{\Delta t}{t_*} \right]^{\lambda}
    \label{e:LamDef}
\end{equation}
where $\lambda$ is the associated decay index.

The fitted decay and growth indices $\ze$ and $\la$ are given in Table~\ref{tab:KxiInd}. The reason for our interest in the power-law behaviour of the kinetic energy and integral scales is that the gravitational wave power spectrum is sourced by the kinetic energy squared, which exhibits power law decay $K^2 \propto t^{-2\zeta}$ at late times. The gravitational wave power is also proportional to the shear stress length scale, and so  growth in the length scale is also relevant~\cite{Dahl:2021wyk,Dahl:2024eup}.

The prediction for a causal velocity spectrum ($k^5$ at low wavenumber \cite{Durrer:2003ja}) is $\zeta = 10/7 \simeq 1.43$~\cite{Dahl:2021wyk,Dahl:2024eup}. 
Our fitted value for the detonation is consistent with the prediction, but the deflagration is quite far away. 
The simulations of Ref.~\cite{Caprini:2024gyk}, which have better coverage of wavenumbers below the peak, have kinetic energy decay indices closer to $10/7$ over a range of wall speeds at $\alpha_\text{n} = 0.5$ (see their Fig.~5).  However, the ``Higgsless'' method for injecting fluid flow into the system does not allow for slowing of the bubble walls in deflagrations as the metastable phase heats up.  It is possible that the kinetic energy decay  is influenced by the rate of disappearance of the hot droplets. We additionally remark that the abovementioned prediction assumes flow purely dominated by compressional modes, which is not the case for our deflagration, but may be a more reasonable approximation for Higgless simulations, where relatively less vorticity is generated. We will return to this point in Section ~\ref{sec:powerSpectra}.

There is also a prediction for the integral scale index of $\lambda = 2/7 \simeq 0.29$~\cite{Dahl:2024eup}. Once more the detonation has an index close to the prediction $0.22\pm0.02$, although the deflagration shows a constant (bar oscillations) integral scale, with corresponding $\lambda$ close to zero. The difference again may be due to the effect
of hot droplets or more relative generation of vorticity, or both.

\begin{table}
    \begin{ruledtabular}
    \begin{tabular}{ccc}
    Phase transition  & $\zeta$ & $\lambda$\\
    \hline
    \rule{0pt}{3ex}{$\alpha_\text{n} = 0.67$, $v_w=0.92$} & $1.48\pm0.14$ &  $0.23\pm0.02$\\ 
    {$\alpha_\text{n} = 0.50$, $v_w=0.44$}  & $0.86\pm0.04$ &  $0.01\pm0.01$  
    \end{tabular}
  \end{ruledtabular}
    \caption{\raggedright Time evolution indices of compressional kinetic energy $\zeta$ and the integral scale $\lambda$, from 
    fits to Equations (\ref{e:ZetDef}, \ref{e:LamDef}), after the maximum RMS weighted compression 4-velocity $\max(\bar{U}_\parallel)$ (in the range $t\in[t_\text{max}, t_\text{end}]$). 
    \label{tab:KxiInd}}
\end{table}

\begin{figure*}[p]
\includegraphics[width=1.0\columnwidth]{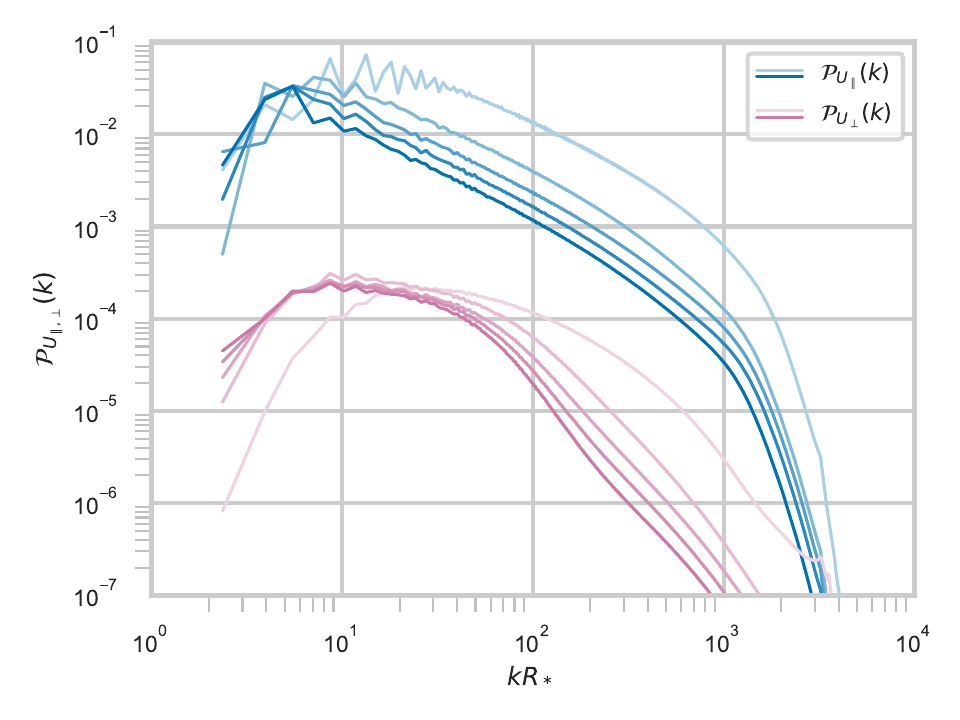}
\includegraphics[width=1.0\columnwidth]{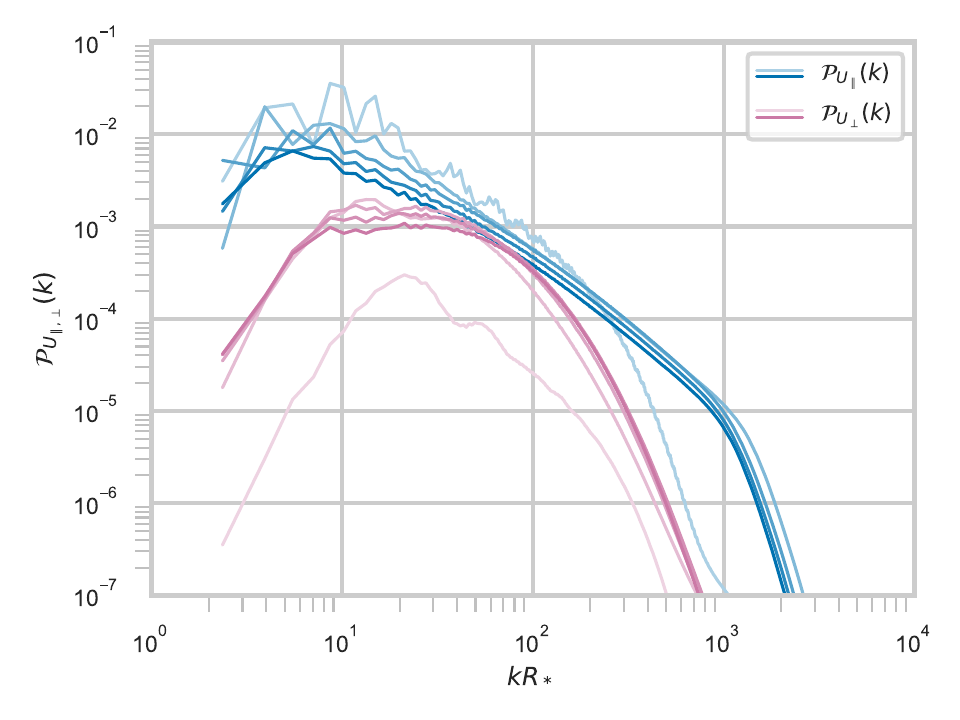}
\includegraphics[width=1.0\columnwidth]{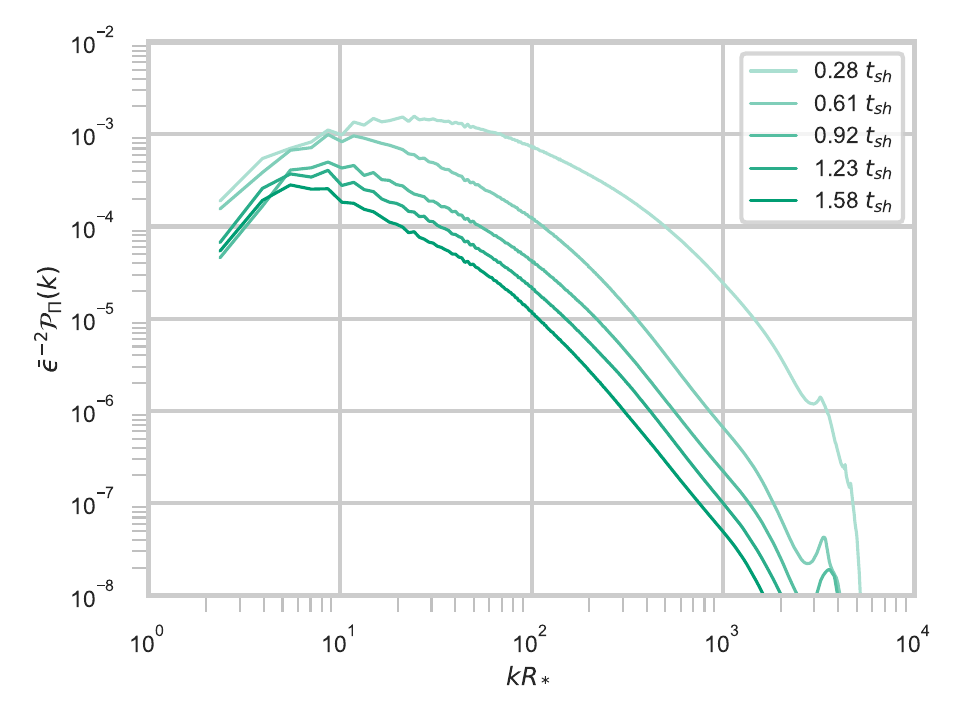}
\includegraphics[width=1.0\columnwidth]{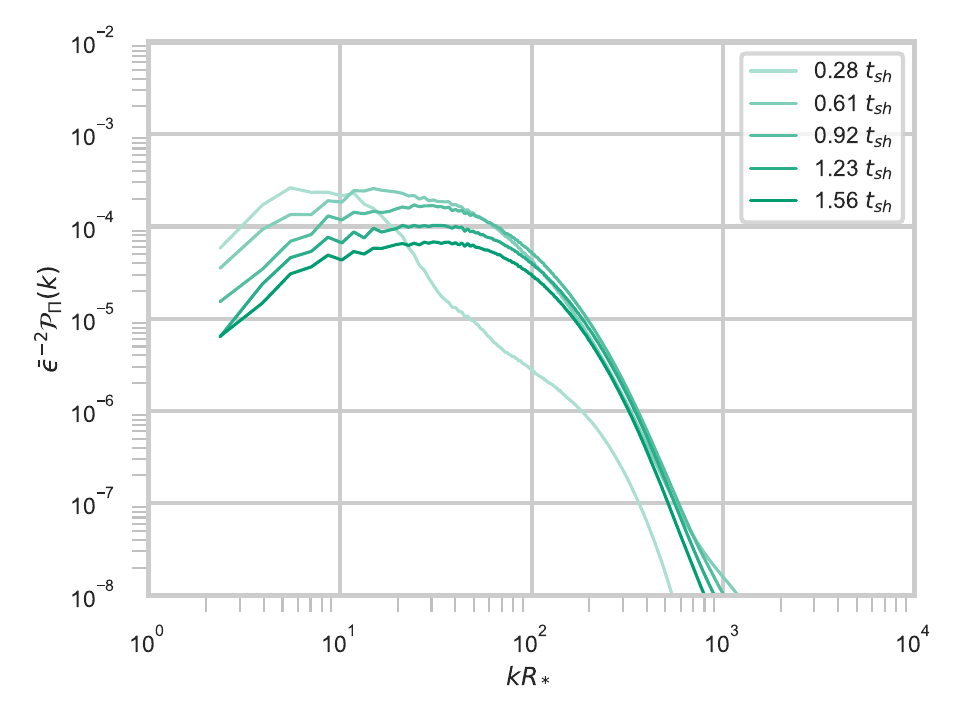}
\includegraphics[width=1.0\columnwidth]{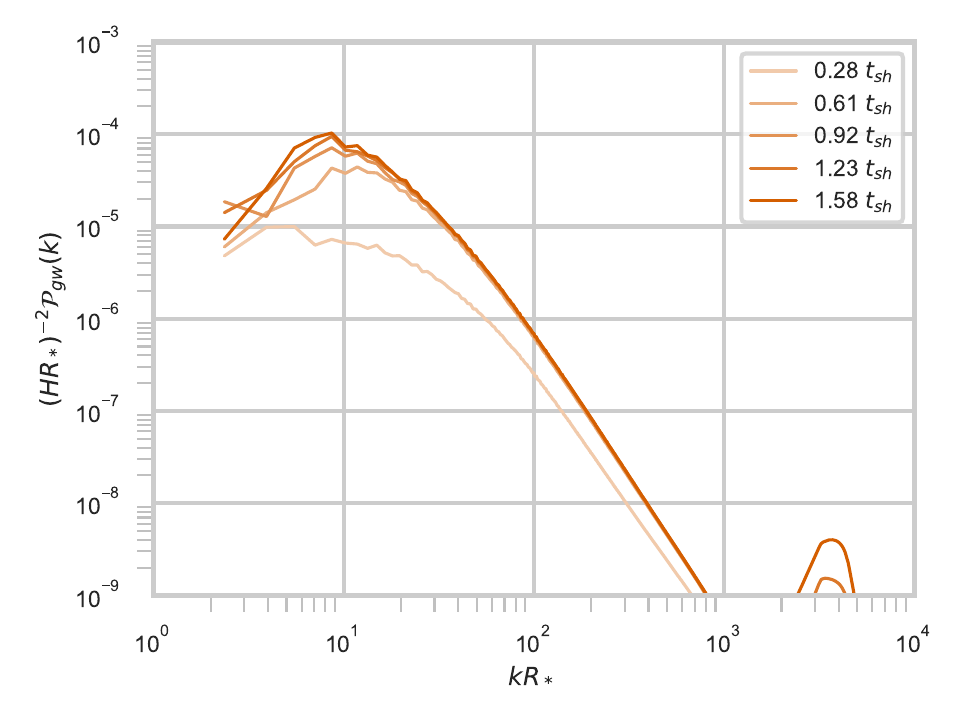}
\includegraphics[width=1.0\columnwidth]{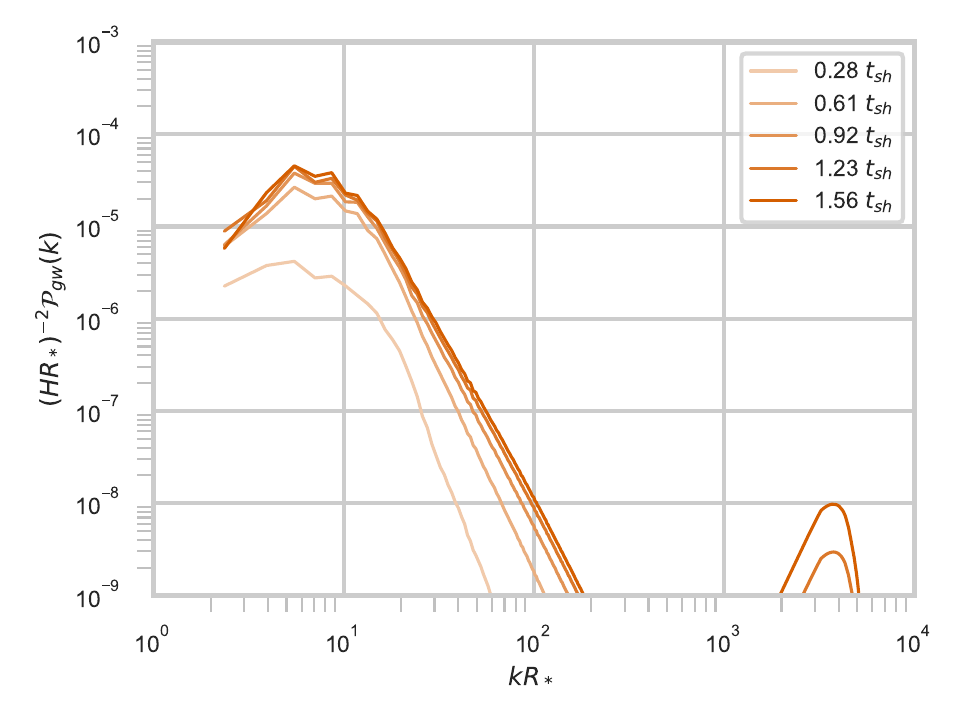}
\caption{\raggedright Power spectra for  the detonation ($\alpha_\text{n}=0.67$, $\vw=0.92$, left column) at times $t\Tc = 880, 1720, 2560, 3400,4240$, 
and the deflagration ($\alpha_\text{n}=0.50$, $\vw=0.44$, right column), at times $t \Tc = 1440, 2800, 4160, 5520, 6880 $.  The times in units of the shock appearance time are given in the legend. 
Top: power spectra of compressional modes $\mathcal{P}_{v_\parallel}$ (pastel blue to cyan) and vortical modes $\mathcal{P}_{v_\perp}$ (pastel pink to pink). 
Middle: power spectra of the shear stress $\mathcal{P}_\Pi$ (light teal to teal) in units of $\bar{\epsilon}^2$. Bottom: the gravitational wave power spectra $\mathcal{P}_\text{gw}$ (pastel orange to orange) in units of $(H_*R_*)^2$. 
\label{fig:power_spectra}}
\end{figure*}

\subsection{Power spectra and time decorrelation \label{sec:powerSpectra}}

We begin this section by centering our discussion on Fig.~\ref{fig:power_spectra}, where power spectra 
are displayed for selected times covering about 1.5 shock evolution times $\tsh$. 

In the top panels of Fig.~\ref{fig:power_spectra} we show power spectra of the enthalpy-weighted 4-velocity components $\mathcal{P}_{U_\parallel}$, $\mathcal{P}_{U_\perp}$.
In the detonation, there is much more power in the compressional modes than in the vortical modes. This is the case at all scales, consistent with our previous analysis of fluid average velocities. The deflagration, for which the average velocities in the compressional and vortical components were closer to one another, exhibits a particular range of scales $20 \lesssim kR_* \lesssim 200$ where there is a comparable amount of power in the compressional and vortical modes at later times in the range shown. In Fig. 16 of Ref.~\cite{Caprini:2024gyk} the maximum ratio of vortical to compressional spectral density power reached is $P_{v_\perp} / P_{v_\parallel} \approx 0.3$ for this deflagration, while we report that the maximum ratio of 3-velocity power is $\mathcal{P}_{v_\perp} / \mathcal{P}_{v_\parallel} \approx 1.11$ and for enthalpy-weighted 4-velocity power $\mathcal{P}_{U_\perp} / \mathcal{P}_{U_\parallel} \approx 0.94$. This signals that Higgsless simulations are relatively less efficient at generating power in vortical modes.

The compressional power spectrum $\mathcal{P}_{U_{\parallel}}$ shows a $k^{-1}$ dependence for $kR_* \gtrsim 10$, as expected for a fluid with shocks~\cite{Dahl:2021wyk,Dahl:2024eup}.
The vortical power spectrum shows a rather broad flat peak, with a noticeable tilt in the case of the detonation. The visual appearance of the vorticity in Fig.~\ref{fig:figureSlices} indicates that vortical turbulence is developing in the deflagration, but not in the detonation. The characteristic Kolmogorov behaviour of $k^{-2/3}$ in the power spectrum at wavenumbers higher than the peak has not yet developed. We note that the simulation evolves for less than one eddy turn-over time (see Table \ref{tab:tsh_teddy}).

The middle panels of Fig.~\ref{fig:power_spectra} show the shear stress power spectra $\mathcal{P}_\Pi (k)$. For the detonation, the shape of shear stress $\mathcal{P}_\Pi (k)$ is consistent with the $k^{-3}$ high-$k$ power law expected for a velocity spectrum going as $k^{-1}$~\cite{Hindmarsh:2016lnk,Hindmarsh:2019phv}. The deflagration, however, has quite a different shape. This can be ascribed to an important contribution from the vortical modes. The apparent impact of vortical modes on the shear stress (and their absence in the gravitational wave power spectra) will be supported by modelling work in Section ~\ref{sec:comparison}.

The bottom panels show the gravitational wave power spectra $\mathcal{P}_\text{gw}(k)$, with a $k^{-3}$ power law in $\mathcal{P}_\text{gw}(k)$ in both cases. 
This suggests that vortical modes are not contributing significantly to the resulting gravitational wave signal. Again, our modelling in Section ~\ref{fig:sound_shell} will support this suggestion.

\begin{figure*}[p]
\includegraphics[width=1.0\columnwidth]{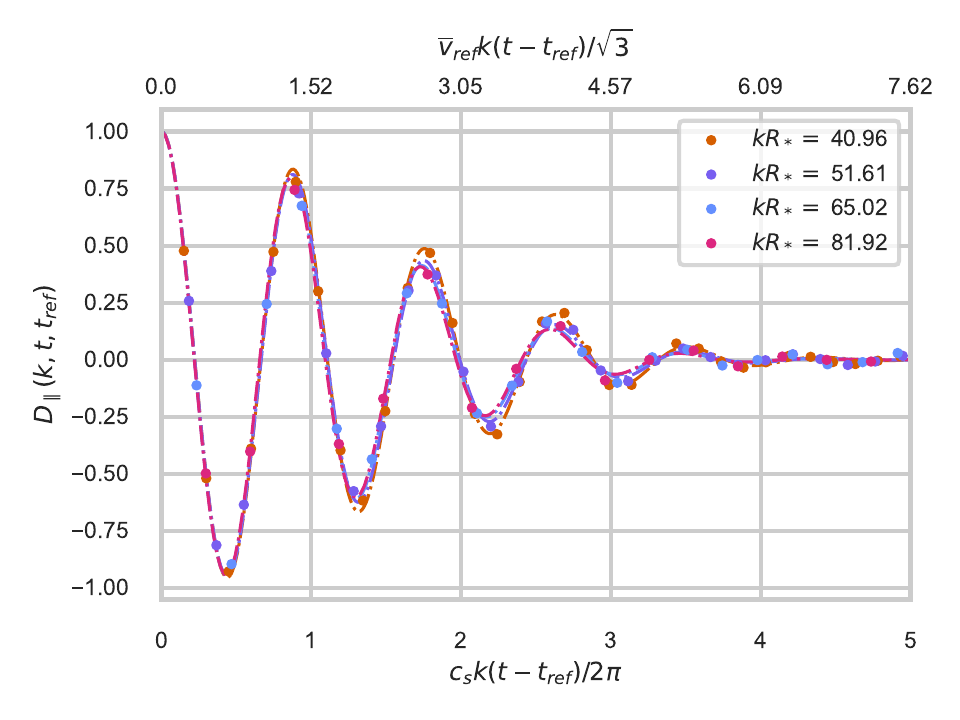}
\includegraphics[width=1.0\columnwidth]{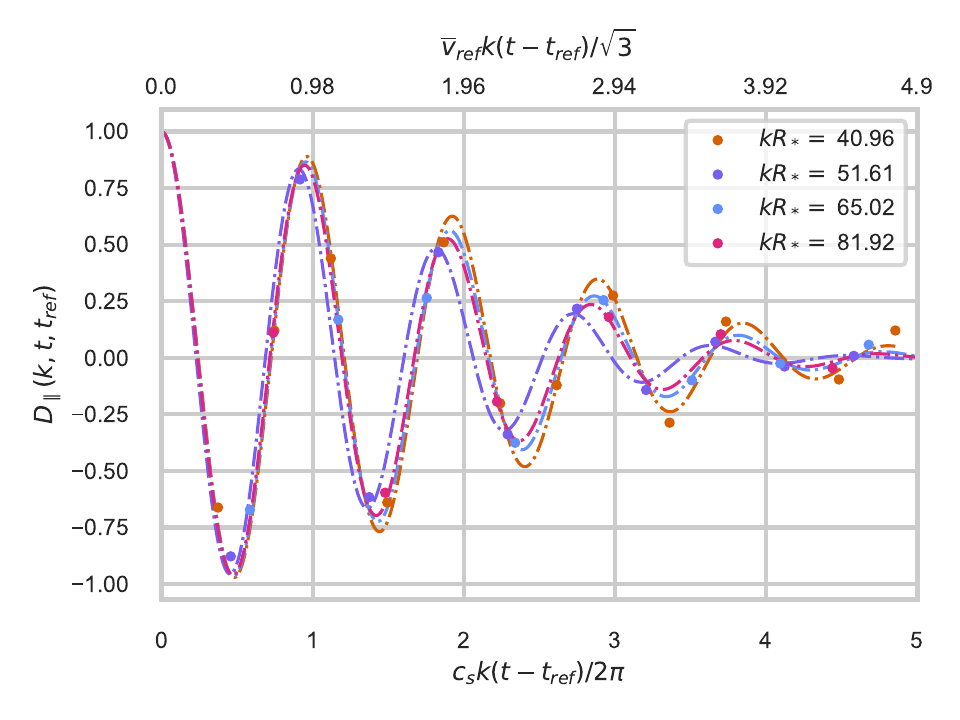}
\includegraphics[width=1.0\columnwidth]{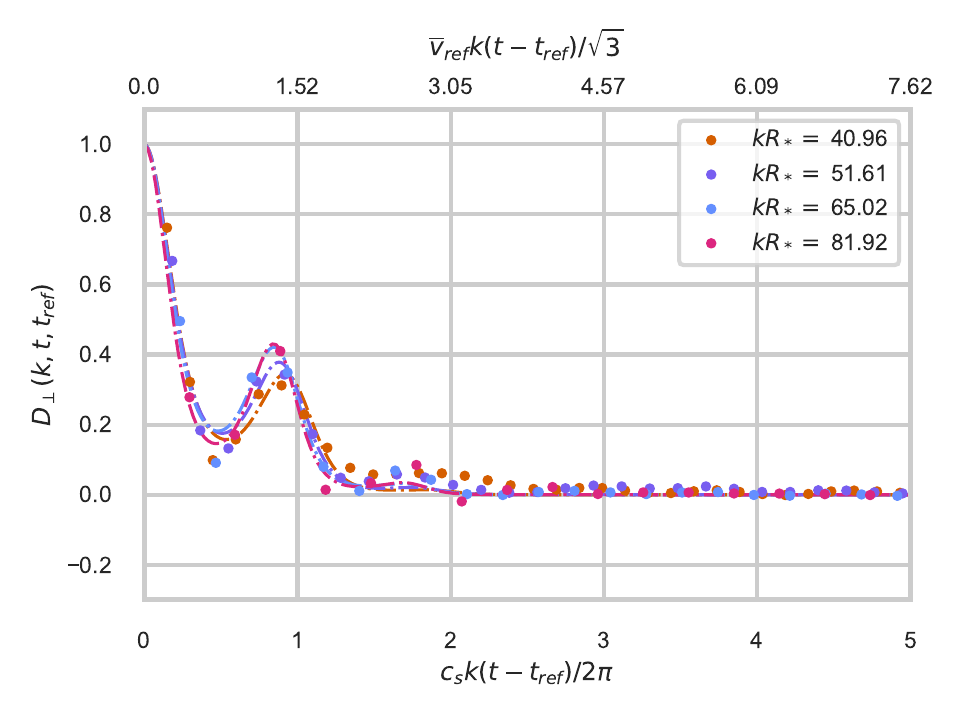}
\includegraphics[width=1.0\columnwidth]{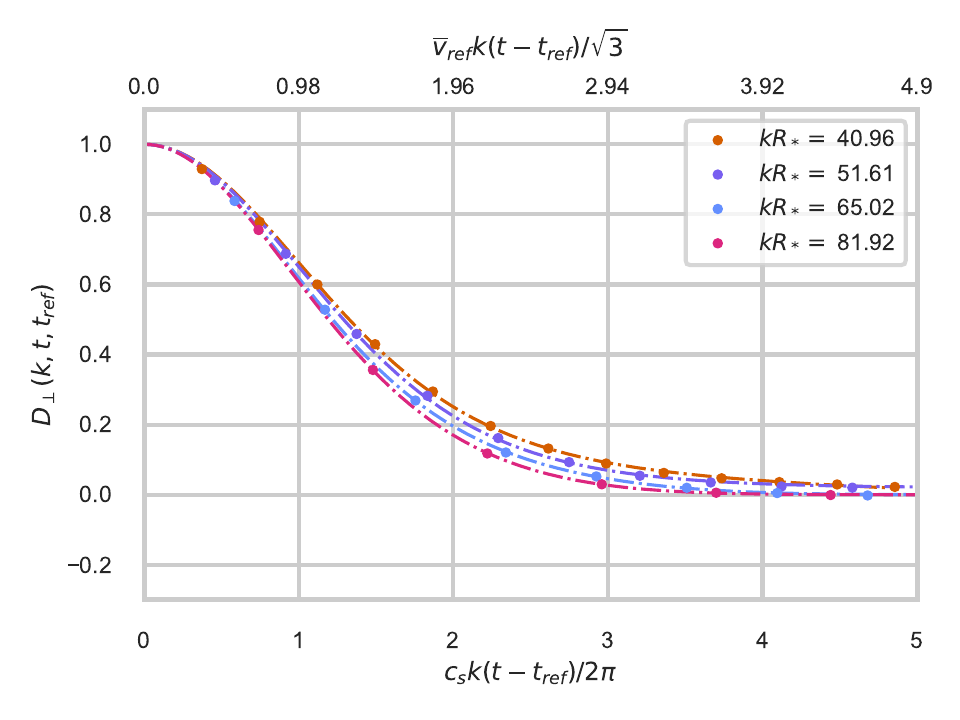}
\includegraphics[width=1.0\columnwidth]{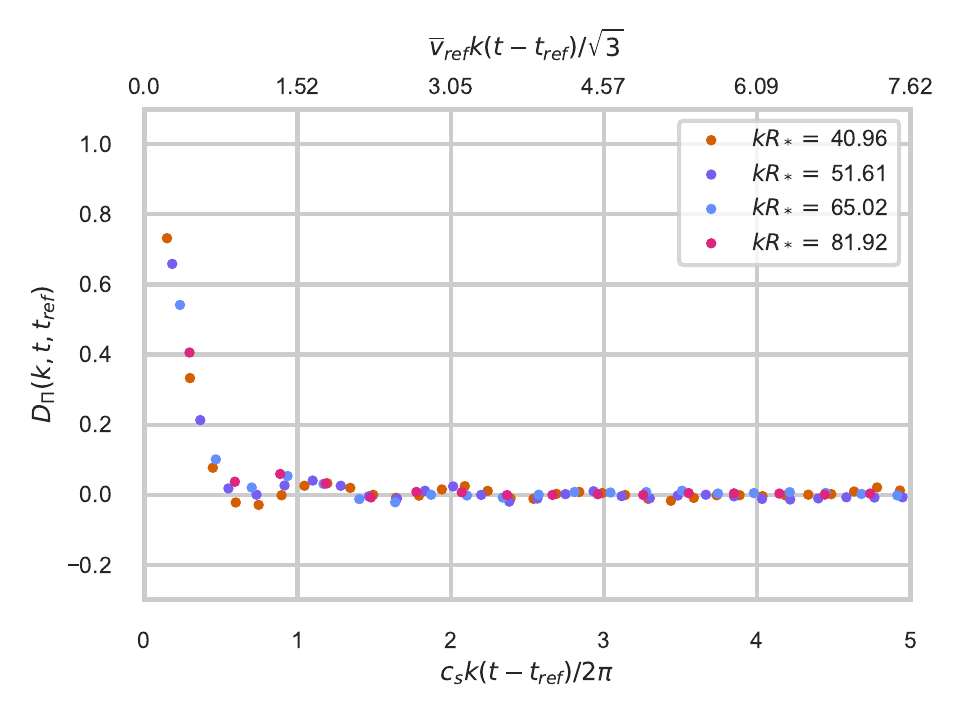}
\includegraphics[width=1.0\columnwidth]{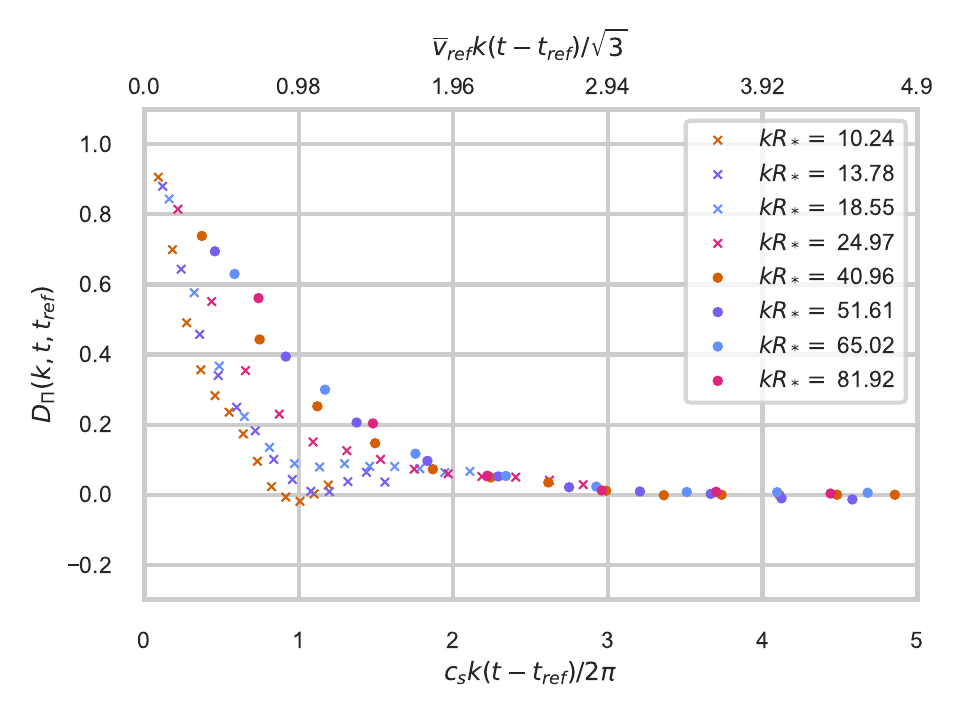}
\caption{\raggedright Decorrelation functions (defined in Eq. \ref{e:DecFunDef}) of $v_\parallel$ (top panels), $v_\perp$   (middle row) and shear stress $\Pi$ (bottom panels) for the detonation ($\alpha_\text{n}=0.67$, $\vw=0.92$, left-hand-side) and deflagration ($\alpha_\text{n}=0.50$, $\vw=0.44$, right-hand-side). $\overline{v}_\text{ref}$ is the volume averaged 3-velocity measured at the UETC reference time.
\label{fig:decorrelation_functions}}
\end{figure*}

We will now look at the time decorrelation of $v_{\parallel}$, $v_{\perp}$ and $\Pi$. 
The resulting decorrelation functions $D_{\parallel}(k)$, $D_{\perp}(k)$  and $D_\Pi (k)$ are presented in Fig.~\ref{fig:decorrelation_functions}, in the top, middle and bottom panels, respectively. 
We show a range of wavenumbers around the inverse integral scales (see Fig.~\ref{fig:U_vel_xi}). 

Beginning with the detonation,  the decorrelation function of longitudinal modes $D_{\parallel}$ shows oscillatory behaviour with period roughly corresponding to that of sound waves, and with decreasing amplitude.
A good fit to this decorrelation function is obtained with a modified version of Eq.~\eqref{e:DparExp}
\begin{multline}
D_{\parallel} (k, t, t_{\text{ref}}) = \cos \bigg( M_\parallel \cs k(t-t_\text{ref}) \bigg) \\
         \times \exp\bigg(-\frac{1}{2} \Vsw{\parallel}^2k^2 (t-t_\text{ref})^2 \bigg) 
\label{e:DecorPar}
\end{multline}
where $\Vsw{\parallel}$ can be thought of as a sweeping velocity driving the decorrelation of the compressional component of velocity, and $M_\parallel$ can be thought of as a Mach number.  We note that this type of decorrelation function has already been seen in other literature, for example in simulations of isotropic compressional turbulence of Ref.~\cite{PhysRevE.88.021001}. The fitted values can be found in the upper half of Table~\ref{tab:velocities}.  We note that the Mach number parameter $M_\parallel$ is slightly larger than one. The compressional modes are organised into shocks and hence can travel slightly faster than $\cs$.

If one were to apply the  Kraichnan random sweeping model, the parameter  $\Vsw{\parallel}$ would be equal to $v_\parallel/\sqrt{3}$, where $v_\parallel$  is the  globally-averaged RMS velocity in compressional modes. Values of the RMS velocity  given in Table~\ref{tab:decorrelation_fits}.  The RMS velocity prediction is about three times the fitted value of $\Vsw{\parallel}$.

\begin{table}
\begin{center}
\begin{ruledtabular}
\begin{tabular}{ D{.}{.}{2} D{.}{.}{3} D{.}{.}{3} D{.}{.}{4} D{.}{.}{3} D{.}{.}{4} }
    \multicolumn{6}{c}{$\alpha = 0.67$, $\vw=0.92$} \\ 
    \\[-1em]
    \multicolumn{1}{c}{$kR_*$} & \multicolumn{1}{c}{$\Vsw{\parallel}$} & \multicolumn{1}{c}{$M_\parallel$} & \multicolumn{1}{c}{$\Vsw{\perp}$} & \multicolumn{1}{c}{$A_\perp$} & \multicolumn{1}{c}{$N_\perp$} \\
    \hline
    40.96  & 0.063 & 1.130 & 0.1406 & 0.877 & 0.9153 \\
    51.61  & 0.067 & 1.130 & 0.1404 & 0.854 & 0.9594 \\
    65.02  & 0.070 & 1.139 & 0.1382 & 0.854 & 1.0055 \\
    81.92  & 0.071 & 1.144 & 0.1376 & 0.920 & 1.0193 \\      
    \hline
    \multicolumn{1}{c}{Average}  & 0.068 & 1.136 & 0.1392 & 0.876 & 0.975 \\      
    \hline
    \hline
    \\[-0.75em]
    \multicolumn{6}{c}{$\alpha = 0.5$, $\vw=0.44$} \\ 
    \multicolumn{1}{c}{$kR_*$} & \multicolumn{1}{c}{$\Vsw{\parallel}$} & \multicolumn{1}{c}{$M_\parallel$} & \multicolumn{1}{c}{$\Vsw{\perp}$} & \multicolumn{1}{c}{$A_\perp$} & \multicolumn{1}{c}{$N_\perp$} \\
    \hline
    40.96 & 0.046 & 1.03 & 0.077 & - & - \\
    51.61 & 0.061 & 1.09 & 0.080 & - & - \\
    65.02 & 0.051 & 1.04 & 0.086 & - & - \\
    81.92 & 0.055 & 1.05 & 0.089 & - & - \\
    \hline
    \multicolumn{1}{c}{Average} & 0.053 & 1.05 & 0.083 & - & - \\
\end{tabular}
 \end{ruledtabular}

    \caption{\raggedright Fitting parameters from analytical models of decorrelation functions for compressional \eqref{e:DecorPar} and vortical \eqref{e:DecorPerp} modes, at selected wavenumbers.  The parameter $C_\perp$ is set to zero for the vortical decorrelation function of the deflagration, as the Gaussian form is already a good fit.  The parameter $N_\perp$ is then irrelevant. An average over the wave numbers is also presented. 
    \label{tab:decorrelation_fits}}
\end{center}
\end{table}

\begin{table}
\begin{center}
\begin{ruledtabular}
  \begin{tabular}{c D{.}{.}{3} D{.}{.}{3}}
Phase transition & \multicolumn{1}{c}{$ \bar{v}_\parallel/\sqrt{3}$} & \multicolumn{1}{c}{$\bar{v}_\perp/\sqrt{3}$} \\
\hline
\rule{0pt}{3ex}{$\alpha = 0.67$, $v_w=0.92$} & 0.140 & 0.028 \\ 
{$\alpha = 0.50$, $v_w=0.44$}  & 0.080 & 0.050  
\end{tabular}
\end{ruledtabular}

    \caption{\raggedright Averages of the root mean square 3-velocities at the UETC reference time $t_\text{ref}$ and the end of the simulation $t_\text{end}$, for compressional and vortical modes ($\parallel, \perp$), divided by $\sqrt{3}$ to give the average in each vector component. 
     The upper row gives values for the detonation, the lower row for the deflagration. 
    \label{tab:velocities}}
\end{center}
\end{table}

Moving to the middle panel of Fig.~\ref{fig:decorrelation_functions}, and still considering the detonation, it is clear that the rotational component oscillates while it decorrelates, with period approximately $2\pi/\cs k$. 
These oscillations are not accounted for in the Kraichnan sweeping model. 
An improved fit can be obtained with
\begin{multline}
D_{\perp} (q, t_1, t_2) = \exp\left( -\frac{1}{2} \Vsw{\perp}^2 q^2 (t_1-t_2)^2 \right. \\
-  \left.  \frac{1}{2} A_\perp^2 \sin^2(N_\perp \cs q (t_1-t_2 ) )\right).   
\label{e:DecorPerp}
\end{multline}

The extra sinusoidal term in the exponent can be motivated by a model in which the vortical modes are ``washed'' back and forth by the oscillatory compressional modes. The fit parameters can again be found in the top half of Table~\ref{tab:velocities}.
The fit parameter $\Vsw{\perp}$ is 
much larger the sweeping velocity obtained via the root mean square velocity of vortical modes (Table~\ref{tab:decorrelation_fits}).
 However, we note that the decorrelation velocity $\Vsw{\perp}$ is quite close to $v_{\parallel}/\sqrt{3}$;
such a feature might not be totally unexpected, as the vorticity lives in a dominantly compressional background.

For the deflagration, similar features can be observed for the compressional decorrelation function, with the same decaying oscillatory behaviour as for the detonation. The vorticity decorrelates in the standard Gaussian fashion of the Kraichnan model. 
The fitted coefficients $\Vsw{\parallel}$ and $\Vsw{\perp}$ are similar in magnitude to the corresponding values for $v_\parallel/\sqrt{3}$ and $v_\perp/\sqrt{3}$.  It is interesting to note that there is closer correspondence between $\Vsw{\parallel}$ and $v_\perp/\sqrt{3}$, and vice versa, with the exception of the detonation, where $\Vsw{\parallel}$ is twice $v_\perp/\sqrt{3}$.  
The correspondence is particularly close for $\Vsw{\perp}$ and $v_\parallel/\sqrt{3}$. 
Inspecting slices of the three velocity in Fig.~\ref{fig:figureSlices}, shocks seem to propagate through a background of eddies while the eddies are washed back and forth by the shocks, which supports the idea of ``swapped" sweeping velocities.

The decorrelation of the shear stress at wavenumber $k$ in the deflagration is much less oscillatory that that of the compressional velocity, even though the flow is dominated by the compressional modes. We can understand this as a consequence of different wavenumbers in the velocity field contributing to the decorrelation, leading to destructive interference. The decorrelation time is around $\pi/\cs k$. 
In the deflagration, the characteristic Gaussian form of vortical mode decorrelation is visible at wavenumbers where the velocity power spectrum is more vortical, in the range  $20 \lesssim kR_* \lesssim 200$.  The decorrelation time is longer, and similar to that of the vortical velocity mode.  At lower wave numbers, where compressional modes dominate, the decorrelation function looks more like that of the detonation.

\section{GW power spectrum from velocity UETCs}
\label{s:GWpsVelps}

\subsection{Theoretical framework}
\label{s:theory}

We continue to suppose that the fluid flow is set up and decays rapidly in comparison with the expansion rate, so that a Minkowski space approximation during the production of gravitational waves is appropriate. The gravitational wave power spectrum then follows directly from the shear stress correlation function
\begin{multline}
\label{e:Pgw1}
\mcPgw(k, t) =  \frac{(16\pi G)^2}{12 \Hn^2} \\ \times \int_{t_*}^t dt_1 \int_{t_*}^t dt_2 \dot G(k, t, t_1)\dot G(k, t, t_2)  \mathcal{P}_\Pi (k, t_1, t_2),
\end{multline}
where we have used the Friedmann equation to replace energy density with the Hubble rate, and $t_*$ is the time the flow reaches its maximum.
We assume this is reached rapidly so that we can neglect gravitational waves generated before then.
The function $\dot G(k,t, t_1) = \cos[k(t - t_1)]$ is the Minkowski space Green's function for $\dot{h}_{ij}$.

In order to compute the shear stress UETC, we assume that $\bU$ is a Gaussian random vector field (see Refs.~\cite{Caprini:2009fx,Hindmarsh:2019phv}) with a characteristic length scale $R$, such that the spectral densities can be written 
\begin{equation}
P_{U_A}(q, t) = R^3(t) \bar{U}_A^2(t) \tilde{P}_{U_A}(qR(t)), 
\end{equation}
where the indices $A,B$ take values $\parallel$ (compressional) and $\perp$ (vortical), and 
where $\int dx \, x^2 \tilde{P}_{U_A}(x)/2\pi^2 = 1$. The scale can be fixed by requiring that the maximum of $\tilde{P}_{U_A}(x)$ is at $x=1$, or some other convenient choice. We will take $R$ to coincide with $R_*$ at the time the kinetic energy density reaches its maximum.

It follows that the shear stress UETC can be written in terms of convolutions of the spatial components of the enthalpy-weighted 4-velocities as 
 \begin{equation}
    {P}_\Pi(k, t_1,t_2) = \bar{R}^3\bar{w}(t_1) \bar{w}(t_2)\sum_{A,B}C[U_A,U_B](k, t_1,t_2),
    \label{e:shstCon}
 \end{equation}
where
\begin{multline}
C[U_A,U_B](k,t_1,t_2) \\
= \bar{R}^{-3}\int \frac{d^3q}{(2\pi)^3}    \GeoFac_{AB}(\mu, \tilde{\mu}) P_{U_A}(q,t_1,t_2) P_{U_B}(\tilde{q},t_1,t_2),
\label{e:ConDef}
\end{multline}
and $\bar{R} = \sqrt{R(t_1)R(t_2)}$.

In the convolutions, $\tilde{q} = |\tilde{\bq}|$ and $\tilde{\bq} = \bq - \bk$,  and the functions $\GeoFac_{AB}$ are geometrical factors from the projector algebra.
They take the form 
\begin{align}
    \GeoFac_{\parallel \parallel} & = (1-\mu^2)(1-\tilde{\mu}^2) \\
    \GeoFac_{\perp \perp} & = (1+\mu^2)(1+\tilde{\mu}^2) \\
    \GeoFac_{\parallel \perp} & = (1-\mu^2)(1+\tilde{\mu}^2),
\end{align}
where $\mu=\bq \cdot \bk/qk$ and $\tilde\mu=\tilde\bq \cdot \bk/\tilde{q}k$.
We note also that 
\begin{align}
\tilde{q}^2 &= q^2 - 2kq\mu + k^2, \\ 
\tilde{\mu}^2 &= 1 - (1-\mu^2)q^2/\tilde{q}^2.
\end{align}
Note that $C$ has the dimensions of velocity to the fourth power.

The computation of the gravitational wave power spectrum is in principle a 4-dimensional integral, after using the azimuthal symmetry of the wavenumber integration in the convolution.  In the following we make some approximations which enable one of the  time integrations to be performed, and we study the rate of change of the gravitational wave spectrum, reducing the integration to a less costly 2-dimensional one, which forms the basis of current modelling~\cite{Hindmarsh:2016lnk,Hindmarsh:2019phv}.

The time dependence of the length scale $R$ and the mean square velocity of a fluid losing energy through shocks follow~\cite{Dahl:2024eup}
\begin{align}
R(t) & = R_*[ 1 + (\Delta t/\tdecay )]^\la, \label{eq:RPowLawInd} \\
\bar{U}^2_A(t) & = \bar{U}^2_{A,*}[ 1 + (\Delta t/\tdecay )]^{-\zeta}, 
\label{eq:UPowLawInd}
\end{align}
where $\la$ and $\ze$ are positive, $\tdecay $ is a constant proportional to the shock decay time $\tsh \equiv R_*/\bar{U}_\parallel$, and $\Delta t$ is the time since the maximum kinetic energy was reached, or $\Delta t = t - t_*$. 

The appearance of power laws suggests that a useful pair of time coordinates are the geometric mean and the ratio of shifted times,  $t ' =  \tdecay  + \Delta t$, that is 
\begin{equation}
\bar{t} = \sqrt{t'_1t'_2}, \qquad r = t'_2/t'_1.
\label{e:TimCoo}
\end{equation}
Then, for example, $\bar{R} = R_* (\bar{t}/\tdecay)^{\lambda}$.

We then define the geometric mean of the equal time correlators
\begin{equation}
\bar{P}_{U_A}(q, t_1, t_2) =   \sqrt{ P_{U_A}(q,t_1)P_{U_A}(q,t_2)},
\end{equation}
so that the velocity UETCs can be written
\begin{equation}
{P}_{U_A}(q, t_1, t_2) =  \bar{P}_{U_A}(q,  t_1, t_2) D_{A}(q,t_1,t_2).
\end{equation}
We now consider wavenumber ranges where the spectral densities are power laws in $k$, which is approximately true for a wide range of our data. 
In this case we can write $ P_{U_A}(q,t_1) = R^3(t_1) \bar{U}_{A,*}^2 \tilde{P}_{U_A}(qR_1)^{n-3}$, with $\tilde{P}_{U_A,*}$ a constant. 
Then we have that the geometric mean of the power spectra is independent of $r$, and 
\begin{equation}
\bar{P}_{U_A}(q, \bar{t}) \simeq R_*^{3} \bar{U}^2_{A,*} (\bar{t}/\tdecay )^{n\la - \ze}  \tilde{P}_{U_A,*}(q R_*)^{n-3} .
\end{equation}

\begin{widetext}
Returning to the gravitational wave power spectrum, and changing integration variables to $\bar{t}$ and $r$, we have that the change in the power spectrum between times $\tinit$ and $t$ ($\tinit' = t_d + \Delta t_i$ defined analogously to $t'$), both assumed to be much greater than $\tdecay $, is 
\begin{equation}
\Delta \mcPgw(k, t, \tinit) \simeq  3  \int_{\tinit'}^{t'} \frac{d\bar{t}}{\bar{R}} (\Hn \bar{R} )^2 \frac{(k\bar{R})^3}{2\pi^2} \Gamma^2   \sum_{A,B} C_\Delta [U_A, U_B](k,\bar{t})
\label{e:PgwOne}
\end{equation}
where $\Gamma = \bar{w}/\bar{\ep}_0$, $\bar{R} \simeq R_*(\bar{t}/\tdecay )^\lambda$, 
and 
\begin{equation}
C_\Delta [U_A, U_B](k,\bar{t}) = \bar{R}^{-3}\int \frac{d^3q}{(2\pi)^3} \; \GeoFac_{AB}(\mu, \tilde{\mu}) \Delta_{AB}(q, \tilde{q}, k)\bar{P}_{U_A}(q,\bar{t})\bar{P}_{U_B}(\tilde{q},\bar{t}).
\end{equation}
The functions $\Delta_{AB}$ are integration kernels, defined from integrating the product of the velocity decorrelation functions \eqref{e:DecorPar} and \eqref{e:DecorPerp} and the gravitational wave Green's functions, with respect to $r$, 
\begin{equation}
\Delta_{AB}(q, \tilde{q}, k) = 
\frac{\bar{t}}{2\bar{R}}
\int_{\rinit^{-1}}^{\rinit} \frac{dr}{r} \cos[k(t_1 - t_2)]
D_{A}(q,t_1,t_2) D_{B}(\tilde{q},t_1,t_2) ,
\label{e:KerDef}
\end{equation}
where $\rinit = t'/\tinit'$.
We show in Appendix~\ref{ap:gwKernels} that, for $k\bar{t} \gg 1$, the resulting kernels are independent of $\bar{t}$ and good approximations are
    \begin{align}
         \Delta_{\perp \perp}(q,\tilde{q},k) &= \frac{1}{2}\sqrt{\frac{2\pi}{{\omega}_{\perp\perp}^2\bar{R}^2}} \exp\bigg( -\frac{1}{2}\frac{k^2}{{\omega}_{\perp\perp}^2}\bigg) \; ,
         \label{e:DPerpPerp}\\
        \Delta_{\parallel \perp}(q,\tilde{q},k) &= \frac{1}{4}\sqrt{\frac{2\pi}{{\omega}_{\parallel\perp}^2\bar{R}^2}} \bigg[  \exp\bigg(  -\frac{1}{2}\frac{(k - \cs' q)^2}{{\omega}_{\parallel\perp}^2} \bigg) + \exp \bigg(  -\frac{1}{2}\frac{(k+\cs' q)^2}{{\omega}_{\parallel\perp}^2} \bigg) \bigg] \; ,
         \label{e:DParPerp}\\
        \Delta_{\parallel \parallel}(q,\tilde{q},k) &= \frac{1}{8} \sqrt{\frac{2\pi}{{\omega}_{\parallel\parallel}^2\bar{R}^2}}\sum_{\pm \pm}\exp\bigg(-\frac{1}{2}\frac{(k \pm \cs'q \pm \cs' \tilde{q})^2}{{\omega}_{\parallel\parallel}^2}\bigg),
        \label{e:DParPar}
    \end{align}
\end{widetext}
where 
\begin{align}
\omega_A &= \Vsw{A} q, \\
\tilde\omega_A &= \Vsw{A} \tilde{q}, \\
\omega_{AB} &= \sqrt{\omega_A^2 + \tilde\omega_B^2 },
\end{align}
and $\cs' = M_\parallel \cs $, 
as an approximation to the fit results in the previous section. 
The leading term is the $--$ term in the sum over signs in $\Delta_{\parallel\parallel}$ \cite{Hindmarsh:2016lnk,Hindmarsh:2019phv,RoperPol:2023dzg,Sharma:2023mao}, 
which produces the characteristic sharply rising ($k^9$) peak in the power spectrum for long-lived flows.  In our simulations, the characteristic scale is initially set by the mean bubble spacing $R_*$.

In order to distinguish the gravitational production by the fluid flow after the transition has completed from that produced during the  bubble collisions, it is useful to look at the rate of increase in the gravitational wave power spectrum. In numerical simulations, we approximate this with a finite difference, which from 
 Eq.~\eqref{e:PgwOne} takes the form
 \begin{equation}
\frac{1}{\Hn}\frac{\Delta \mcPgw}{\Delta t}(k, \bar{t}) \simeq  3 \Gamma^2 (\Hn \bar{R})  \frac{(k\bar{R})^3}{2\pi^2} \sum_{A,B} C_\Delta [U_A, U_B](k, \bar{t}).  
\label{e:dGWdt}
 \end{equation}
The integral in the convolution is carried out at fixed $\bar{t} = \sqrt{t'\tinit'}$, with the geometric mean spectral densities $\bar{P}_{U_A}$ calculated from spectra at times $\tinit$ and $t$.

The dependence of the right hand side on $k$ is purely through the combination $k\bar{R}$.

The time dependence of the right hand side of Eq.~\eqref{e:dGWdt}, in the wavenumber range where the velocity spectral densities are pure power laws, is through the RMS velocities $\bar{U}_A$ and the characteristic length scale $\bar{R}$, which we have assumed in the analysis grow as power laws.  Then, the dominant term amongst the three convolutions depends on wavenumber $k$, length scale $R$ and RMS 4-velocity $\bar{U}_\parallel$ as 
\begin{equation}
C_\Delta [U_\parallel, U_\parallel](k, \bar{t}) \sim \ (k \bar{R})^{2(n - 2)} \bar{U}_\parallel^4
\end{equation}
from which we deduce 
\begin{equation}
\frac{1}{\Hn}\frac{\Delta \mcPgw}{\Delta t}(k, \bar{t})  \sim (\Hn \bar{R}) (k\bar{R})^{2n - 1} (\Gamma\bar{U}_\parallel^2)^2 .
\label{e:dGWdtPowLaw}
\end{equation}
Inserting the power-law time dependences of $\bar{R}$ and $\bar{U}_\parallel$ from Eqs.~(\ref{eq:RPowLawInd}, ~\ref{eq:UPowLawInd}), we have
\begin{multline}
\frac{1}{\Hn}\frac{\Delta \mcPgw}{\Delta t}(k, \bar{t}) \\
  \sim (\Hn R_*) (kR_*)^{2n - 1} (\Gamma\bar{U}_{\parallel,*}^2)^2 \left( \frac{\bar{t}}{\tdecay }\right)^{2n\la - 2\ze}.
\label{e:dGWdtPowLawTim}
\end{multline}
At wavenumbers in the range between the peak wavenumber and the inverse shock width we expect the velocity power spectrum to behave as $k^{-1}$ (i.e.~with power law index $n = -1$), characteristic of shocks~\cite{Dahl:2021wyk,Dahl:2024eup}.  

Hence, provided $2n\la - 2\ze < -1$, we expect the gravitational wave power spectrum to approach an asymptote as 
\begin{equation}
\mcPgw(k, t) = \mcPgw^\infty(k)\left[ 1  - \left(\frac{t}{\tdecay }\right)^{1 - 2\la - 2\ze} \right],
\end{equation}
with  $\mcPgw^\infty(k) \sim (\Hn \tsh ) (\Hn R_*)  (\Gamma\bar{U}_{\parallel,*}^2)^2 (kR_*)^{-3}$, and we have used the assumption that the flow decay timescale is of order the shock decay timescale, $\tdecay \sim \tsh$.  The $k^{-3}$ power law is the imprint of shocks in the gravitational wave power spectrum~\cite{Hindmarsh:2016lnk,Hindmarsh:2019phv}. 

Note that the first investigation into kinetic energy decay and emisted gravitational wave power came from Ref.~\cite{Caprini:2024gyk}, and it was conducted using the Higgsless simulations on a wider range of values of phase transition strength $\alpha_n$ and wall velocity $v_w$. However, the integral scale evolution was not taken into account.

In Refs.~\cite{Dahl:2021wyk,Dahl:2024eup} it was shown that decaying non-relativistic acoustic turbulence with a long-distance velocity power spectrum going as $k^{\beta + 1}$ has 
\begin{equation}
\la = 2 / (\beta + 3), \quad \ze = 2(\beta+1)/(\beta + 3).  
\end{equation}
The analyticity requirement of the Fourier transform in infinite volume gives $\beta = 4$ (see Ref.~\cite{Durrer:2003ja}).  Our simulations do not have good scale separation between the box size and the peak wavelength, and are mildly relativistic, so we do not necessarily expect to see the $\la = 2/7$ and $\zeta = 10/7$ power laws emerging.   We do however expect that the parametric dependence of the gravitational wave power spectrum on $R_*$ and $\bar{U}_{\parallel,*}$ is correctly captured by these arguments. Recalling that the velocity decay timescale $t_*$ is proportional to the shock timescale $\tsh \equiv R_*/\bar{U}_{\parallel,*}$, and so 
\begin{equation}
\mcPgw^\infty(k) = (\Hn \tsh) (\Hn R_*)  (\Gamma\bar{U}_{\parallel}^2)_*^2 \frac{(kR_*)^3}{2\pi^2} 3 \tilde{P}_\text{gw}(kR_*),
\end{equation}
where $ \tilde{P}_\text{gw}$ is a dimensionless function of its argument, and the other factors have been inserted for consistency with Ref.~\cite{Hindmarsh:2019phv}. The gravitational wave density fraction is then
\begin{equation}
\Omgw^\infty = (\Hn \tsh) (\Hn R_*)  (\Gamma\bar{U}_{\parallel}^2)_*^2  3 \tilde\Omega_\text{gw}^\infty. 
\label{e:OmegaGWDef}
\end{equation}
The factor 3, while somewhat redundant, is included for consistency with earlier work~\cite{Hindmarsh:2015qta,Hindmarsh:2017gnf}.

\subsection{Comparison to simulations}
\label{sec:comparison}

\begin{figure*}
\includegraphics[width=1.0\columnwidth]{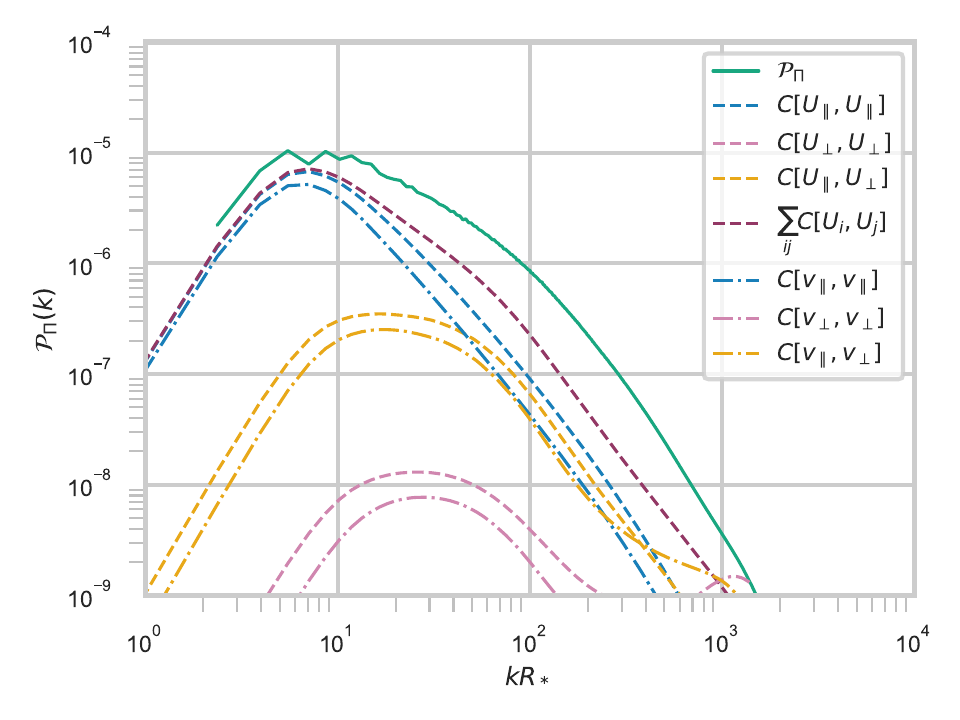}
\includegraphics[width=1.0\columnwidth]{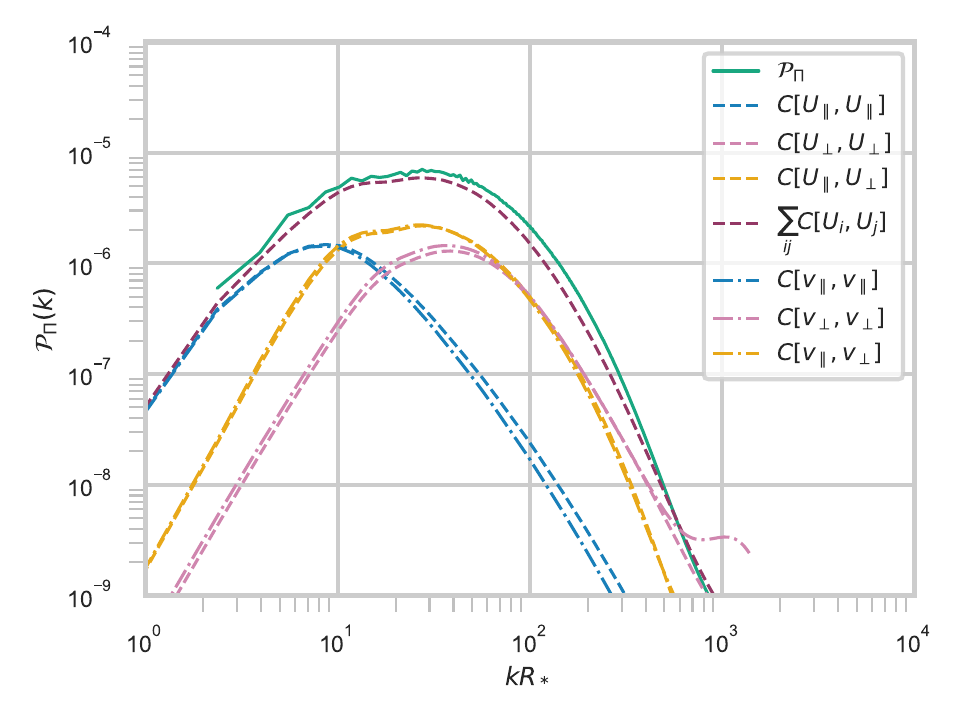}
\includegraphics[width=1.0\columnwidth]{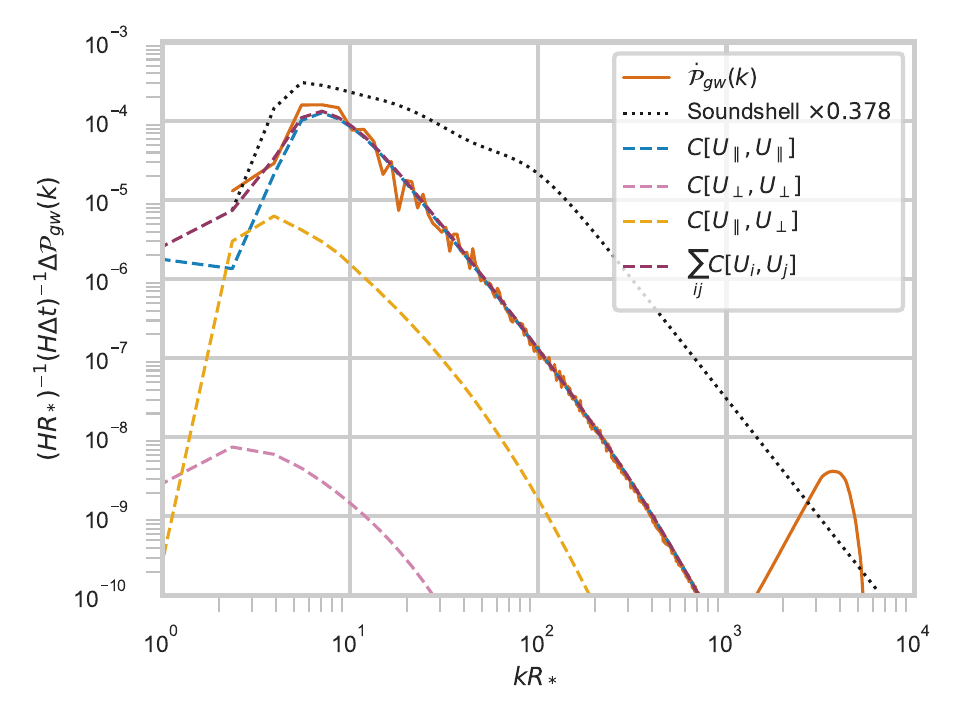}
\includegraphics[width=1.0\columnwidth]{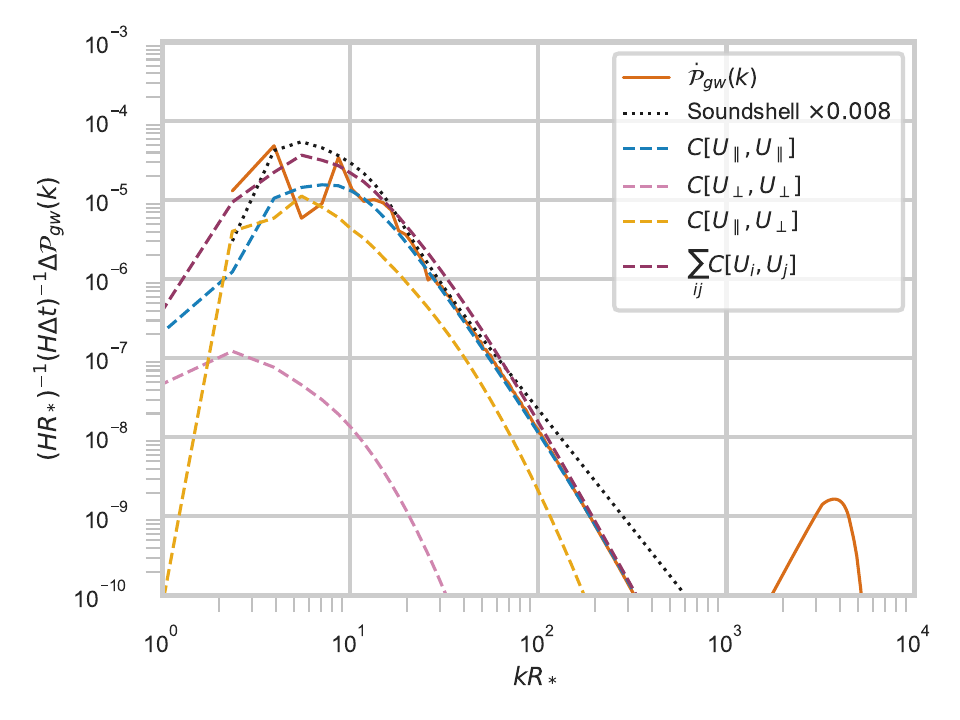}
\caption{\raggedright
Top row: comparison of the measured shear stress power spectrum (solid lines) normalised by energy density squared $\mcP_\Pi/\bar{\ep}^2$ with that obtained by using Eqs.~(\ref{e:shstCon}, ~\ref{e:ConDef}) to convolve the measured weighted 4-velocity power spectra (dashed lines) and the 3-velocity power spectra (dash-dotted lines). Left panel:  detonation ($\alpha_\text{n}=0.67$, $\vw=0.92$) at time $3600/T_c$; right panel: deflagration ($\alpha_\text{n}=0.5$, $\vw=0.44$) at time $4400/T_c$.
Bottom row: comparison of the measured instantaneous rate of change of the gravitational wave power spectra $\dot{\mathcal{P}}_\text{gw}$ (solid lines) with that computed via the convolution model prediction, Eq.~\eqref{e:dGWdt} (dashed lines). Left panel:  detonation ($\alpha_\text{n}=0.67$, $\vw=0.92$) with growth between times $2400/T_c$ and $3600/T_c$ ; right panel: deflagration ($\alpha_\text{n}=0.5$, $\vw=0.44$) with growth between times $3600/T_c$ and $4400/T_c$. For both the top and bottom rows, the convolution predictions are shown with dashed lines in four colours, representing the three terms (purely compressional, mixed compressional-vortical, and purely vortical) and their sum.  In the bottom row we show with a dotted line the sound shell model prediction, including the phenomenological suppression factor~\cite{Cutting:2019zws}, whose value is given in the legend. 
\label{fig:sound_shell}}
\end{figure*}

In Fig.~\ref{fig:sound_shell} (top row) we present the shear stress power spectra at a single time, along with those obtained from the convolutions of velocity power spectra, Eqs.~(\ref{e:shstCon},~\ref{e:ConDef}). We convolve with 3-velocity spectra and weighted 4-velocity spectra.  The sum of the convolution contributions is close in shape to the measured spectra, but the weighted 4-velocity spectra are closer to the measured amplitude, particularly for the detonation. This motivates the use of the weighted 4-velocity in the gravitational wave production calculation.  
The remaining mismatch signals a departure from Gaussianity for the weighted 4-velocity field.

In the bottom row of Fig.~\ref{fig:sound_shell}, we present the approximate rate of change of the gravitational wave power spectra $\Delta \mcPgw / \Delta t$ along the convolution model prediction (\ref{e:dGWdt}), broken down by components. The parameters for the kernels are summarised in Table \ref{tab:decorrelation_fits} in the row labelled ``Average". 

The rate of change of $\mcPgw$ is computed from a first order difference in time, between times given in the caption, and compared to the sum of convolutions~\eqref{e:dGWdt}, which uses the geometric mean of the $U$-velocity power spectra at the two different times. 
We divide the spectra by $\Hn^2 \bar{R}$, where $\bar{R}$ is given by 
\begin{equation}
\bar{R} = R_* \left[{\xi_{U_\parallel}(t_1)\xi_{U_\parallel}(t_2)}/ \xi^2_{U_\parallel}(t_*) \right]^{1/2}.
\label{e:RbarDef}
\end{equation}
We plot against $kR_*$ to make explicit the scale evolution. 

We see that both the amplitude and shape are in good agreement with the convolution prediction, indeed perhaps better agreement than the shear stress power spectrum. We might expect to see the same differences between convolution prediction and simulation as in the shear stress power spectrum, as the GW spectrum is obtained by a linear operation from the shear stress UETC.  There are several approximations made to go from the shear stress UETC to the GW spectrum, so the better agreement is perhaps a fortuitous cancellation of the non-Gaussianity with errors in the approximations.

A noticeable feature is that the GW power is almost entirely accounted for by $C_\Delta[U_\parallel,U_\parallel]$, the pure compressional emission, even in the deflagration where the vortical modes dominate in the range $20 \lesssim kR_* \lesssim 200$.  The next most significant contribution is the mixed compressional-vortical emission $C_\Delta[U_\parallel,U_\perp]$.  Pure vortical emission is subdominant in both cases, but with a smaller slope at high wavenumbers, and thus may dominate at high wavenumbers, albeit where the signal is very weak.

\begin{figure*}[p]
\includegraphics[width=1.0\columnwidth]{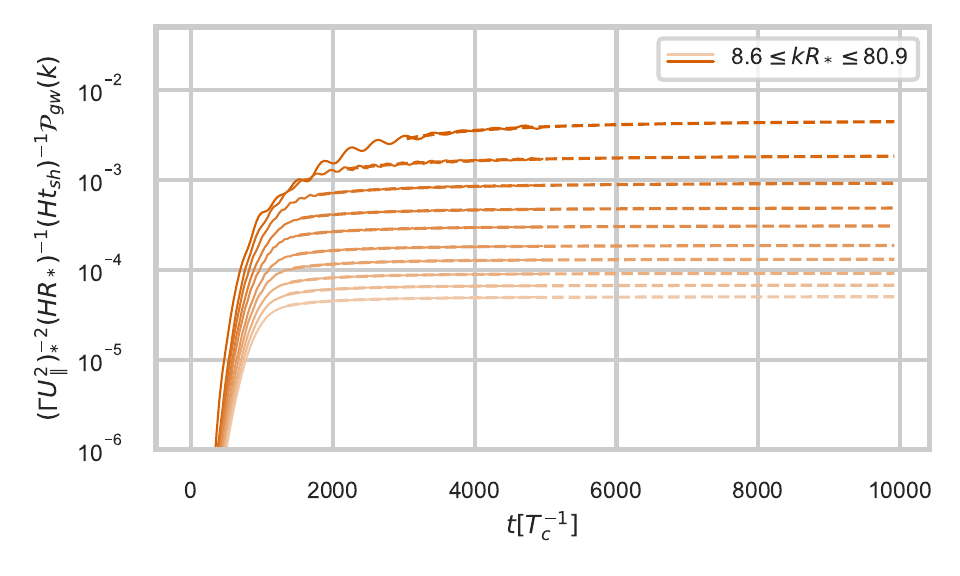}
\includegraphics[width=1.0\columnwidth]{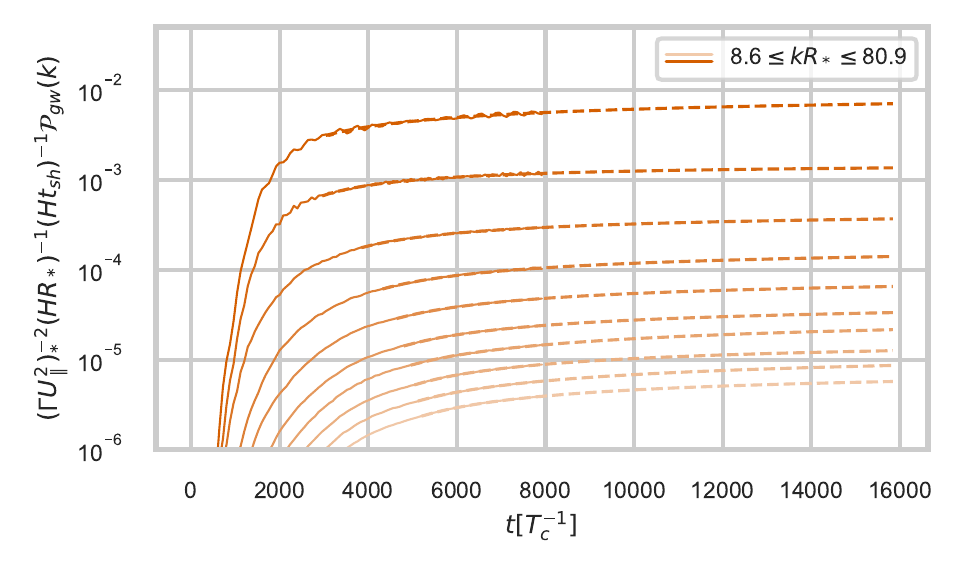}
\includegraphics[width=1.0\columnwidth]{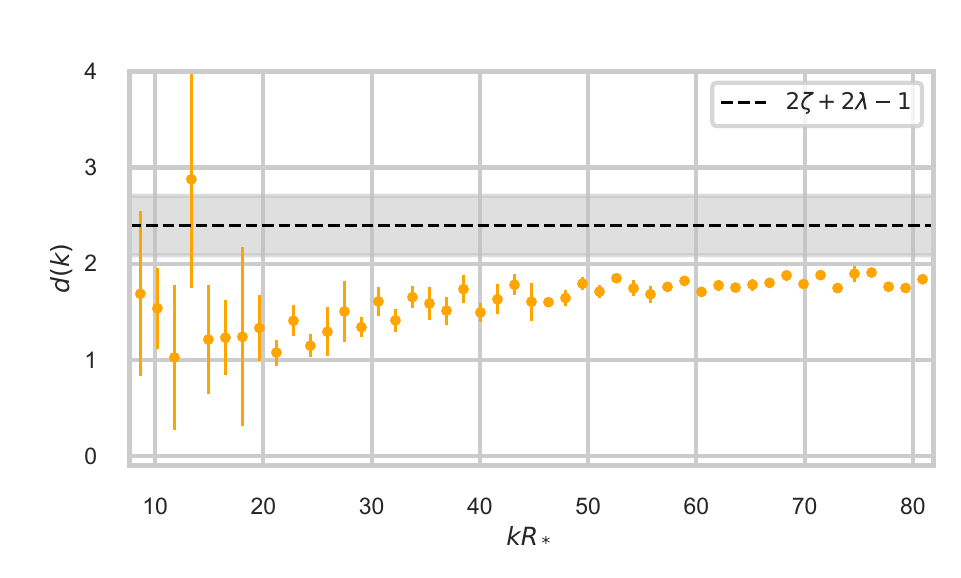}
 \includegraphics[width=1.0\columnwidth]{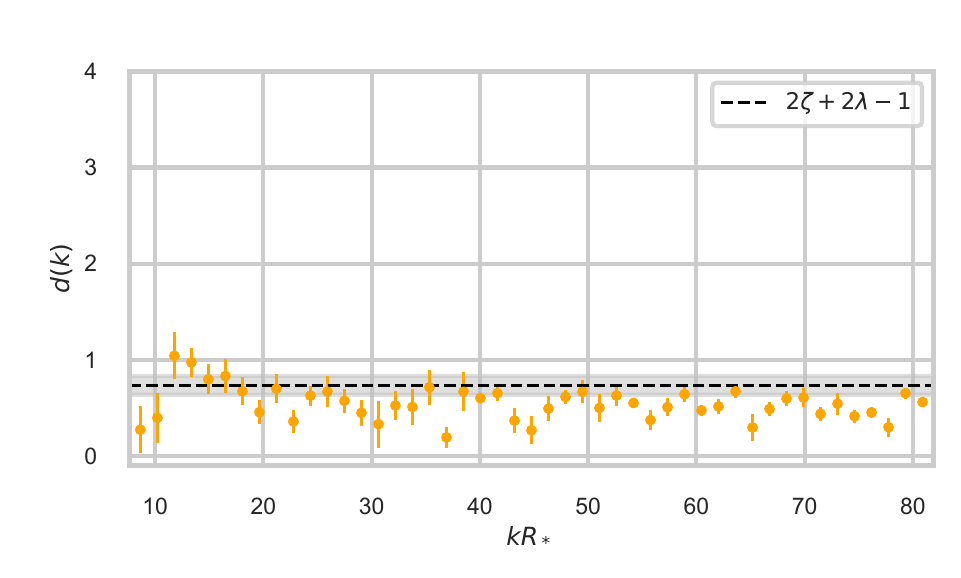}
\includegraphics[width=1.0\columnwidth]{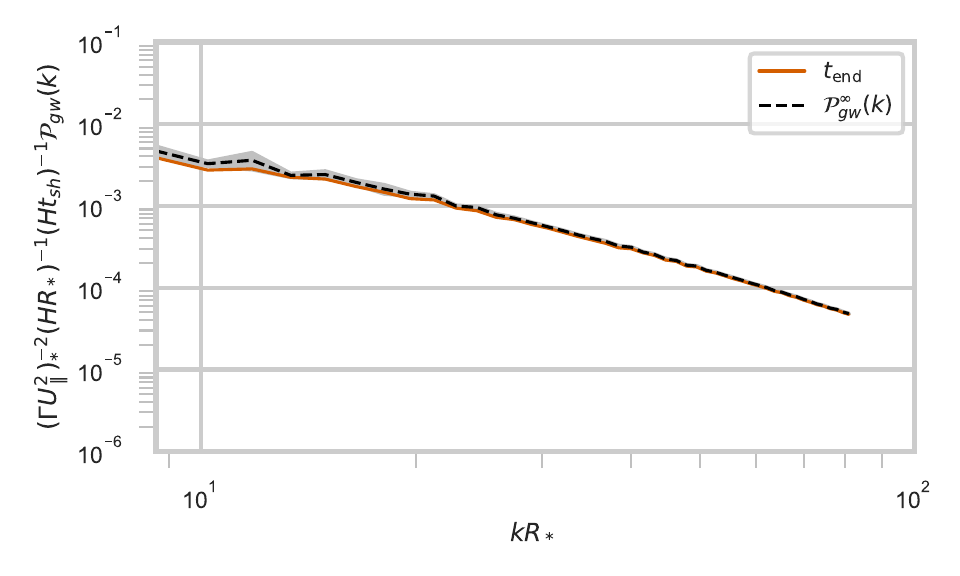}
 \includegraphics[width=1.0\columnwidth]{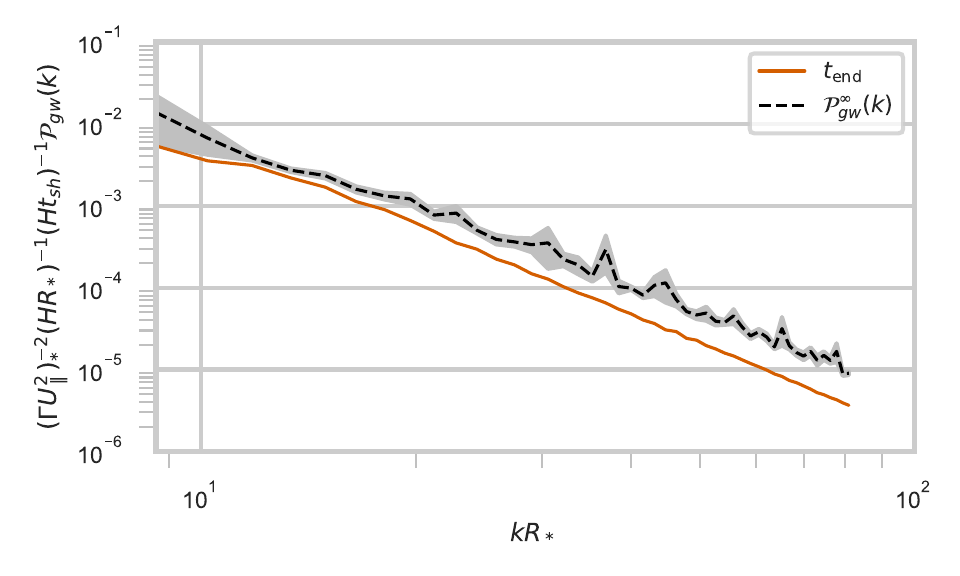}
\includegraphics[width=1.0\columnwidth]{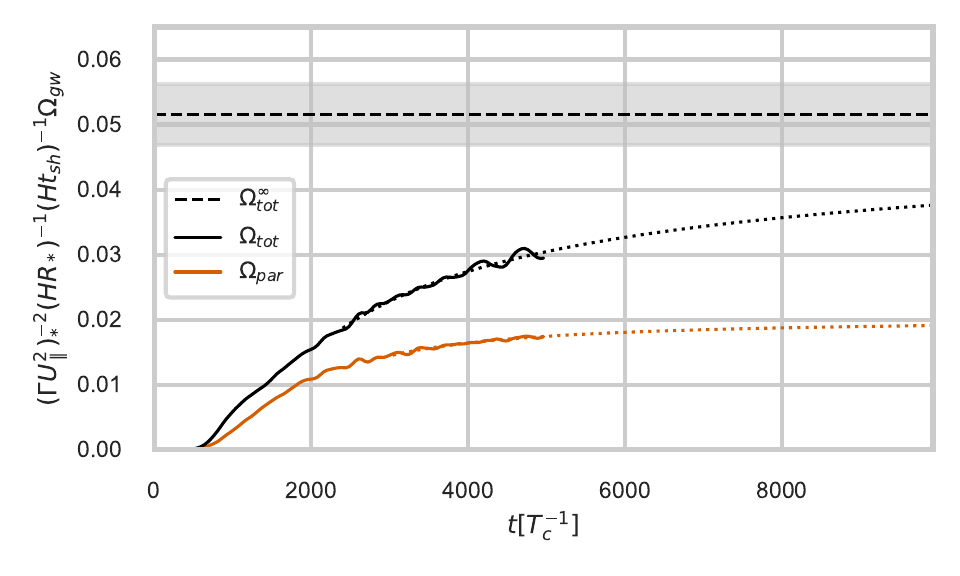}
 \includegraphics[width=1.0\columnwidth]{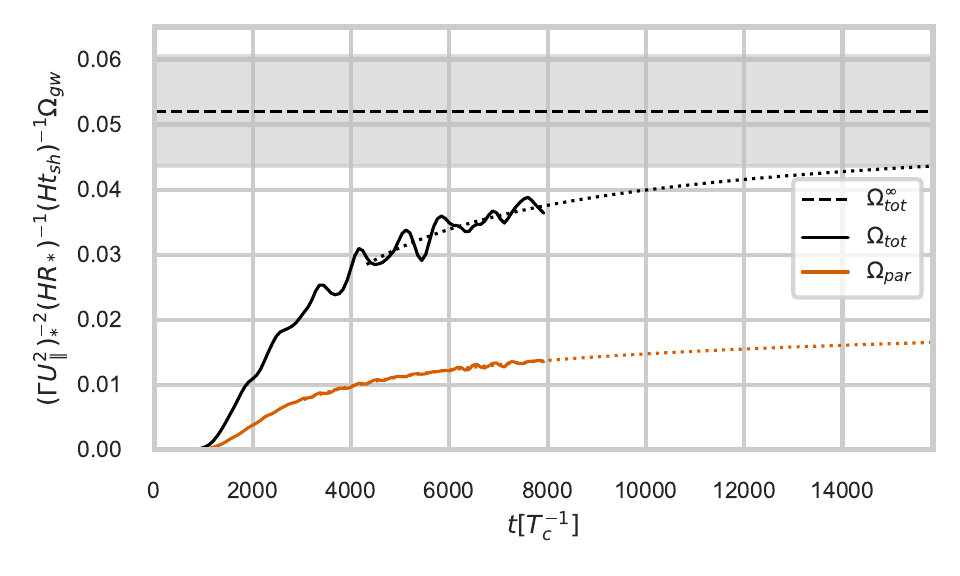}
\caption{\raggedright Extrapolations of power spectra of gravitational waves over time. Left-hand-side panels correspond to the detonation  ($\alpha_n=0.67$, $v_w=0.92$); the right-hand side ones to the deflagration ($\alpha_n=0.5$, $v_w=0.44$). Top panels show fits of $\mcPgw(k,t)$  (solid orange)  to Eq.~\eqref{e:GWPStFit} (faded orange)  over time for selected $kR_*$ in the interval $kR_* \in [8.6, 80.9]$. The fitted growth index $d(k)$ in Eq.~\eqref{e:GWPStFit} for each gravitational wave power at each individual $k$-mode in the range $kR_* \in [8,80]$ is shown in the panels on the second row. 
In the third row is the extrapolated late-time power $\mcPgw(k)$ (dashed black, with fitting uncertainties added) along with the power spectra at the last available time in the simulation (orange). In the bottom panels are shown the total (solid black) and partial (solid orange) gravitational wave power, fits (dotted lines) to the function in Eq.~\eqref{e:GWPStFit_total}, and the extrapolated total power  (dashed black, with grey uncertainty band). 
\label{fig:psGW}
}
\end{figure*}

In the same figure we also show the sound shell model prediction of each transition computed via the PTtools package (see Ref.~\cite{PTTools}) at its default precision level, with the suppression factor of Ref.~\cite{Cutting:2019zws} included. In the case of the detonation, we see the model does not accurately predict the growth rate of the gravitational wave power spectrum, signaling a breakdown in assumptions behind the model.  The deflagration however, shows very good agreement between the sound shell model and the simulation for $kR_* \leq 10^2$, provided the phenomenological supression factor from Ref.~\cite{Cutting:2019zws} is applied. This result validates the supression factor, although we note that we do not have new physical insight into its origin (its relationship with droplets was investigated in Ref.~\cite{Cutting:2022zgd}). 

To further test the sound shell model, we compared its growth rate prediction for the weak phase transitions studied in Ref.~\cite{Hindmarsh:2017gnf} ($\alpha_\text{n} = 0.0046$, and the same wall speeds) with the numerical simulations, shown in Fig.~\ref{fig:sound_shell_prace} in Appendix~\ref{ap:prace}.  The agreement at wavenumbers around the peak and above is very good, both in amplitude and shape.  At lower wavenumbers the simulations do not have enough volume or duration to be safely compared. 

We suspect therefore that the main cause of disagreement in the shear stress lies in the linearity assumption, which does not hold when fluid velocities become relativistic and enthalpy perturbations become of order one.  The difference between the linear prediction and the numerical simulation is strong evidence that the collision of the rarefaction waves behind the detonation front is a non-linear process.

To conclude this section, we examine the convergence of gravitational wave power spectrum $\mcPgw$ as $t \to \infty$. 

In Fig.~\ref{fig:psGW} (top row) we show the time evolution of selected wave numbers, along with fits to the form
\begin{equation}
\mcPgw(k,t) = \mcPgw^\infty(k)\left[ 1 - \left(\frac{t}{\tdecay(k)}\right)^{-d(k)}\right],
\label{e:GWPStFit}
\end{equation}
scaled by the factor $ (\Gamma \bar{U}_{\parallel}^2)_*^{2} (\Hn R_*)^{2}$. 
The fits are taken over the interval where the power spectrum exceeds 85\% of its final value.  A power law approach to an asymptote is expected for RMS weighted 4-velocities $\bar{U}_\parallel$ decreasing faster than $t^{-1/4}$ (see Section ~\ref{s:theory}). 

In Fig.~\ref{fig:psGW} (middle row) we show the extrapolation of the fits as $t \to \infty$ (that is, the parameter $\mcPgw^\infty$), along with a fit uncertainty estimated from the parameter covariance matrix. For low wavenumbers, where there are fewer modes in the wavenumber bin, the oscillations in the gravitational wave power are larger and the estimate more uncertain.  At higher wavenumbers ($kR_* \gg 1$), the asymptote is well determined, and the power spectrum is close to convergence.

\begin{table}
    \begin{center}
    \begin{ruledtabular}
\begin{tabular}{ccc}
Phase transition  & $d_\text{tot}$ & $2\zeta + 2\lambda-1$ \\
\hline
\rule{0pt}{3ex}{$\alpha_\text{n} = 0.67$, $v_w=0.92$} & $0.61\pm0.04$ & $2.40\pm0.31$ \\ 
{$\alpha_\text{n} = 0.50$, $v_w=0.44$}  & $0.79\pm0.10$ & $0.74\pm0.09$
\end{tabular}  \end{ruledtabular}
    \caption{\raggedright Power-law indices of the approach of the 
    total gravitational wave density to its asymptote 
    (see Eq.~\eqref{e:GWPStFit_total}), 
    compared to the prediction in terms of the kinetic energy decay index $\zeta$ and the integral scale growth index $\lambda$. Fits are obtained in the range $t \in [3t_\text{max}, t_\text{end}]$.
    \label{tab:pl_indices}}
\end{center}
\end{table}

In Fig.~\ref{fig:psGW} (bottom row) we show the growth of the partial gravitational wave power $\Omega_{par}(t)$ (obtained from integrating data over the restricted wavemode range $kR_*\in[8.6,80.9]$) along with the prediction obtained by integrating the fitted form of the power spectrum, Eq.~\eqref{e:GWPStFit}.

Here the expectation is that the power law index from kinetic energy fraction decay ($K^2 \propto t^{-2\zeta}$) and integral scale growth ($\xi \propto t^\lambda$) can be used to predict the asymptoting index of gravitational wave emission. To do so we fit the following,
\begin{equation}
\Omega_\text{gw}(t) = \Omega_\text{gw}^\infty\left[ 1 - (t/\tdecay)^{-d_\text{tot} }\right],
\label{e:GWPStFit_total}
\end{equation}
to the total emitted power to the measured $\Omega_\text{gw}$. 
From these we present the fitted parameters in Table \ref{tab:pl_indices} and the extrapolation itself in the lower panels of Fig.~\ref{fig:psGW}. 
The fit shown uses data in the range $t\in[3\tmax, \tend]$, and the uncertainties are the square root of the parameter covariances.

The theoretical expectation is that the approach indexes $2\zeta + 2\lambda -1$ should be in agreement with the fitted $d_\text{tot}$. 
The agreement is reasonably good for the deflagration, but very poor for the detonation. However, the power law approach indices of individual modes are in much better agreement with  $2\zeta + 2\lambda -1$  (see Fig.~\ref{fig:psGW}).  
We note that the total power is dominated by modes which are close to the box size in wavelength, and the power law indices of volume-averaged quantities are likely to suffer from significant finite-volume effects.

The extrapolations indicate that the simulations have reached a stage where approximately $60\%$ (detonation) or $70\%$ (deflagration) of the total power has been generated. The predicted density fraction in gravitational waves, scaled by $(\Hn \tsh)(\Hn R_*)(\Gamma \Upar^2)_*^{2}$, is given in Table~\ref{t:OmgwInf}. 

The uncertainties include the parameter covariances, the standard deviation of the best fit parameter over fit start times of $2.5\tmax$, $3\tmax$ and $3.5\tmax$, and an energy non-conservation error, added in quadrature.
The energy non-conservation error is derived from the relative change in the total energy over the course of the simulation (which should be conserved), $0.08$ for the deflagration, and $0.008$ for the detonation. The gravitational wave efficiency parameter includes a division by the square of the total energy (one factor to convert to fractional energy density, and one to produce a division by $\Hn^2$). 

\begin{table}
    \begin{ruledtabular}
\begin{tabular}{cc}
Phase transition  & $\tilde\Omega_\text{gw}^\infty$   \\
\hline
\rule{0pt}{3ex}{$\alpha_\text{n} = 0.67$, $\vw=0.92$} & $0.017 \pm 0.004$ \\ 
{$\alpha_\text{n} = 0.50$, $\vw=0.44$}  & $0.017 \pm 0.003$ \\
\end{tabular}  
\end{ruledtabular}
    \caption{\raggedright 
    Fitted asymptotic density fractions in gravitational waves, scaled with $\Hn \tsh$, the shock decay time in Hubble time units, $\Hn R_*$, the mean bubble spacing in Hubble length units, and $(\Gamma \Upar^2)_*^{2}$, the maximum compressional kinetic energy fraction squared (see Eq.~\eqref{e:OmegaGWDef}). 
        \label{t:OmgwInf}}
\end{table}

The similar magnitudes of the dimensionless constants $\tilde{\Omega}_\text{gw}^\infty$ suggests that there is an approximate efficiency factor of gravitational wave production of order $10^{-2}$ throughout the lifetime of the flow.  A factor with a similar order of magnitude was observed in the early growth rates of gravitational wave power for transitions of weaker strength ($\alpha_\text{n} = 0.0046,\, 0.05$) in Ref.~\cite{Hindmarsh:2017gnf}.  This supports the idea that, in the case $\Hn\tsh \ll 1$ studied here, a reasonable estimate of the gravitational wave power can be obtained by multiplying the initial growth rate by the shock decay time $\tsh = R_*/\bar{U}_{\parallel,*}$ \cite{Caprini:2019egz}. 
A similar efficiency factor, although about a factor 2 larger, was obtained at $\alpha_\text{n} = 0.5$ in Ref.~\cite{Caprini:2024gyk}.  
Reasons for the difference include: smaller simulations ($512^3$), a different definition of shock decay time (called the source duration $\tau_\text{sw}$ there), and a different definition of the reference kinetic energy fraction.  The simulations were also carried out for wall speeds $\vw \le 0.8$ at which the bubble solutions are deflagrations or hybrids. For these solutions, the Higgsless method misses the wall deceleration effect when compression waves start to interact, and overestimates the kinetic energy fraction.  Nonetheless, one might expect to see the same efficiency factor if one normalises with the observed kinetic energy.

\section{\label{sec:conc}Conclusions}

In this paper we have explored the evolution of the fluid flow and gravitational wave production in strong first order phase transitions. In such transitions, the effects of fluid non-linearities in the form of acoustic and vortical turbulence are important, and have been shown to substantially change the kinetic energy generated by the transition and hence the rate at which gravitational waves are produced~\cite{Cutting:2019zws}. Furthermore, the lifetime of the fluid flow, and hence the total gravitational wave production, is controlled by the non-linear dissipation of energy~\cite{Caprini:2019egz,Dahl:2021wyk,Dahl:2024eup,Caprini:2024gyk}.

We studied two cases, a detonation with wall speed $\vw=0.92$ and transition strength parameter of $\aln=0.67$, and a deflagration characterised by $\vw=0.44$ and $\aln=0.5$. These two cases provide contrasting examples of phase transitions, as they have different proportions of compressional and vortical modes in the total fluid velocity, as was seen in Ref.~\cite{Cutting:2019zws}. Here, we evolved on much larger grids, and for a sufficiently long time after the phase transitions to allow the non-linearities to affect the evolution. 

We investigated the evolution of the velocity power spectra---separating compressional and vortical components---and the shear stress spectrum, the source of the gravitational wave power spectrum.  We also measured separately, for the first time, the spatial components of the 4-velocity weighted by the enthalpy density.  The shear stress is quadratic in these components. Another first is the study of the unequal time correlations of the velocity field and the shear stress, allowing us to study the time decorrelation functions. The decorrelation functions are an important part of the modelling of the gravitational wave power spectrum, which as the key observable, we also followed.  

In the detonation, non-linearities become clear even before collisions of the sound shells: the enthalpy-weighted RMS 4-velocity is approximately 40\% higher than the RMS 3-velocity. The collisions generate vorticity, but the RMS velocity in the  vortical modes is roughly an order of magnitude smaller than that of the compressional modes.

In the deflagration, the RMS velocity is smaller, and below the naive value estimated from the sound shells, as observed in Ref.~\cite{Cutting:2019zws}.  The RMS 3-velocity and  RMS weighted 4-velocity are similar, and 
there is more equal division between compressional and vortical modes, which appears to evolve towards a constant ratio, around $1.4$ in this case, with the compressional mode dominating.
Turbulence seems to be developing here, although as the simulation does not run for much more than one eddy turn-over time, we do not see the emergence of the $k^{-2/3}$ Kolmogorov spectrum, observed in freely decaying vortical turbulence~\cite{Auclair:2022jod}.  Instead, the vortical velocity spectrum shows a broad and approximately flat ($k^0$) plateau, which is still quite different from the characteristic $k^{-1}$ of the compressional modes.  The larger length scale in the plateau is the mean bubble spacing, and we speculate that the smaller length scale is the thickness of the vortices produced by the collision of the sound shells.  

The decorrelation function of the compressional modes of wavenumber $q$ shows a decaying oscillation with angular frequency $M_\parallel\cs q$, with $M_\parallel > 1$.  In the detonation, the clear difference of the parameter $M_\parallel$ from unity indicates that the decorrelation is not due to oscillating linear sound waves.  We interpret the parameter as the Mach number of the prominent shock waves.  The shock waves are not so strong in the deflagration, and travel closer to the sound speed, so there is not such a clear distinction between shocks and sound waves.

The envelope of the compressional mode oscillations is well fitted by a Gaussian in the wave number multiplied by the time difference, with a coefficient $V_\parallel$ with dimensions of velocity.  In the deflagration, this parameter is around $0.06$, which is close to $v_\perp/\sqrt{3}$, the RMS of the vector components of the vortical flow velocity.  This would be the expectation in a model where the compressional modes are ``swept'' by the vortical modes~\cite{kraichnan:1964}.  On the other had, in the detonation, the coefficient $V_\parallel$  is twice $v_\perp/\sqrt{3}$.  It is also about half $v_\parallel/\sqrt{3}$, so the origin of the Gaussian envelope is unclear. 
 One possibility is through interactions of shocks themselves, which are non-trivial (see e.g. Refs.~\cite{courant1999supersonic,PhysRevA.45.6130,2016ApJ...823..148H}). We see $\la$-configurations, three-way junctions of shocks, are clearly visible in the slices through the simulation volume of the detonation (Fig.~\ref{fig:figureSlices}).

The decorrelation function of the subdominant vortical velocity modes of the detonation shows a novel form: instead of the purely Gaussian decorrelation observed in purely vortical flows~\cite{kraichnan:1964,Auclair:2022jod}, we see a Gaussian with an oscillatory argument.  The frequency of the oscillation of mode with wavenumber $q$ is $\cs q$, consistent with that of the sound waves.  

In the deflagration, the vortical velocity decorrelation function shows the standard Kraichnan (Gaussian) sweeping form, where the value of fit parameter $V_\perp$ coincides with the component-wise RMS of the compressional velocities $v_\parallel/\sqrt{3}$. In animations of snapshots of the velocity field, there is a strong impression of vortical modes being washed back and forth by the compressional modes, supporting a model where the  vortical modes are being swept by oscillatory compressional modes.  

The shear stress power spectrum has quite different shapes between the detonation and deflagration, as a consequence of the importance of the vortical modes in the latter. The shapes are very close to those obtained by a convolution of the weighted 4-velocity spectra, except that the overall amplitude is higher
This points to significant non-Gaussianity in the flow, particularly for the detonation.   
There, the convolution with the weighted 4-velocity gives a significant improvement in the agreement over convolution with the 3-velocity. 
The decorrelation time of the shear stress at wavenumber $k$ is controlled by the dominant fluid oscillation mode at that wavenumber: 
around $\pi/\cs k$ for flow which is mostly compressional at that wavenumber, and around $1/k\bar{U}$ for mostly vortical.  The shear stress in the deflagration is mostly vortical in the wavenumber range  $20 \lesssim kR_* \lesssim 200$.

The shapes of the gravitational wave spectra are quite similar in the two cases simulated, despite the important contribution of the vortical modes in the deflagration. They show quite a sharp peak at $kR_* \simeq 10$, with the $k^{-3}$ produced by compressional flows with shocks.  This suggests that the gravitational wave production in the deflagration case is dominated by the compressional modes.

To compare with the gravitational wave power spectrum obtained directly from simulations, in Section~\ref{s:GWpsVelps} we computed the gravitational wave power spectrum with the assumption of a Gaussian velocity field.  The computation convolves products of two velocity power spectra with kernels derived from integrating the gravitational wave Green's function with the decorrelation functions.   There are three possible contributions: pure compressional, mixed compressional-vortical, and pure vortical. The result is a prediction for the growth rate of the gravitational wave power spectrum $\Delta \mathcal{P}_\text{gw} / \Delta t$ between the two times at which the velocity field is evaluated. 
The result in both the detonation and the deflagration is that both the amplitude and the shape of the growth rate are well predicted by the compressional modes alone. This is perhaps a surprise given the importance of the vortical modes in the shear stress spectrum.  We draw the conclusion that vortical motion does not seem to source a significant contribution to gravitational wave production in first order phase transitions of strength $\alpha_\text{n} \lesssim 1$.  

We can understand the relative efficiency of acoustic production as a resonance effect: there is a range of sound wave wavenumber pairs $(\bq, \tilde{\bq})$ for which the shear stress oscillates at the same frequency $\cs(q + \tilde{q})$ as the resulting gravitational wave, $|\bq + \tilde{\bq}|$. The shear stress for a vortical mode of a given wavenumber has a spread of frequencies, of order $\Vsw{\perp} q$, which are well below resonance for the gravitational wave.

Our work has important implications for 
 theoretical modelling, for example the sound shell model of Refs.~\cite{Hindmarsh:2019phv,Hindmarsh:2016lnk}.  While considering only the compressional modes appears to be a good approximation, the assumption that sound shells do not interact is not correct.  This is already known for deflagrations, where the interaction leads to a slowing of the phase boundary, and the eventual formation of shrinking hot droplets, which do not leave much kinetic energy behind~\cite{Cutting:2019zws,Cutting:2022zgd}.  Here, a phenomenological suppression factor (from Ref.~\cite{Cutting:2019zws}), which needs further investigation, gives a reasonable approximation. The presence of non-linearities in the detonation post-collision is also suggested by the disappearance of the "knee" at $kR_* \geq 10^2$ present in the Sound Shell Model prediction \ref{fig:sound_shell}, a scale associated with the thickness of the rarefaction wave.

Theoretical modelling also needs to account for the kinetic energy decay, and further non-linear shape evolution, rather than assuming a stationary flow.  Kinetic energy decay has recently been studied numerically~\cite{Caprini:2024gyk}, and power laws close to the predicted $t^{-10/7}$ have been observed in Ref.~\cite{Dahl:2024eup}.  There is also predicted to be an associated $t^{2/7}$ evolution of the integral scale, which was not included in Ref.~\cite{Caprini:2024gyk}. Other simulations (see Ref.~\cite{Dahl:2024eup}) show that the compressional mode spectrum evolves towards a self-similar form, which can be fitted by a broken power law with asymptotic indices $k^5$ at low wave numbers and $k^{-1}$ at high wave numbers. The evolution towards this shape also needs to be investigated, as the majority of the gravitational wave emission will probably occur during the evolutionary phase.

Our results for 
the decorrelation functions also have implications 
for future modelling. 
In current implementations, 
 it is assumed that the decorrelation functions of compressional modes oscillate indefinitely with period set by sound speed, leading to a kernel which (in the limit $kR_* \to \infty$) can be approximated by a $\delta$-function. 
The new form given in Eq.~\ref{e:DParPar} improves the prediction of the growth rate of gravitational wave power spectra.
The ratio between a $M_\parallel=1.14$ and $M_\parallel$ for the $C[U_\parallel, U_\parallel]$ is contained in the range $[0.4, 2.0]$, where the lowest ratios occur for the lowest values of $kR_*$ (below order 10), while the largest are closer to $kR^*\approx 9\times10^2$.

Returning to the total gravitational wave fractional energy density, 
the theoretical expectation for flows lasting less than a Hubble time is that it should be proportional to the flow lifetime in Hubble units, the mean bubble spacing in Hubble units, and the peak kinetic energy squared~\cite{Hindmarsh:2013xza,Hindmarsh:2016lnk,Caprini:2019egz}. 
Taking the lifetime to be the shock decay time, we 
found that our extrapolated, asymptotic gravitational wave power can be characterised by a dimensionless efficiency factor (see Eq.~\ref{e:OmegaGWDef}) 
$\tilde\Omega_\text{gw}^\infty \simeq 0.017$ in both simulations, even though they have quite different kinetic energy densities.

We can translate the extrapolated total power to its present day value $\Omega_{\text{gw},0}$ 
by multiplying by the transfer function  $F_{\text{gw},0} = (3.57 \pm 0.05) \times 10^{-5} (100/g_*)^{1/3}$, where we take the FIRAS temperature $T_{\gamma, 0}$ for the Cosmic Microwave Background (Ref.~\cite{Mather:1998gm}) and the Planck best-fit value for $H_0$~\cite{Planck:2015fie}.
We find, for a phase transition where the fluid has O($100$) effective relativistic degrees of freedom,
\begin{multline} 
\Omega_{\text{gw},0}/(\Hn R_*)^2  \\
= \left\{
\begin{array}{cc}
(4.8\pm1.1)\times 10^{-8} & (\vw = 0.92, \aln = 0.67)\\
(1.3\pm0.2)\times 10^{-8} & (\vw = 0.44, \aln = 0.50)\\
\end{array}
\right. .
\end{multline}
The gravitational wave power peaks at a frequency
\begin{equation}
f_\text{p} \simeq 26\, (\Hn R_*)^{-1} (\Tn /100\,\text{GeV})\; \mu\text{Hz}.
\end{equation}

We conclude with a remark on the deflagration. As previously mentioned, due to reheating effects in front of the bubble wall, the wall speed is reduced from $\vw\approx0.4 \rightarrow$  to $\vw\approx0.1$ in the course of the simulation. Despite this reducing the observed gravitational wave power, as  explained in Ref.~\cite{Cutting:2019zws}, this may provide a way to enact baryogenesis in the latter part of the phase transition. Electroweak baryogenesis \cite{Huber:2013kj} in its original form was argued to be most efficient in transitions with low asymptotic wall speeds,  at odds with strong gravitational wave production. However, this reheating implies that electroweak baryogenesis and gravitational wave production need not be mutually exclusive: the two different regimes (prior and after reheating) could allow for a sufficiently strong (observable) gravitational wave imprint, while still allowing for baryogenesis to occur. For example, the deflagration we simulate produces a loud signal for sufficiently large $\Hn R_*$, while approximately 30\% of the universe is swept by walls with speed $\vw \approx 0.1$, which if maintained from the beginning would produce a much weaker signal. 
A detailed exploration of this effect as function of transition strength, wall speed and bubble size deserves further investigation.

\section*{Acknowledgments}
JC acknowledges useful and fruitful discussions with members of the Computational Field Theory group of Helsinki, including Lauri Niemi, Lorenzo Giombi, Deanna Hooper, Jani Dahl. We thank Jani Dahl for comments on the manuscript. All data collected for this manuscript with the exception of material in Appendix \ref{ap:prace} was produced using resources from CSC IT Center for Science via project 2008297 (codenamed GRAPE). Earlier stages of this project (with smaller grid sizes) used projects 2003695 and hy5787 on the same machine. We acknowledge PRACE for previously awarding access to the HAZEL-HEN facility in Germany at the High-Performance Computing Center Stuttgart (HLRS), where the material from Appendix \ref{ap:prace} was used. JC (ORCID ID 0000-0002-3375-0997) acknowledges support from Research Council Finland grant 354572 and ERC grant CoCoS 101142449; MH (ORCID
ID 0000-0002-9307-437X) from Academy of Finland grant 333609, Research Council of Finland grant 363676 and STFC grant ST/X000796/1; KR (ORCID
ID 0000-0003-2266-4716) from the ERC grant CoCoS
101142449 and the Research Council of Finland grant
354572; and DJW (ORCID ID 0000-0001-6986-0517) from Research Council of Finland grants 324882, 328958, 349865 and 353131.

\appendix

\section{Advection routines\label{ap:advect}}

Our implementation of one of the update operations, $\dot{W} = -\partial_i (W v^i)$, where $W$ is a state variable, such as $E$ or $Z_j$, 
supports high-order, low-order, and hybrid versions of the flux computation. 
State variables are defined as cell-centered or edge-centred.  Velocities such as the 3-velocity $v^i$ are generally face-centered. The flux $f(W)$ of a cell-centred variable through face $i$ is $W v^i$. 
The rate of change of a cell-centred state variable at site $k$ is 
\begin{equation}
\frac{dW_{[k]}}{dt} + \frac{1}{\delta x} \bigg[ f(W_{[{k+1/2}]}) - f(W_{[{k-1/2}]}) \bigg] = 0, 
\end{equation}
where $f(W_{[k+1/2]})$ is the flux of the state variable through the face in the positive $i$ direction, and  $f(W_{[k-1/2]})$ is the flux in the negative $i$ direction. 
The updates of edge-centred variables are calculated similarly, but with velocities computed as a average over 8 faces. 

The low-order version of flux computation (denoted $f^L$) corresponds to donor cell advection~\cite{DONOR}. For donor cell, if the flux is positive, $W_{[{k+1/2}]} = W_{[k]}$, or if it is negative, $W_{[{k+1/2}]} = W_{[k+1]}$.

For the higher-order scheme (where the flux is now denoted $f^H$), a function $\Phi(r_k)$ is introduced, where $r_k = (W_{[k]} - W_{[k-1]}) / (W_{[k+1]} - W_{[k]})$ is a ratio of successive gradients in direction $i$, and the flux in direction $i$ defined as 
\begin{multline}
f(W_{[k+1/2]}) \\ 
= f^L (W_{[k+1/2]}) -\Phi(r_k) \bigg( f^L(W_{[k+1/2]}) - f^H(W_{[k+1/2]}) \bigg).
\end{multline}
The function $\Phi(r_k)$ is a flux limiter.  For pure high-order advection, $\Phi$ can be set to unity independent of $r_k$, and for donor cell, $\Phi$ is set to zero. The limiter function can then be set to some function which forces the flux to be low-order when $r_k$ is near zero (sharp gradients), or non-zero in regions with smooth flow.

Besides Donor cell, we have also examined MinMod and VanLeer~\cite{VANLEER1979101} flux limiters,
\begin{equation}
\Phi(r) 
=
\begin{cases} 
\max[0, \min(1,r)], & \text{(MinMod)}, \\[10pt]
{\displaystyle \frac{r + |r|}{1 + |r|}}, &\text{(Van Leer)}. 
\end{cases}
 \label{e:FluxLimFun}
\end{equation}

\begin{figure}
\includegraphics[width=1.0\columnwidth]{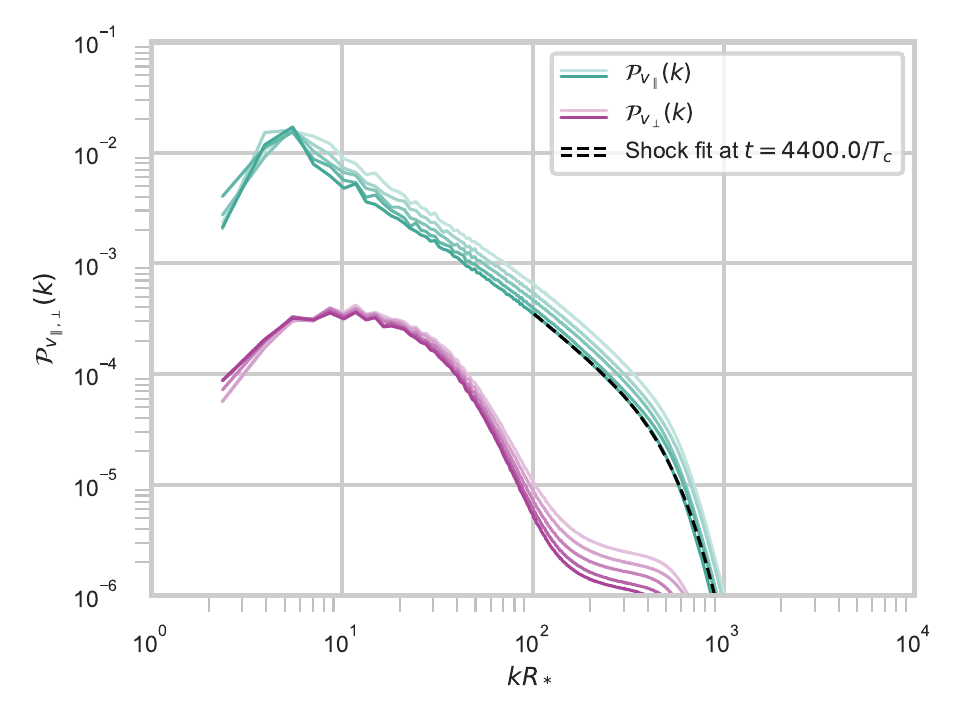}
\includegraphics[width=1.0\columnwidth]{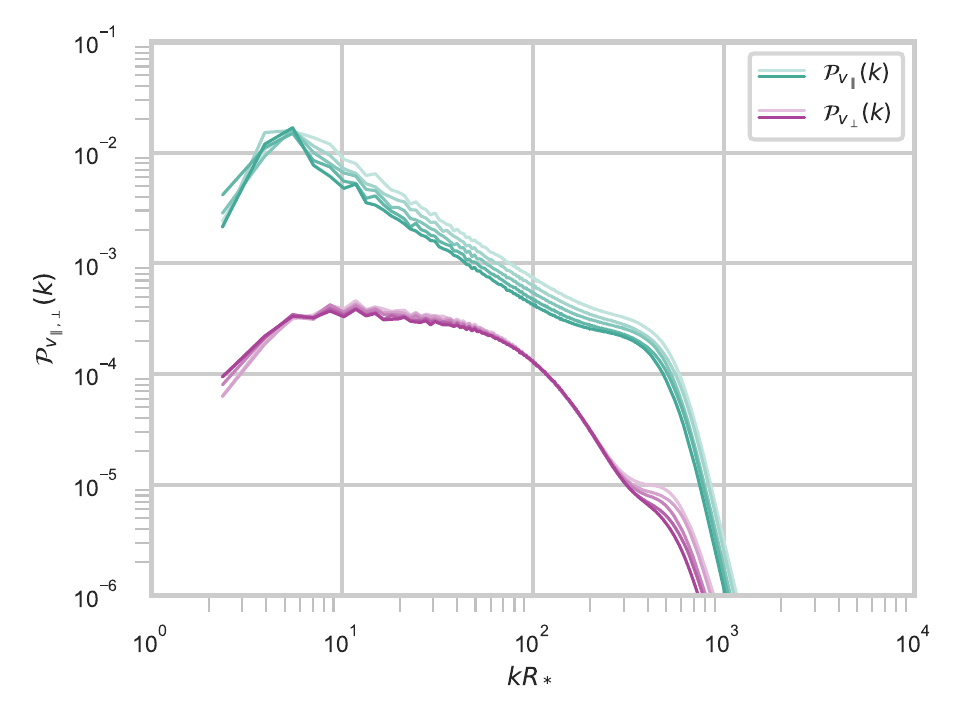}
\includegraphics[width=1.0\columnwidth]{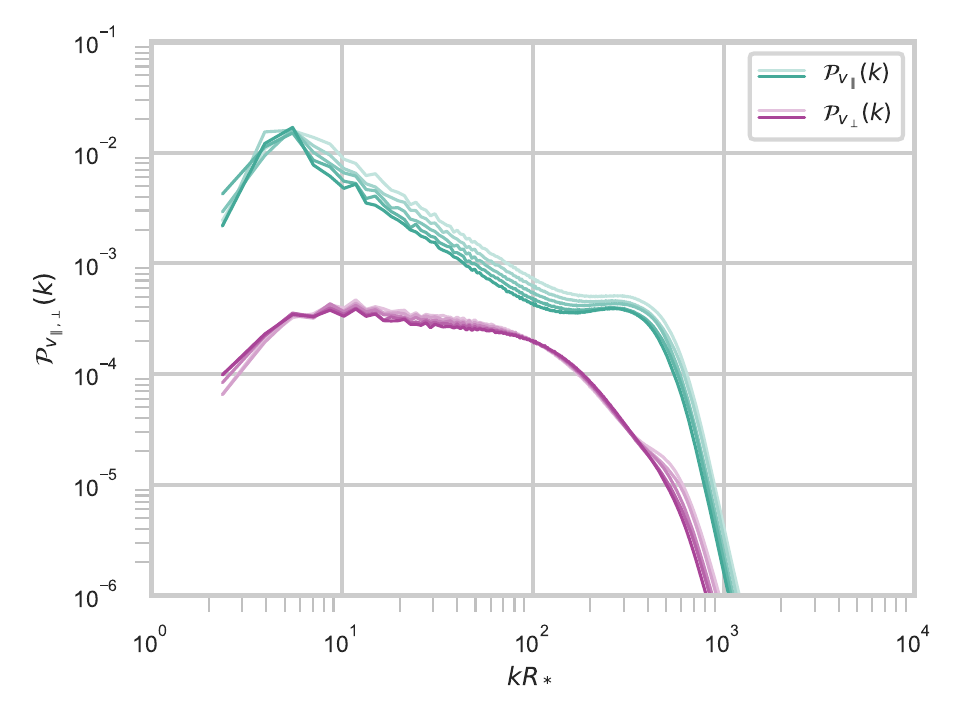}
\caption{\raggedright The compressional and vortical velocity power spectra $\mathcal{P}_{v_\parallel,\perp}$ (in shades of green and pink, respectively) for three smaller volume ($2048^3$) test runs of the detonation case ($\vw=0.92$, $\alpha_\text{n}=0.67$), with three different choice of flux limiter. From top to bottom: donor cell, MidMod, and Van Leer (see discussion around Eq.~\ref{e:FluxLimFun}).
The runs are initialised with $N_b=64$ bubbles, giving mean bubble spacing $R_* = 512$. The bubbles have the same positions, and 
the times plotted  are $t=2800, 3200, 3600, 4000, 4400 T_c^{-1}$ in all three. 
We additionally overlay a shock power spectra fit (shown in dashed black, see Eq.~\ref{e:ShockFit}) to the high-$kR_*$ region of the donor cell plot.
\label{fig:VanLeer}}
\end{figure}

In our case, we have observed that for the velocity flows set up by the phase transitions we simulated, spurious oscillations were still present in the compressional mode with either Van Leer or MinMod limiters. These were not only visible when rendering visualization slices, but also on velocity power spectra, shown in Fig.~\ref{fig:VanLeer}.

This figures show the results of test runs with the detonation simulation parameters (albeit at smaller volume $2048^3$). At roughly $kR_* \approx 2 \times 10^2$, a `bump' appears in the compressional velocity power spectra obtained with the higher-order schemes, which is not present with the low-order scheme.

The vortical component seems to lack the bump, and an approximate power law persists to higher wavenumber. For this reason Van Leer was chosen in the vortical mode simulations of Ref.~\cite{Auclair:2022jod}. In the current simulations, where compressional modes were dominant, the wish to remove unphysical oscillations behind shocks motivated the choice of donor cell advection.

To further justify our choice, and demonstrate that the downturn in the velocity power spectrum away from a power law at high wavenumber in donor cell advection has a physical explanation, we fit the analytical formula for a power spectrum dominated by shocks from Ref.~\cite{Dahl:2024eup}. The following behaviour is expected at higher wavenumbers,
\begin{equation}
\mathcal{P}_v(z) = \mathcal{A} z^{n} \mathcal{M}\bigg( \frac{\pi z}{2\kappa_s} \bigg)
\end{equation}
where $\mathcal{A}$, $n$, and $\kappa_s$ are free parameters, $z=kR_*$ and $\mathcal{M}$ is a modulating function given by
\begin{equation}
\mathcal{M}(P) = P^3 \frac{\cosh(P)}{\sinh^3(P)} .
\label{e:ShockFit}
\end{equation}
The fitting results for $kR_*\in [101.3, 1609.3]$ are presented in Fig.~\ref{fig:VanLeer} (top) for the latest velocity power spectrum. The fit itself is excellent and we report the following best-fit parameters: 
$\mathcal{A}=0.26$, $n=-1.43$ and $\kappa_s=3.33\times10^{2}$.
The parameter $\kappa_s$ gives the shock width, $\delta_s = R_*/\kappa_s \simeq 1.53/\Tc$.

\section{\label{ap:Econs} Energy evolution}

\begin{figure*}
\includegraphics[width=1.0\columnwidth]{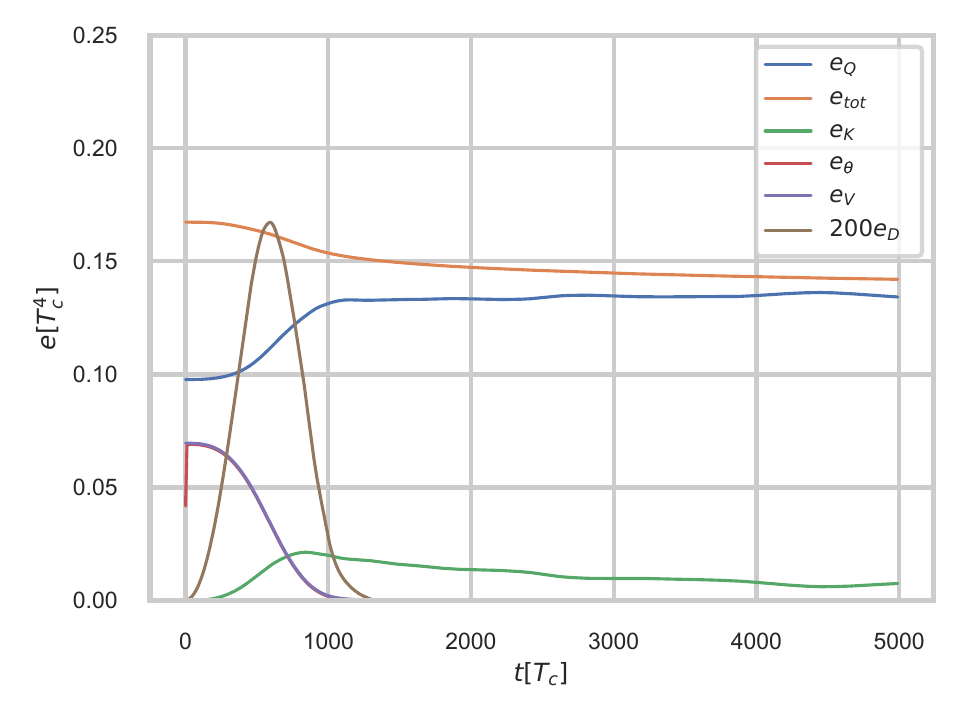}
 \includegraphics[width=1.0\columnwidth]{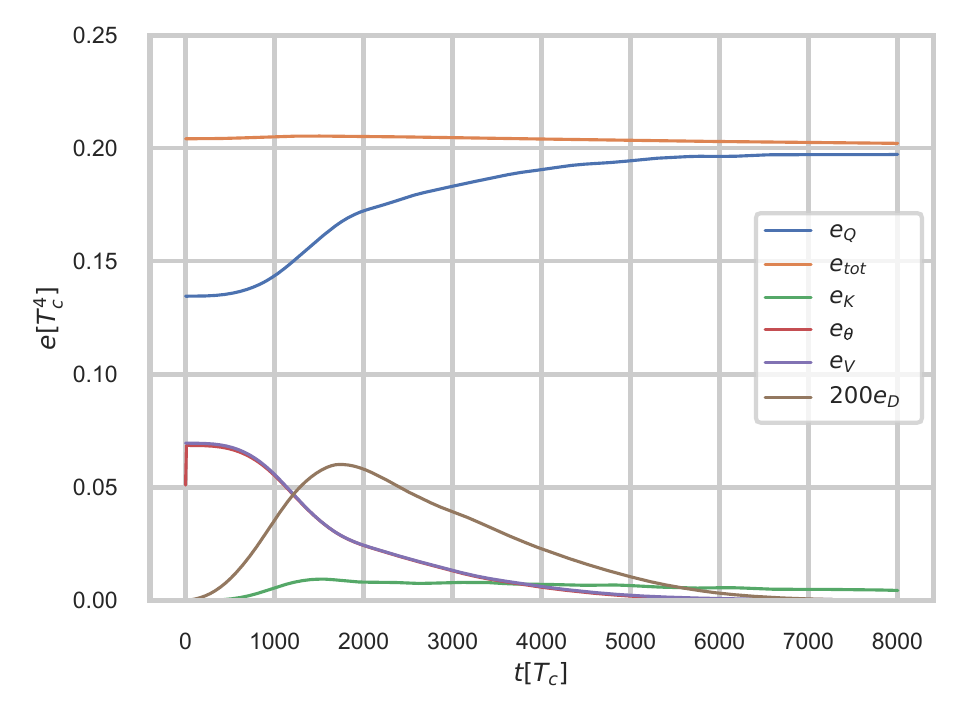}
\caption{\raggedright The behaviour of volume averaged energies in each test case, with the left panel corresponding to the detonation ($\vw=0.92$, $\alpha_\text{n}=0.67$) and the right one to the deflagration ($\vw=0.44$, $\alpha_\text{n}=0.5$). We present the total fluid-scalar energy in $e_{tot}$, the thermal energy of the fluid the $e_Q$, the fluid kinetic energy $e_K$, 
the trace energy $e_\theta$, the scalar field potential and gradient energies in $e_V$ and $e_{D}$. 
\label{fig:energy_plots}}
\end{figure*}

The time evolution of various volume averaged energy components is shown in Fig.~\ref{fig:energy_plots}. There we present fluid kinetic energy density $e_K$, the potential energy density of the scalar field $e_V$, the gradient energy density of the scalar field $e_D$, the trace anomaly of the fluid $e_\theta$, and the thermal energy density $e_Q$, defined as 
\begin{align}
e_K &= \vev{w \gamma^2 v^2 }_\bx\\
e_V &= \vev{V(\phi,T) }_\bx\\
e_D &= \half\vev{  (\nabla \phi)^2 }_\bx\\
e_\theta &= \frac{1}{4} \vev{ \ep - 3p }_\bx\\
e_Q &= \vev{ \ep - V(\phi,T)}_\bx.  
\end{align} 
where $\ep$ is the fluid rest energy, $w = \ep + p$ is the enthalpy density, and the angle brackets with the subscript $\bx$ denote a volume average. The total energy is also presented. 

As can be seen total energy conservation (of both fluid and scalar field combined) is impacted in the initial stages of the run when the bubbles 
are expanding. 
Note that the bubble wall with higher $\vw$ (the detonation) is more affected. The bubble walls undergo Lorentz contraction, 
 indicating that the energy loss is due to the lack of resolution at the faster wall speed. 
Following the completion of the transition, where the scalar field relaxes to its ground state, the energy conservation is much better.

We will also highlight the evolution of $e_D$ in the abovementioned figure. As the Gaussian blobs (from the initial conditions) relax towards the spherically symmetric bubbles, not only will the scalar field profile from inside to outside of the bubble sharpen, but also the effective surface area of bubble walls will increase over time (increase in spatial gradient energy) which is translated in the initial increase of $e_D$, until the transition places half of the box in the new phase. From this point onwards, the surface area will decrease, until the transition becomes complete, and so $e_D$ will decline to zero. We note that the slowing down of walls in the deflagration is evident in the slower drop-off of $e_D$ (compared to its initial rise).

\section{\label{ap:profile} Fluid profile from one bubble simulations}

\begin{figure*}
\includegraphics[width=1.0\columnwidth]{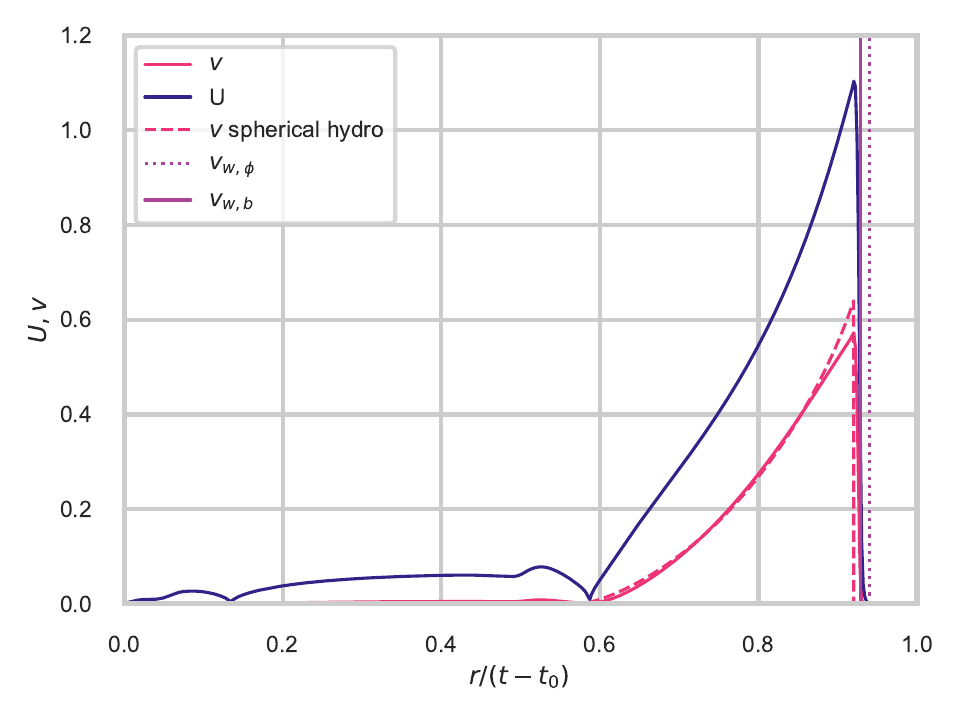}
\includegraphics[width=1.0\columnwidth]{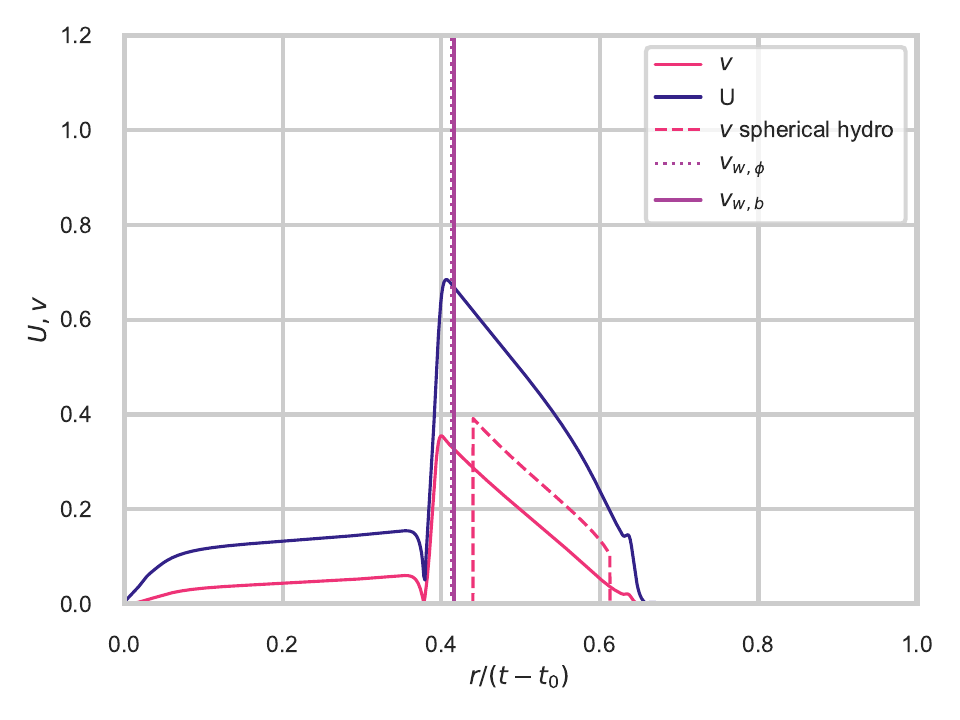}
\caption{\raggedright \label{fig:profile_plots} The 3-velocity and weighted 4-velocity fluid profiles measured from 3d one bubble simulations at $1024^3$ lattice size before the collision of the fluid shells. The times displayed correspond to $t=500 T_c^{-1}$ and for the detonation (left-hand-side panel), and $t=740 T_c^{-1}$ for the deflagration (right-hand-side panel). Note that the time variable ($r/t$) is offset by time $t_0$ spent in the initial acceleration phase (ie. $r/(t-t_0)$). For the detonation $t_0=8.46 T_c^{-1}$ and for the deflagration $t_0=32.50 T_c^{-1}$. We also plot the wall velocity estimates $v_{w,b}$ and $v_{w,\phi}$.
}
\end{figure*}

We present here the results of one-bubble 3d simulations, where a single bubble is placed in a smaller lattice ($1024^3$) and allowed to collide with itself around the periodic volume.  All other parameters are the same as the full $4096^3$ grid simulations, as is the effective $R*$ (bubble centre spacing). The resulting fluid profiles at position $(x, y=0,z=0)$ before the fluid shells interact  are shown in Fig.~\ref{fig:profile_plots}, along with the ideal hydrodynamic solutions for the expanding bubbles. These simulations indicate how close the fluid profiles are to their asymptotic form when they collide. We remark that the intial acceleration phase when the fluid relaxes towards its self-similar solution implies that computing $r/t$ using total simulation time will mildly underpredict $r/t$. As a result we use linear fitting of $r$ to $t$ to calculate the time offset $t_0$ and present the corrected measure in Fig.~\ref{fig:profile_plots}. There we can see that the detonation fluid shell is very close to its asymptotic profile, while the deflagration fluid shell is still undergoing acceleration. To further support this observation we additionally plot the wall velocity estimates ($v_{w,b}$ and $v_{w,\phi}$ defined in Section ~\ref{sec:sim_setup}) measured in the 3-dimensional one bubble simulations.

We plot both the magnitudes of the 3-velocity $v$ and the enthalpy-weighted 4-velocity $U$ \eqref{e:Udef},  The strong peak in the enthalpy in the fluid profiles is responsible for the much larger value of $U$.

\begin{widetext}

\section{Derivation of GW kernels \label{ap:gwKernels}}

Here we show how to obtain  Eqs.~\eqref{e:DPerpPerp},  \eqref{e:DParPerp}, \eqref{e:DParPar}. We start with the definition of the kernel \eqref{e:KerDef}, 
\ben
\Delta_{AB}(q, \tilde{q}, k) = 
\frac{\bar{t}}{2\bar{R}}
\int_{\rinit^{-1}}^{\rinit} \frac{dr}{r} \cos[k(t_1 - t_2)]
D_{A}(q,t_1,t_2) D_{B}(\tilde{q},t_1,t_2),
\een 
where the model decorrelation functions are given in Eqs.~(\ref{e:DecorPar}, \ref{e:DecorPerp}), $r = {t_2/t_1}$,  $\bar{t} = \sqrt{t_1t_2}$, and $\rinit = t/\tinit$. Hence 
$t_2 = \bar{t} r^{1/2}$ and $t_1 = \bar{t} r^{-1/2}$.

We then make the further transformation $s = 2 \sinh ( \ln(r)/2)$, so that $t_2 - t_1 = \bar{t}s$, to reach
\bea
\Delta_{\perp\perp}(q, \tilde{q}, k) &=& 
\frac{\bar{t}}{2\bar{R}}
\int_{-\sinit}^{\sinit} \frac{ds}{\sqrt{1 + s^2/4}} \cos(k\tbar s)
\exp\left( -\half \Vsw{\perp}^2 (q^2 + \tilde{q}^2)\tbar^2 s^2 \right) \\
\Delta_{\parallel\perp}(q, \tilde{q}, k) &=& 
\frac{\bar{t}}{2\bar{R}}
\int_{-\sinit}^{\sinit}  \frac{ds}{\sqrt{1 + s^2/4}}  \cos(k\tbar s)  \cos(\cs' q \tbar s)
\exp\left( -\half (\Vsw\parallel^2 q^2 + \Vsw\perp^2 \tilde{q}^2)\tbar^2 s^2 \right) \\
\Delta_{\parallel\parallel}(q, \tilde{q}, k) &=& 
\frac{\bar{t}}{2\bar{R}}
\int_{-\sinit}^{\sinit}  \frac{ds}{\sqrt{1 + s^2/4}}  \cos(k\tbar s)  \cos(\cs' q \tbar s)  \cos(\cs' \tilde{q} \tbar s)
\exp\left( -\half \Vsw\parallel^2 (q^2 + \tilde{q}^2)\tbar^2 s^2 \right) \\
\eea 
where $\cs' = M_\parallel \cs$ and $\sinit = 2 \sinh ( \ln(\rinit)/2)$. 
We write 
\ben
\omega_A = \Vsw{A} q, \qquad
\tilde\omega_A = \Vsw{A} \tilde{q}, \qquad
\omega_{AB} = \sqrt{\omega_A^2 + \tilde\omega_B^2 }.
\een
In the integrations, we assume that $\om_{AB} \bar{t} \gg 1$, which is the case as wavenumbers greater than about $10/R_*$ and times $\bar{t} \gtrsim R_*$. 
Then, the limits on $s$ can be taken to infinity, giving
    \begin{align}
         \Delta_{\perp \perp}(q,\tilde{q},k) &= \frac{1}{2}\sqrt{\frac{2\pi}{{\omega}_{\perp\perp}^2\bar{R}^2}} \exp\bigg( -\frac{1}{2}\frac{k^2}{{\omega}_{\perp\perp}^2}\bigg) 
         \\
        \Delta_{\parallel \perp}(q,\tilde{q},k) &= \frac{1}{4}\sqrt{\frac{2\pi}{{\omega}_{\parallel\perp}^2\bar{R}^2}} \bigg[  \exp\bigg(  -\frac{1}{2}\frac{(k-\cs' q)^2}{{\omega}_{\parallel\perp}^2} \bigg) + \exp \bigg(  -\frac{1}{2}\frac{(k+\cs' q)^2}{{\omega}_{\parallel\perp}^2} \bigg) \bigg] \; .
        \\
        \Delta_{\parallel \parallel}(q,\tilde{q},k) &= \frac{1}{8} \sqrt{\frac{2\pi}{{\omega}_{\parallel\parallel}^2\bar{R}^2}}\sum_{\pm \pm}\exp\bigg(-\frac{1}{2}\frac{(k \pm \cs' q \pm \cs' \tilde{q})^2}{{\omega}_{\parallel\parallel}^2}\bigg) 
    \end{align}
\end{widetext}

\begin{figure*}[ht]
\includegraphics[width=1.0\columnwidth]{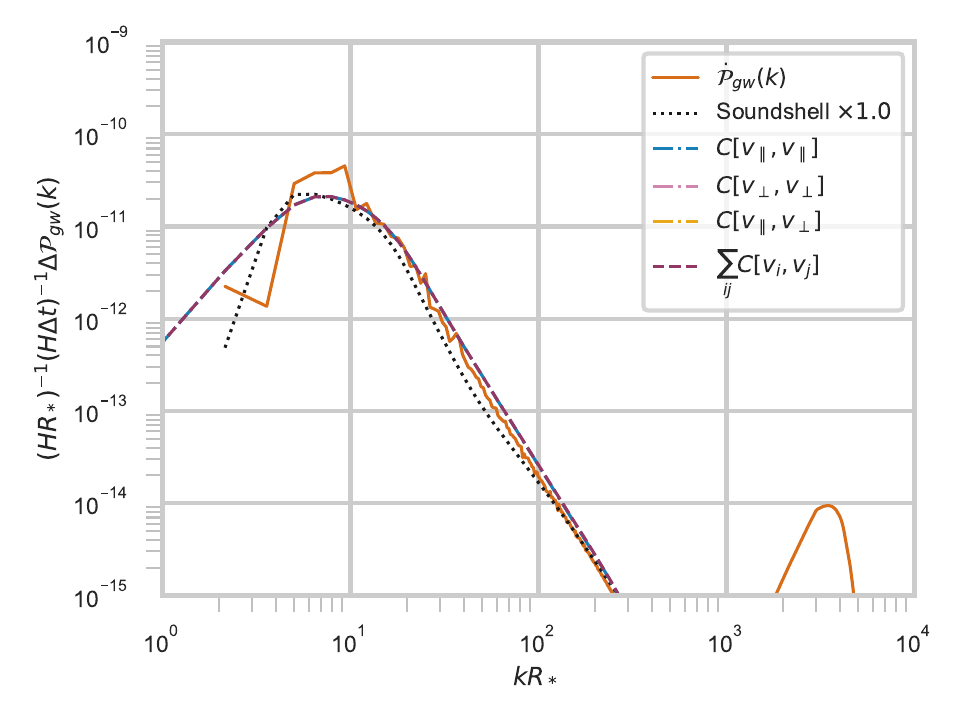}
 \includegraphics[width=1.0\columnwidth]{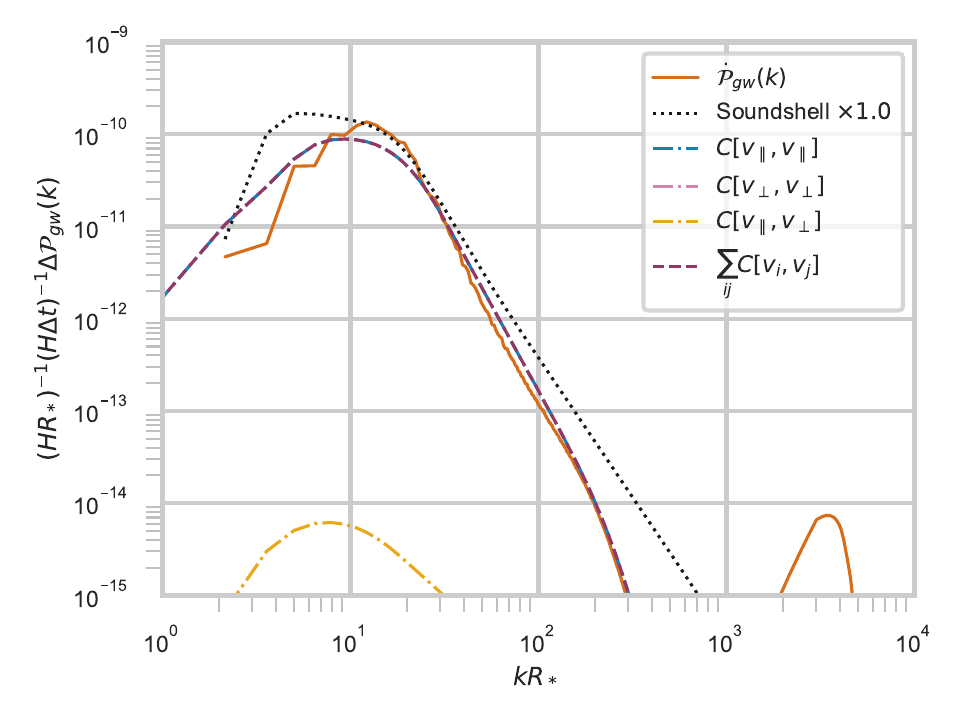}
\caption{\raggedright Comparison of numerical and theoretical gravitational wave spectra for weak phase transitions. We show the instantaneous rate of change of the gravitational wave power spectra $\dot{\mathcal{P}}_\text{gw}$ computed via the convolution of velocity power spectra of compressional modes, $C_\Delta[{v_\parallel},{v_\parallel}]$, and estimated from simulation data with a numerical difference. Both are given in dimensionless form, normalised by a factor of $\Hn^2R_*$. Left: detonation ($\alpha_\text{n}=0.0046$, $\vw=0.92$), with differences taken between times $2400/T_c$ and $2800/T_c$. Right: deflagration ($\alpha_\text{n}=0.0046$, $\vw=0.44$) using times $3800/T_c$ and $4200/T_c$. \label{fig:sound_shell_prace}}
\end{figure*}

\section{\label{ap:prace} Weak phase transitions and sound shell model prediction}

In this appendix we compare previous data for weak phase transitions ($\aln = 0.0046$) \cite{Hindmarsh:2017gnf} to the predictions from the velocity convolution formula \ref{e:dGWdt} and to the sound shell model \cite{Hindmarsh:2016lnk,Hindmarsh:2019phv}, as realised in the PTtools package (see Fig.~\ref{fig:sound_shell_prace}).  Wall speeds are $\vw=0.92$ and $\vw=0.44$. 
More precisely, we compare the growth rate of the spectrum, as computed from a numerical difference  $\Delta \mathcal{P}_\text{gw}(k)\Delta t$  between spectra at times given in the caption, scaled by $\Hn^2 R_*$. 
The sound shell model computation is carried out at the default precision, and with the simultaneous nucleation option. 
Excellent agreement is found.

\bibliographystyle{apsrev4-1.bst}
\bibliography{article}

\end{document}